\documentclass[%
 reprint,
 amsmath,amssymb,amsfonts,
 aps,
]{revtex4-1}
\usepackage{graphicx} 
\usepackage{rotating}
\usepackage{amsmath}
\usepackage{subfig}
\usepackage{bm}

\begin{document}

\newcommand{\pr}[1]{\mathrm{Pr}\left( #1 \right)}
\newcommand{\mat}[1]{\mathbf{#1}}
\title{A hyper--efficient model--independent bayesian method for the analysis of pulsar timing data.}


\author{Lindley Lentati}
\email[]{ltl21@cam.ac.uk}
\affiliation{Astrophysics Group, Cavendish Laboratory, JJ Thomson Avenue,  Cambridge, CB3 0HE, UK}
\author{P. Alexander}
\affiliation{Astrophysics Group, Cavendish Laboratory, JJ Thomson Avenue,  Cambridge, CB3 0HE, UK}
\author{M. P. Hobson}
\affiliation{Astrophysics Group, Cavendish Laboratory, JJ Thomson Avenue,  Cambridge, CB3 0HE, UK}
\author{S. Taylor}
\affiliation{Institute of Astronomy, University of Cambridge, Madingley Road, Cambridge, CB3 0HA, UK}
\author{S. T. Balan}
\affiliation{Astrophysics Group, Cavendish Laboratory, JJ Thomson Avenue,  Cambridge, CB3 0HE, UK}
\affiliation{Department of Physics and Astronomy,University College London, Gower Street, London, WC1E 6BT, UK}
\author{J. Gair}
\affiliation{Institute of Astronomy, University of Cambridge, Madingley Road, Cambridge, CB3 0HA, UK}
\author{R. van Haasteren}
\affiliation{Max-Planck-Institut fur Gravitationsphysik (Albert-Einstein-Institut), D-30167 Hannover, Germany}


\date{\today}
\begin{abstract}
A new model--independent method is presented for the analysis of pulsar timing data and the estimation of the spectral properties of an isotropic gravitational wave background (GWB).
Taking a Bayesian approach, we show that by rephrasing the likelihood we are able to eliminate the most costly aspects of computation normally associated with this type of data analysis.  When applied to the IPTA Mock Data Challenge datasets this results in speedups of approximately two to three orders of magnitude compared to established methods, in the most extreme cases reducing the run time from several hours on the high performance computer 'DARWIN' to less than a minute on a normal work station.
Due to the versatility of this approach we present three applications of the new likelihood.  In the low signal--to--noise regime we sample directly from the power spectrum coefficients of the GWB signal realisation.  In the high signal--to--noise regime, where the data can support a large number of coefficients, we sample from the joint probability density of the power spectrum coefficients for the individual pulsars and the GWB signal realisation using a `Guided Hamiltonian Sampler' to sample efficiently from this high dimensional ($\sim$ 1000) space.  Critically in both these cases we need make no assumptions about the form of the power spectrum of the GWB, or the individual pulsars.  Finally where one wishes however, we show a power-law model can still be fitted at the point of sampling.
We then apply this method to a more complex dataset designed to represent better a future IPTA or EPTA data release.  We show that even in challenging cases where the data features large jumps of the order 5 years, with observations spanning between 4 and 18 years for different pulsars and including steep red noise processes we are able to parameterise the underlying GWB signal correctly.  Finally we present a method for characterising the spatial correlation between pulsars on the sky, making no assumptions about the form of that correlation, therefore providing the only truly general Bayesian method of confirming a GWB detection from pulsar timing data.
\newline
\newline
\end{abstract}


\maketitle

\section{Introduction}

Millisecond pulsars (MSPs) have for some time been known to exhibit exceptional rotational stability, with decade long observations providing timing measurements with accuracies similar to atomic clocks (e.g. \cite{1994ApJ...428..713K,1997A&A...326..924M}).  Such stability lends itself well to the pursuit of a wide range of scientific goals, e.g. observations of the pulsar PSR B1913+16 showed a loss of energy at a rate consistent with that predicted for gravitational waves \citep{1989ApJ...345..434T}, whilst the double pulsar system PSR J0737-3039A/B has provided precise measurements of several `post Keplerian' parameters allowing for additional stringent tests of general relativity \citep{2006Sci...314...97K}.  

By measuring the arrival times (TOAs) of the radio pulses to high precision it is possible to construct a timing model: a deterministic model that describes the physical properties of the pulsar e.g. its binary period and spin evolution, its trajectory, post-Keplerian terms and so on.  A detailed description of this process is available in the Tempo2 series of papers \cite{2006MNRAS.369..655H, 2006MNRAS.372.1549E, 2009MNRAS.394.1945H}.  The timing model can then be subtracted from the TOAs resulting in a set of residuals that contain within them any physical effects not correctly accounted for by the timing model.  

In this paper we will be concerned with extracting information from these residuals that results from time-correlated stochastic signals.  These can include additional red noise terms due to rotational irregularities in the neutron star \citep{2010ApJ...725.1607S} or correlated noise between the pulsars due to a stochastic gravitational wave background (GWB) generated by, for example, coalescing black holes (e.g. \cite{2003ApJ...583..616J,2001astro.ph..8028P}) or cosmic strings (e.g. \cite{2010PhRvD..81j3523K,2010PhRvD..81j4028O,2012PhRvD..85l2003S}).  These could be detected using a pulsar timing array (PTA), a collection of Galactic millisecond pulsars from which the cross correlated signal induced by a GWB could be extracted.
Current methods for the analysis of PTA data are for the most part extremely computationally expensive.  This is particularly true for existing Bayesian methods (\cite{2009MNRAS.395.1005V,2013MNRAS.429...55V} henceforth vH2009, vHL2013) with large dense matrix inversions resulting in a scaling with the number of data points of approximately O($n^3$).  Recently new methods have been proposed to speed up this analysis.  In \cite{2013MNRAS.429...55V} (henceforth vH2013), lossy data compression is used to reduce the time these matrix inversions require, resulting in a speed up of $\sim$ 3--6 orders of magnitude over previous methods, whilst \cite{2013arXiv1302.1903E} make an approximation to the likelihood function that allows speedups proportional to the square of the number of pulsars in the array.  As with other existing Bayesian techniques, however, these methods still assume specific models for the properties of both the GWB and the intrinsic pulsar noise, a statement of prior knowledge whose validity is unknown, since as yet any GWB remains undetected.  

In this paper we present an alternative, model independent approach to performing a Bayesian analysis of PTA data that results in a speed up of between two and three orders of magnitude when compared to vHL2013, is not limited by the number of free parameters fitted or system memory, using $<$ 1GB of system memory for the analysis of the IPTA Datasets, and critically at no stage requires the specification of any prior form for the shape of the correlated power spectrum induced by a GWB, or the red noise present in a particular pulsar at the point of sampling.  This represents a true model--independent means of performing inference on the shape of the power spectrum of a gravitational wave background, where we do not know the form that background will take.  We accomplish this in two ways.  In the low signal--to--noise regime (Section \ref{Section:Marginalisation}) we sample directly from the power spectrum coefficients of the GWB signal realisation.  We show that for the IPTA data challenges, the number of coefficients required to describe the signal is roughly an order of magnitude less than the number of data points in the time domain, and so correspondingly the matrix inversions required in the likelihood are $\sim10^{3}$ times faster to compute.  

In the high signal--to--noise regime, when the number of coefficients to be sampled is larger, these matrix inversions once again become untenable, and so we sample from the joint probability density of the power spectrum coefficients for the individual pulsars and the GWB signal realisation.  This allows us to eliminate all matrix-matrix multiplications and costly matrix inversions from the likelihood calculation entirely, replacing them with matrix-vector operations and sparse, banded matrix inversions, so that this new likelihood scales as O$(n\times n_p^3)$ with the number of frequencies sampled $n$, and number of pulsars $n_p$ whilst still retaining the ability to make robust statistical inferences about the white and red noise present in the PTA data with the same precision as in vH2009/vHL2013.  We perform the sampling process in this case using a Guided Hamiltonian Sampler (GHS) (Balan, Ashdown $\&$ Hobson, in prep, henceforth B13) which provides an efficient means of sampling in large numbers of dimensions (potentially $> 10^6$).  This method of sampling, in combination with the new, simpler likelihood function, allows us to greatly extend what is computationally feasible from a Bayesian analysis of pulsar timing data.  This includes the ability to parameterise the spatial correlations between pulsars directly, without having to assume anything about the form it might take.  This spatial correlation is the `smoking gun' of a signal from a gravitational wave background, and so the ability to extract it directly from the data is crucial for the credibility of any future detections from pulsar timing data.

Finally, due to the versatility of this approach we show that where desired, models for the power spectrum of the GWB and additional red noise processes such as a single power law can still be applied at the point of sampling.  

In Sections \ref{Section:Power} and \ref{Section:Marginalisation} we derive the new likelihood functions.  In Section \ref{Section:Sampling} we describe the guided Hamiltonian sampler and how it can be applied to PTA data analysis.  In Section \ref{Section:FindCoeffs} we provide a way of estimating the number of coefficients that are supported by the data in both the low and high signal--to--noise cases.  In Section \ref{Section:Results} we apply the three different methods described thus far to the first IPTA data challenge and compare the results with both the established method described in vHL2013 and the updated method described in vH2013.  In Section \ref{Section:MySims} we then describe and analyse a set of more challenging simulated datasets designed to represent better a future IPTA data release.  Finally in Section \ref{Section:CrossCorr} we describe our method of parameterising the spatial correlation between pulsars.

This research is the result of the common effort to directly detect gravitational waves using pulsar timing, known as the European Pulsar Timing Array (EPTA) \citet{2008AIPC..983..633J} \footnote{www.epta.eu.org/}.

\section{Estimating the Power Spectrum}
\label{Section:Power}

For any pulsar we can write the TOAs for the pulses as a sum of both a deterministic and a stochastic component:

\begin{equation}
\mathbf{t}_{\mathrm{tot}} = \mathbf{t}_{\mathrm{det}} + \mathbf{t}_{\mathrm{sto}},
\end{equation}
where $\mathbf{t}_{\mathrm{tot}}$ represents the $n$ TOAs for a single pulsar, with $\mathbf{t}_{\mathrm{det}}$ and $\mathbf{t}_{\mathrm{sto}}$ the deterministic and stochastic contributions to the total respectively, where any contributions to the latter will be modelled as random Gaussian processes.  In estimating the timing model parameters for the pulsar, a standard weighted least-squares fit, as performed in packages such as Tempo2, will model the stochastic contributions purely as white noise characterised by the TOA uncertainties.  In doing so, a set of pre-fit timing residuals $\mathbf{\delta t}_{\mathrm{pre}}$ are produced using an initial estimate of the $m$ timing model parameters $\beta_{\mathrm{0}i}$ such that:
\begin{equation}
\mathbf{\delta t}_{\mathrm{pre}} = \mathbf{t}_{\mathrm{tot}} -  \mathbf{t}_{\mathrm{det}}(\bm{\beta}_{\mathrm{0}}).
\end{equation}
From here a linear approximation of the timing model can be used such that any deviations from the initial guess of the timing model parameters are encapsulated using the $m$ parameters $\epsilon_i$ such that:

\begin{equation}
\epsilon_i = \beta_i - \beta_{\mathrm{0}i}.
\end{equation}
We can therefore write the set of post--fit residuals $\mathbf{\delta t}$ that arise from this fitting process as:
\begin{equation}
\mathbf{\delta t} = \mathbf{\delta t}_{\mathrm{pre}} +  \mathrm{\bf{M}}\bm{\epsilon},
\end{equation}
where $\mathbf{M}$ is the $n\times m$ `design matrix' which describes the dependence of the timing residuals on the model parameters.
Thus any contribution to $\mathbf{t}_{\mathrm{sto}}$ not described by the TOA uncertainties, such as the signal from a GWB, will be absorbed by the timing model fit and so when the timing model is subtracted from the data, any attempt to characterise the power spectrum of the resulting post--fit residuals will be incorrect.  Whilst some methods exist to model the intrinsic red noise at the point of fitting the timing model (e.g. \cite{2011MNRAS.418..561C}, and indeed, one can use Tempo2 in conjunction with the methods described in this paper to simultaneously fit for the red noise and the non-linear timing model, this is not an approach we pursue in the following work.

In order to account for this, we instead begin by following the approach of vHL2013 which we describe in brief here so as to aid subsequent discussion.  We begin by assuming that the effect of the additional noise processes beyond the TOA uncertainties on the timing model fit will be small, so that the linear approximation will still hold even in their presence.  By refitting for the set of parameters $\bm{\epsilon}$ we can therefore write the stochastic component of the residuals as:
\begin{equation}
\mathbf{\delta t}_{\mathrm{sto}} = \mathbf{\delta t} -  \mathrm{\bf{M}}\bm{\epsilon}.
\end{equation}
We can then write the likelihood for the timing residuals as (vH2009):

\begin{eqnarray}
\mathrm{Pr}(\mathbf{\delta t} | \bm{\epsilon},\bm{\phi})& =& \frac{1}{\sqrt{(2\pi)^n\mathrm{det}\mathbf{C}}} \\
&\times& \exp{\left(-\frac{1}{2}(\mathbf{\delta t} - \mathbf{M}\bm{\epsilon})^T\mathbf{C}^{-1}(\mathbf{\delta t} - \mathbf{M}\bm{\epsilon})\right)} \nonumber
\end{eqnarray}
where the $n\times n$ covariance matrix $\mathbf{C}$ describes the stochastic contributions to the timing residuals such that

\begin{equation}
\left< \delta t_{\mathrm{sto_i}}\delta t_{\mathrm{sto_j}}\right> = C_{ij}
\end{equation}
and is described by a set of parameters $\bm{\phi}$.

We can then marginalise over all variables $\bm{\epsilon}$ in order to calculate the likelihood of a particular set of parameters $\bm{\phi}$ for the stochastic contributions to the residuals, i.e.

\begin{eqnarray}
\mathrm{Pr}(\mathbf{\delta t} | \bm{\phi}) &=& \int \; \mathrm{d}^m\bm{\epsilon} \; \mathrm{Pr}(\bm{\epsilon}) \;\mathrm{Pr}(\mathbf{\delta t} | \bm{\epsilon},\bm{\phi})
\end{eqnarray}

In vHL2013 this marginalisation is performed analytically assuming a uniform prior on $\bm{\epsilon}$ to give:

\begin{eqnarray}
\label{Eq:GEq}
\mathrm{Pr}(\mathbf{\delta t} | \bm{\phi}) &=& \frac{1}{\sqrt{(2\pi)^{(n-m)}\mathrm{det}(\mathbf{G}^T\mathbf{C}\mathbf{G})}} \nonumber \\
&\times&\exp{\left(-\frac{1}{2}\mathbf{\delta t}^T\mathbf{G}(\mathbf{G}^T\mathbf{C}\mathbf{G})^{-1}\mathbf{G}^T\mathbf{\delta t}\right)},
\end{eqnarray}
where $\mathbf{G}$ is a positive definite symmetric $n\times(n-m)$ matrix, the derivation of which will not be described here.  

For the IPTA Data Challenge, data sets consisted of 130 residuals for 36 pulsars such that $n=4680$.  $\mathbf{G}$ therefore is $\sim 4500\times4500$, and so the bottleneck in this calculation comes from the matrix inversion that must occur for every likelihood calculation, along with the set of matrix-matrix multiplications required to calculate $\mathbf{G}^T\mathbf{C}\mathbf{G}$.

Our goal is to remove this obstacle by rephrasing the likelihood such that its evaluation requires no matrix-matrix multiplications and to either eliminate the need to perform computationally intensive (i.e. O($n^3$)) dense matrix inversions, or to reduce the size of these matrices sufficiently such that their their inversion no longer dominates the evaluation time of the likelihood function, whilst retaining the ability to determine the power spectrum of the stochastic contributions to the residuals.

We do this by first writing our timing residuals $\mathbf{\delta t}$ as the sum of a signal $\mathbf{s}$ that we are interested in parameterising which will include contributions from both intrinsic red noise and the GWB signal, and some additional white noise $\mathbf{n}$ so that we have

\begin{equation}
\mathbf{\delta t} = \mathbf{s} + \mathbf{n}.
\end{equation}

We can expand $\mathbf{s}$ in terms of its Fourier coefficients $\mathbf{a}$ so that $\mathbf{s} = \mathbf{F}\mathbf{a}$ where $\mathbf{F}$ denotes the Fourier transform such that for frequency $\nu$ and time $t$ we will have both:

\begin{equation}
F_{\nu,t} = \sin\left(\frac{2\pi}{T}\nu t\right),
\end{equation}
and an equivalent cosine term.  
For a single pulsar the covariance matrix $\bm{\varphi}$ of the Fourier coefficients $\mathbf{a}$ will be diagonal, with components

\begin{equation}
\label{Eq:BPrior}
\varphi_{ij} = \left< a_ia_j^*\right> = \varphi_{i}\delta_{ij},
\end{equation}
where there is no sum over $i$, and the set of coefficients $\{\varphi_{i}\}$ represent the theoretical power spectrum for the residuals.  

Note that, whilst this equation states that the Fourier modes are orthogonal to one another, this does not mean that we assume they are orthogonal in the time domain where they are sampled, and will show explicitly later that this non-orthogonality is accounted for within the likelihood.  Instead, in Bayesian terms, Eq. \ref{Eq:BPrior} represents our prior knowledge of the power spectrum coefficients within the data.  We are therefore stating that, whilst we do not know the form the power spectrum will take, we know that the underlying Fourier modes are still orthogonal by definition, regardless of how they are sampled in the time domain.  It is here then that, should one wish to fit a specific model to the power spectrum coefficients at the point of sampling, such as a broken, or single power law, the set of coefficients $\{\varphi_{i}\}$ should be given by some function $f(\Theta)$, where we sample from the parameters $\Theta$ from which the power spectrum coefficients $\{\varphi_{i}\}$ can then be derived.

When dealing with a signal from a stochastic gravitational wave background, however, it is crucial to include the cross correlated signal between the pulsars on the sky.  We do this by using the Hellings-Downs relation \citep{1983ApJ...265L..39H}:

\begin{eqnarray}
\alpha_{mn} &= & \frac{3}{2}\frac{1-\cos(\theta_{mn})}{2}\ln\left(\frac{1-\cos(\theta_{mn})}{2}\right) \nonumber\\ 
			&-&   \frac{1}{4}\frac{1-\cos(\theta_{mn})}{2}+\frac{1}{2} + \frac{1}{2}\delta_{mn},
\end{eqnarray}
where $\theta_{mn}$ is the angle between the pulsars $m$ and $n$ on the sky and $\alpha_{mn}$ represents the expected correlation between the TOAs given an isotropic background.  With this addition our covariance matrix for the Fourier coefficients becomes
 
\begin{equation}
\label{Eq:FreqMatrix}
\varphi_{mi,nj} = \left< a_{mi}a_{nj}^*\right> = \alpha_{mn}\varphi_{i}\delta_{ij},
\end{equation}
where there is no sum over $i$, which results in a band diagonal matrix for which calculating the inverse is extremely computationally efficient.

We then write the joint probability density of the power spectrum coefficients and the signal realisation Pr$(\{\varphi_i\}, \mathbf{a} \;|\; \mathbf{\delta t})$, where here $\mathbf{a}$ refers to the concatenated vector of all coefficients $a_i$ for all pulsars, as:

\begin{equation}
\label{Eq:Prob}
\mathrm{Pr}(\{\varphi_i\}, \mathbf{a} \;|\; \mathbf{\delta t}) \; \propto \; \mathrm{Pr}(\mathbf{\delta t} | \mathbf{a}) \; \mathrm{Pr}(\mathbf{a} | \{\varphi_i\}) \; \mathrm{Pr}(\{\varphi_i\})
\end{equation}
and then marginalise over all $\mathbf{a}$ in order to find the posterior for the parameters $\{\varphi_i\}$ alone.
For our choice of $\mathrm{Pr}(\{\varphi_i\})$ we use a uniform prior in $\log_{10}$ space as the scale of the coefficients is largely unknown below some upper limit, and draw our samples from the parameter $\rho_i = \log_{10}(\varphi_i)$ instead of $\varphi_i$ which has the added advantage that we avoid unnecessary rejections due to samples which have negative coefficients in the sampling process.  Given this choice of prior the conditional distributions that make up Eq. \ref{Eq:Prob} can be written:

\begin{eqnarray}
\label{Eq:ProbTime}
\mathrm{Pr}(\mathbf{\delta t} | \mathbf{a}) \; &\propto& \; \frac{1}{\sqrt{\mathrm{det}(\mathbf{G}^T\mathbf{N}\mathbf{G})}}  \\
& \times & \exp\left[-\frac{1}{2}(\mathbf{\delta t} - \mathbf{F}\mathbf{a})^T\mathbf{G}(\mathbf{G}^T\mathbf{N}\mathbf{G})^{-1}\mathbf{G}^T(\mathbf{\delta t} - \mathbf{F}\mathbf{a})\right] \nonumber
\end{eqnarray}
where $\mathbf{N} = \left<\mathbf{n}\mathbf{n}^T\right>$ and represents the white noise errors in the residuals, which follows from Eq. \ref{Eq:GEq} with $\mathbf{N}$ replacing $\mathbf{C}$, and substituting $\mathbf{\delta t} - \mathbf{F}\mathbf{a}$ for $\mathbf{\delta t}$, and:

\begin{equation}
\label{Eq:ProbFreq}
\mathrm{Pr}(\mathbf{a} | \{\rho_i\}) \; \propto \; \frac{1}{\sqrt{\mathrm{det}\bm{\varphi}}} \exp\left[-\frac{1}{2}\mathbf{a}^{*T}\bm{\varphi}^{-1}\mathbf{a}\right].
\end{equation}
Note that we can calculate $\mathbf{G}(\mathbf{G}^T\mathbf{N}\mathbf{G})^{-1}\mathbf{G}^T$ before the sampling starts and store it in memory which eliminates the need for any dense matrix inversions, or matrix multiplications within the likelihood calculation.  

\subsection{Estimating the white noise properties}
\label{Section:White}
When dealing with realistic pulsar timing data, the properties of the white noise can be split into two components. 

\begin{description}
  \item[1] For a given pulsar, each TOA has an associated error bar, the size of which will vary across a set of observations.  We can therefore introduce an extra free parameter, an EFAC value, to account for possible mis-calibration of this radiometer noise \cite{2006MNRAS.369..655H}.  The EFAC parameter therefore acts as a multiplier for all the TOA error bars for a given pulsar, observed with a particular system. \\
 \item[2] A second white noise component, independent of the size of the error bars is also used to represent some additional source of time independent noise.  We call this parameter EQUAD.
\end{description} 

In both the IPTA data challenges, and the simulations in Section \ref{Section:MySims}, the TOAs for a given pulsar are all assigned a single value for the size of their error bars and so there is no need to include both an EFAC and EQUAD in their analysis, requiring only a single EFAC value per pulsar.  Using the likelihood in Eq. \ref{Eq:ProbTime}, despite pre-calculating the product $\mathbf{G}(\mathbf{G}^T\mathbf{N}\mathbf{G})^{-1}\mathbf{G}^T$ we are still able to make inferences about the properties of this scaling factor.  Denoting the EFAC parameter for each pulsar $p$ as $w_p$,  we can define a diagonal matrix $\mathbf{W}$ such that, if pulsar $p$ has a set of $o_p$ residuals, and a timing model described by $m_p$ model fit parameters, the first $o_1$ diagonal elements of $\mathbf{W}$ will equal $w_1$, the next $o_2$ diagonal elements will equal $w_2$ and so on, we can rewrite the product $\mathbf{G}(\mathbf{G}^T\mathbf{W}\mathbf{N}\mathbf{G})^{-1}\mathbf{G}^T$.  Exploiting the fact that the $\mathbf{G}$ are block diagonal, we can then rewrite this as:
\begin{eqnarray}
\mathbf{G}(\mathbf{G}^T\mathbf{W}\mathbf{N}\mathbf{G})^{-1}\mathbf{G}^T &=& \mathbf{G}(\mathbf{W'}\mathbf{G}^T\mathbf{N}\mathbf{G})^{-1}\mathbf{G}^T \nonumber \\
&=& \mathbf{G}\mathbf{W'}^{-1}(\mathbf{G}^T\mathbf{N}\mathbf{G})^{-1}\mathbf{G}^T \nonumber \\
&=& \mathbf{W}^{-1}\mathbf{G}(\mathbf{G}^T\mathbf{N}\mathbf{G})^{-1}\mathbf{G}^T
\end{eqnarray}
where $\mathbf{W'}$ will be a diagonal matrix where the first $(o_1 - m_1)$ entries are equal to $w_1$, the next $(o_2 - m_2)$ entries will be equal to $w_2$ and so on.  The determinant of the inverted matrix is then given by:

\begin{equation}
\mathrm{det}(\mathbf{W'}\mathbf{G}^T\mathbf{N}\mathbf{G}) = \prod_{p=1}^{N_p}w_p^{(o_p-m_p)}\mathrm{det}(\mathbf{G}^T\mathbf{N}\mathbf{G})
\end{equation}
where $N_p$ is the total number of pulsars in the dataset. Thus we can store $\mathbf{G}(\mathbf{G}^T\mathbf{N}\mathbf{G})^{-1}\mathbf{G}^T$ and the determinant $\mathrm{det}(\mathbf{G}^T\mathbf{N}\mathbf{G})$ in memory and the only additional overhead in the likelihood calculation is the calculation of $\mathrm{det}(\mathbf{W'})$ which is negligible. 

For the sake of simplifying our notation we now redefine

\begin{equation}
\tilde{\mathbf{N}}^{-1} = \mathbf{W}^{-1}\mathbf{G}(\mathbf{G}^T\mathbf{N}\mathbf{G})^{-1}\mathbf{G}^T.
\end{equation}
For more realistic data, where the size of the TOA error bars vary across an observation, and different observing systems are used such that multiple EQUAD and EFAC parameters are desired for the analysis, a slightly different approach is required.  Rather than marginalising over the timing model parameters for each pulsar analytically as in Eq. \ref{Eq:ProbTime}, we can simply perform that marginalisation process numerically and so write: 

\begin{eqnarray}
\mathrm{Pr}(\mathbf{\delta t} | \mathbf{a}, \bm{\epsilon}) &=& \frac{1}{\sqrt{(2\pi)^n\mathrm{det}\mathbf{N}}} \times \\
&\exp&{\left(-\frac{1}{2}(\mathbf{\delta t} - \mathbf{M}\bm{\epsilon} - \mathbf{F}\mathbf{a})^T\mathbf{N}^{-1}(\mathbf{\delta t} - \mathbf{M}\bm{\epsilon}-\mathbf{F}\mathbf{a})\right)}. \nonumber
\end{eqnarray}
In this way as many white noise parameters can be included as needed, however, this approach will not be pursued further in this paper given, as mentioned previously, the datasets under considering can be analysed fully using Eq.\ref{Eq:ProbTime}.

\subsection{Including additional red noise}

In order to account for uncorrelated red noise in the pulsar timing residuals we need only modify the covariance matrix $\bm{\varphi}$ in Eq.\ref{Eq:FreqMatrix} by introducing an additional set of parameters $\kappa_{p\nu}$ along the diagonal such that:

\begin{equation}
\varphi_{mi,nj} = \alpha_{mn}10^{\rho_{i}}\delta_{ij} + 10^{\kappa_{mi}}\delta_{mn}\delta_{ij}
\end{equation}
where we then marginalise over all $\kappa_{p\nu}$.

\subsection{Performing the sampling}

How we now perform the sampling depends entirely on the number of Fourier coefficients we will be using to describe the stochastic signal in the timing residuals.  As we shall see in Sections \ref{Section:Results} and \ref{Section:MySims}, even in datasets which exhibit extremely high signal to noise, the number of coefficients required to adequately describe the system is much less than the number of data--points in the time domain, often by more than an order of magnitude.  This is because practically all the power in the datasets analysed in these sections comes from only a few low frequency modes which are heavily over--sampled in the time domain.  In this situation we can marginalise over the Fourier coefficients $\mathbf{a}$ analytically and sample directly from the power spectrum coefficients $\{\bm{\rho},\bm{\kappa}\}$, a process we describe in Section \ref{Section:Marginalisation}.  Whilst this marginalised likelihood function will still include the inversion of a dense matrix, if the number of coefficients sampled is an order of magnitude less than the number of time series data--points, then the matrix to be inverted will be an order of magnitude smaller than that in Eq. \ref{Eq:GEq} and will thus take a factor 1000 less time to be inverted.

If however we wish to sample over a larger number of Fourier coefficients, to include, for example, higher frequencies where we might expect to observe gravitational wave signals from bright individual sources, then in the limit that we wish to extend our analysis to all frequencies that are Nyquist sampled in the data, the matrix to be inverted when performing the marginalisation analytically will be of the same size as that in Eq. \ref{Eq:GEq} and we will have the same computational burden as when performing the analysis in the time domain.  In this situation we can perform the marginalisation numerically, sampling directly from the high dimension, joint probability distribution described in Eq \ref{Eq:Prob}, a process made possible through the use of a GHS (B13) which we describe in the Section \ref{Section:Sampling}

\section{The Low Signal--To--Noise Regime: Analytical Marginalisation over the Fourier Coefficients}
\label{Section:Marginalisation}

In order to perform the marginalisation over the Fourier coefficients $\mathbf{a}$, we first write the log of the likelihood in Eq \ref{Eq:Prob}, which denoting  $(\mathbf{F}^T\tilde{\mathbf{N}}^{-1}\mathbf{F} + \bm{\varphi}^{-1})$ as $\bm{\Sigma}$ and $\mathbf{F}^T\tilde{\mathbf{N}}^{-1}\mathbf{\delta t}$ as $\mathbf{d}$ is given by:
\begin{equation}
\label{Eq:LogL}
\log \mathrm{L} = -\frac{1}{2} \mathbf{\delta t}^T\tilde{\mathbf{N}}^{-1} \mathbf{\delta t} - \frac{1}{2}\mathbf{a}^T\mathbf{\Sigma}\mathbf{a} + \mathbf{d}^T\mathbf{a}.
\end{equation}
Taking the derivitive of $\log \mathrm{L}$ with respect to $\mathbf{a}$ gives us:
\begin{equation}
\label{Eq:Grada}
\frac{\partial \log \mathrm{L}}{\partial \mathbf{a}} =  -\mathbf{\Sigma}\mathbf{a} + \mathbf{d}^T,
\end{equation}
which can be solved to give us the maximum likelihood vector of coefficients $\hat{\mathbf{a}}$:
\begin{equation}
\label{Eq:amax}
\hat{\mathbf{a}} = \bm{\Sigma}^{-1}\mathbf{d^T}.
\end{equation}
Re-expressing Eq. \ref{Eq:LogL} in terms of $\hat{\mathbf{a}}$:

\begin{eqnarray}
\log \mathrm{L} &=& -\frac{1}{2} \mathbf{\delta t}^T\tilde{\mathbf{N}}^{-1} \mathbf{\delta t} + \frac{1}{2}\hat{\mathbf{a}}^T\mathbf{\Sigma}\hat{\mathbf{a}} \nonumber \\
& - & \frac{1}{2}(\mathbf{a} - \hat{\mathbf{a}})^T\mathbf{\Sigma}(\mathbf{a} - \hat{\mathbf{a}}),
\end{eqnarray}
the 3rd term in this expression can then be integrated with respect to the $m$ elements in $\mathbf{a}$ to give:
\begin{eqnarray}
I &=& \int_{-\infty}^{+\infty}\mathrm{d}\mathbf{a}\exp\left[-\frac{1}{2}(\mathbf{a} - \hat{\mathbf{a}})^T\mathbf{\Sigma}(\mathbf{a} - \hat{\mathbf{a}})\right] \nonumber \\
&=& (2\pi)^m~\mathrm{det} ~ \mathbf{\Sigma}^{-\frac{1}{2}}.
\end{eqnarray}
Our marginalised probability distribution for a set of GWB coefficients is then given as:
\begin{eqnarray}
\label{Eq:Margin}
\mathrm{Pr}(\{\varphi_i\} \;|\; \mathbf{\delta t}) &\propto& \frac{\mathrm{det} \left(\mathbf{\Sigma}\right)^{-\frac{1}{2}}}{\sqrt{\mathrm{det} \left(\bm{\varphi}\right)~\mathrm{det}\left(\tilde{\mathbf{N}}\right)}} \\
&\times&\exp\left[-\frac{1}{2}\left(\mathbf{\delta t}^T\tilde{\mathbf{N}}^{-1} \mathbf{\delta t} - \mathbf{d}^T\mathbf{\Sigma}^{-1}\mathbf{d}\right)\right], \nonumber
\end{eqnarray}
where we can still pre-calculate both $\mathbf{F}^T\tilde{\mathbf{N}}^{-1}\mathbf{F}$ and $\mathbf{F}^T\tilde{\mathbf{N}}^{-1} \mathbf{\delta t}$.  

Eq. \ref{Eq:Margin} shows that the covariance matrix $\bm{\Sigma}$ both acts to whiten residuals, and fully describes the non-orthogonality in the Fourier modes due to uneven sampling in the time domain.  This, in combination with the marginalisation over the timing model parameters included in $\tilde{\mathbf{N}}$, which includes a quadratic in $t$ that describes the pulsar spin-down, and acts to project out any contribution from those frequencies lower than we can properly sample in the data means that no additional pre--whitening steps are required by this method.  Demonstrably this will be shown to have the desired effect;  even for the datasets described in the Section \ref{Section:MySims}, where we have large gaps in the data ($\sim$ 5 year gaps in a 20 year dataset) we extract the correct power spectrum.

To perform the parameter estimation with this method we will then use the MULTINEST algorithm \cite{2009MNRAS.398.1601F, 2008MNRAS.384..449F}, which will simultaneously allow us to calculate the evidence for increasing numbers of Fourier modes until a maximum is reached, and to test whether or not the data supports the inclusion of additional red noise parameters.  

For large numbers of Fourier modes, however, performing this marginalisation analytically and sampling using MULTINEST no longer remains a viable option due to both the scaling of the matrix inversions required, and the performance scaling of MULTINEST with dimensionality.  In the following section we therefore describe a method for performing this marginalisation numerically using a GHS, whilst in Section \ref{Section:FindCoeffs} we describe two possible options for estimating the evidence for different numbers of Fourier modes in order to find the optimal set.  

We note that, in principle one could also use the GHS when marginalising analytically, where the superior scaling of the GHS with dimensionality when compared to MULTINEST could allow for the inclusion of  greater numbers of power spectrum coefficients.  Ultimately however this approach is still limited by the scaling of the matrix inversions and so we do not pursue this idea further.

\section{Guided Hamiltonian Sampling}
\label{Section:Sampling} 

For a detailed account of both Hamiltonian Monte Carlo (HMC) and GHS refer to B13, or Appendix \ref{Appendix:GHS}, here we will describe only the key aspects of each.
HMC sampling \citep{1987PhLB..195..216D}  has been widely applied in Bayesian computation \cite{Neal1993}, and has been successfully applied to problems with extremely large numbers of dimensions ($\sim 10^6$ see e.g. \cite{2008MNRAS.389.1284T}).  Where conventional MCMC methods move through the
parameter space by a random walk and therefore require a prohibitive
number of samples to explore-high dimensional spaces, HMC draws
parallels between sampling and classical dynamics. By exploiting
techniques developed for describing the motion of particles in
potentials it is possible to suppress random walk
behaviour. Introducing persistent motion of the chain through the
parameter space allows HMC to maintain a reasonable efficiency even
for high-dimensional problems.

We define a `potential energy' $\Psi$ which is related to our posterior distribution Pr($\mathbf{x}$) by:

\begin{equation}
\Psi(\mathbf{x}) = -\ln(\mathrm{Pr}(\mathbf{x}))
\end{equation}
where $\mathbf{x}$ is the $N$ dimensional vector of parameters to be sampled.  Each parameter $x_i$ must be assigned a mass $m_i$ and a momentum $p_i$ so that we can write our Hamiltonian as:

\begin{equation}
H = \sum_i \frac{p_i^2}{2m_i} + \Psi(\mathbf{x}).
\end{equation}

The sampler is given a start point $\mathbf{x}$ and a set of initial momenta $\mathbf{p}$, which are drawn from a set of $N$ uncorrelated Gaussian distributions of width $m_i$ in dimension $i$.  The system can then evolve deterministically from then for some length of time $\tau$ using Hamilton's equations.

After it has reached its new position ($\mathbf{x'},\mathbf{p'}$) that point will be accepted with a probability

\begin{equation}
p = \mathrm{min}\left[1,\exp(-\delta H)\right]
\end{equation}
where $\delta H = H(\mathbf{x'},\mathbf{p'}) - H(\mathbf{x},\mathbf{p})$.  A new set of momenta can then be drawn and the process repeats. This implies that if we are able to integrate Hamilton's equations
exactly then, as energy is conserved along such a trajectory, the
probability of acceptance is unity. In practice, however, numerical
inaccuracies mean that this is not the case.

In order to perform the integration along the systems trajectory at each state we use a `leapfrog' method as is common practice.  Here $n_{\mathrm{s}}$ steps are taken of size $\lambda$ such that $n_{\mathrm{s}}\lambda = \tau$ such that:
\begin{equation}
p_i\left(t+\frac{\lambda}{2}\right) = p_i(t) - \left. \frac{\lambda}{2}\frac{\partial\Psi(\mathbf{x})}{\partial x_i} \right|_{\mathbf{x(t)}}
\end{equation}

\begin{equation}
x_i(t+\lambda) = x_i(t) + \frac{\lambda}{m_i}p_i\left(t+\frac{\lambda}{2}\right)
\end{equation}

\begin{equation}
p_i\left(t+\lambda\right) = p_i\left(t+\frac{\lambda}{2}\right) - \left. \frac{\lambda}{2}\frac{\partial\Psi(\mathbf{x})}{\partial x_i} \right|_{\mathbf{x}(t+\lambda)}
\end{equation}
until $t=\tau$ where $\tau$ is varied to avoid resonant trajectories.  HMC thus requires a large number of adjustable parameters, the mass $m_i$, step size $\lambda_i$ and the number of steps $n_{\mathrm{s}}$ in the trajectory.  Adjusting the step size or the mass produces similar effects \cite{Neal1996} and so one is usually fixed and the other tuned during sampling.

GHS is designed to eliminate much of the remaining tuning aspect by using the Hessian $\hat{\mathbf{H}}$ of the joint probability distribution calculated at its peak to set the step size $\lambda$ for each parameter.  The masses $m_i$ are then set to unity and the only tuneable parameter that remains is a global scaling parameter for the step size $\eta$ which is chosen such that the acceptance rate for the GHS is $\sim 68\%$.   

Therefore in order to perform sampling we need the following:

\begin{itemize}
\item The gradient of $\Psi$ for each parameter $x_i$
\item The peak of the joint distribution
\item The Hessian at that peak
\end{itemize}
The gradients of our parameters are given by the following:

\begin{equation}
\frac{\partial \Psi}{\partial \mathbf{a}} = -(\mathbf{\delta t} - \mathbf{F}\mathbf{a})^T\tilde{\mathbf{N}}^{-1}\mathbf{F} + \mathbf{a}^T\bm{\varphi}^{-1}
\end{equation}

\begin{equation}
\label{Eq:GradWhite}
\frac{\partial \Psi}{\partial w_i} = \frac{1}{2w_i}(o_i - m_i) - \frac{1}{w_i}(\mathbf{\delta t}_i - \mathbf{F}_i\mathbf{a}_i)^T\tilde{\mathbf{N}}^{-1}(\mathbf{\delta t}_i - \mathbf{F}_i\mathbf{a}_i )
\end{equation}

\begin{equation}
\label{Eq:rhoderiv}
\frac{\partial \Psi}{\partial \rho_i} = \frac{1}{2}\mathrm{Tr}\left(\bm{\varphi}^{-1} \frac{\partial \bm{\varphi}}{\partial \rho_i} \right) - \frac{1}{2}\mathbf{a}^T\bm{\varphi}^{-1}\frac{\partial \bm{\varphi}}{\partial \rho_i}\bm{\varphi}^{-1}\mathbf{a}
\end{equation}
and the components of the Hessian are:

\begin{equation}
\frac{\partial^2 \Psi}{\partial \mathbf{a}^2} = \mathbf{F}^T\tilde{\mathbf{N}}^{-1}\mathbf{F} + \bm{\varphi}^{-1}
\end{equation}

\begin{equation}
\frac{\partial^2 \Psi}{\partial w_i^2} = \frac{1}{w_i^2}(o_i - m_i) + \frac{2}{w_i^2}(\mathbf{\delta t}_i - \mathbf{F}_i\mathbf{a}_i)^T\tilde{\mathbf{N}}^{-1}(\mathbf{\delta t}_i - \mathbf{F}_i\mathbf{a}_i)
\end{equation}

\begin{equation}
\label{Eq:rhodderiv}
\frac{\partial^2 \Psi}{\partial \rho_i^2} = \mathbf{a}^{T}\bm{\varphi}^{-1}\frac{\partial \bm{\varphi}}{\partial \rho_i}\bm{\varphi}^{-1}\frac{\partial \bm{\varphi}}{\partial \rho_i}\bm{\varphi}^{-1}\mathbf{a}  - \frac{1}{2}\mathbf{a}^{T}\bm{\varphi}^{-1}\frac{\partial^2 \bm{\varphi}}{\partial \rho_i^2}\bm{\varphi}^{-1}\mathbf{a}
\end{equation}

\begin{equation}
\frac{\partial^2 \Psi}{\partial\rho_i\partial \mathbf{a}} = - \bm{\varphi}^{-1}\frac{\partial \bm{\varphi}}{\partial\rho_i}\bm{\varphi}^{-1}\mathbf{a}
\end{equation}

For a set of power spectrum coefficients $\{\rho_i,\kappa_i\}$ and white noise coefficients $\{\Sigma_i\}$ we can solve for the maximum set of Fourier coefficients $\mathbf{a}_{\mathrm{max}}$ analytically using Eq. \ref{Eq:amax} so when searching for the global maximum we need only search over the subset of parameters $\{\rho_i,\Sigma_i,\kappa_i\}$.  This is achieved by using either a particle swarm algorithm (\cite{Kennedy1,Kennedy2} and for uses in cosmological parameter estimation see e.g. \cite{2012PhRvD..85l3008P}, and for a description of the particle swarm method applied to PTA data in this context see \cite{2012arXiv1210.3489T}) or using a gradient search optimisation \cite{Gilbert1989}.  In the work to follow we use the former method, and take an iterative approach, passing the maximum likelihood value at the end of a search to one of the particles as a start point for the next iteration, enabling us to find the maximum using only 1 core per $\sim$ 10 free parameters.

\section{Determining the Optimal number of Fourier Modes}
\label{Section:FindCoeffs}
Whilst in the low signal--to--noise regime, sampling only small numbers of Fourier coefficients,  we are able to use MULTINEST to calculate the evidence directly and thus determine the optimal number of frequencies to describe the data by choosing the set for which the evidence is maximised, when we wish to sample greater numbers of Fourier coefficients, so the dimensionality of the problem is large, this approach is no longer computationally practical.  Whilst in principle we could ensure that we always include a sufficient number of coefficients so that our model is able to correctly describe the data simply by including all possible Fourier coefficients, this will in most cases be sub-optimal.  Therefore we would like to perform model selection between models where we include different sets of frequencies $\{\mathbf{w}\}$ prior to sampling by maximising an approximation to the evidence with respect to the set $\{\mathbf{w}\}$, and use that set for the analysis that follows.  

We do this in two ways, first by considering the Laplace approximation (e.g. \cite{Azevedo1994}) of the marginalised posterior given by Eq. \ref{Eq:Margin}, and second by considering the analytical evaluation of the evidence for an approximate likelihood function.  We then compare the results of applying these two approaches to the result calculated using MULTINEST for each of the IPTA data challenges in Section \ref{Section:Results}.

\subsection{Laplace Approximation}
\label{Section:Laplace}

Given a model with a set of $m$ maximum likelihood parameters $\bm{\hat{\rho}}$ we can approximate the likelihood around the peak using a Gaussian such that given a different set of parameters $\bm{\rho}$ we can write:

\begin{eqnarray}
\mathrm{Pr}(\mathbf{\delta t} | \bm{\rho},m)\mathrm{Pr}({\bm{\rho}},m) &\approx& \mathrm{P}(\bm{\hat{\rho}})\mathrm{Pr}({\bm{\hat{\rho}}},m)  \\
&\times&\exp\left[-\frac{1}{2}(\bm{\rho} - \bm{\hat{\rho}})^T\hat{{\mathbf{H}}}(\bm{\rho} - \bm{\hat{\rho}})\right], \nonumber
\end{eqnarray}
where $\hat{{\mathbf{H}}}$ is the hessian of the negative log likelihood evaluated at the peak as before.  This can be integrated with respect to $\bm{\rho}$  to give the Laplace approximation to the evidence given the set of model parameters $m$:
\begin{equation}
\label{Eq:Laplace}
\mathrm{Pr}(\mathbf{\delta t} | m) \propto (2\pi)^{m/2}\mathrm{det}\hat{\mathbf{H}}^{-1/2}\;\mathrm{P}(\bm{\hat{\rho}})\mathrm{Pr}({\bm{\hat{\rho}}},m)
\end{equation}
Denoting $(\mathbf{F}^T\tilde{\mathbf{N}}^{-1}\mathbf{F} + \bm{\varphi}^{-1})^{-1}$ as $\bm{\Sigma}^{-1}$ and $\mathbf{F}^T\tilde{\mathbf{N}}^{-1}\mathbf{\delta t}$ as $\mathbf{d}$ as before we can write the first derivative of $\Psi = -\log\mathrm{Pr}(\{\rho_i\} \;|\; \mathbf{\delta t})$ as:

\begin{eqnarray}
\frac{\partial \Psi}{\partial \rho_i} &=& \frac{1}{2}\mathrm{Tr}\left(\bm{\varphi}^{-1} \frac{\partial \bm{\varphi}}{\partial \rho_i} - \mathbf{\Sigma}^{-1}\bm{\varphi}^{-1}\frac{\partial \bm{\varphi}}{\partial \rho_i} \bm{\varphi}^{-1}\right) \nonumber \\
&-&\frac{1}{2}\mathbf{d}^T\mathbf{\Sigma}^{-1}\bm{\varphi}^{-1}\frac{\partial \bm{\varphi}}{\partial \rho_i}\bm{\varphi}^{-1}\mathbf{\Sigma}^{-1}\mathbf{d}.
\end{eqnarray}
In order to estimate the number of coefficients $\mathbf{\rho}$ to be used, we then assume that all the signal in the data for the set of $N_p$ pulsars is the result of a GWB so that this simplifies slightly to:
\begin{eqnarray}
\frac{\partial \Psi}{\partial \rho_i} &=& \frac{1}{2}\log(10)N_p - \frac{1}{2}\mathrm{Tr}\left(\mathbf{\Sigma}^{-1}\bm{\varphi}^{-1}\frac{\partial \bm{\varphi}}{\partial \rho_i} \bm{\varphi}^{-1}\right) \nonumber \\
&-&\frac{1}{2}\mathbf{d}^T\mathbf{\Sigma}^{-1}\bm{\varphi}^{-1}\frac{\partial \bm{\varphi}}{\partial \rho_i}\bm{\varphi}^{-1}\mathbf{\Sigma}^{-1}\mathbf{d}.
\end{eqnarray}
Writing $\mathbf{\bar{d}}^T=\mathbf{d}^T\mathbf{\Sigma}^{-1}\bm{\varphi}^{-1}$ our hessian is therefore given by:
\begin{eqnarray}
\label{Eq:Laplace1}
\frac{\partial^2 \Psi}{\partial \rho_i^2} &=& \frac{1}{2}\mathrm{Tr}\left(-\mathbf{\Sigma}^{-1}\bm{\varphi}^{-1}\frac{\partial \bm{\varphi}}{\partial \rho_i}\bm{\varphi}^{-1}\mathbf{\Sigma}^{-1}\bm{\varphi}^{-1}\frac{\partial \bm{\varphi}}{\partial \rho_i}\bm{\varphi}^{-1}\right. \nonumber \\
&+&\left. \mathbf{\Sigma}^{-1}\bm{\varphi}^{-1}\frac{\partial^2 \bm{\varphi}}{\partial \rho_i^2}\bm{\varphi}^{-1}\right) \nonumber \\
&-&\mathbf{\bar{d}}^T\frac{\partial \bm{\varphi}}{\partial \rho_i}\bm{\varphi}^{-1}\mathbf{\Sigma}^{-1}\bm{\varphi}^{-1}\frac{\partial \bm{\varphi}}{\partial \rho_i}\mathbf{\bar{d}} \nonumber \\
&+&\mathbf{\bar{d}}^T\frac{\partial \bm{\varphi}}{\partial \rho_i}\bm{\varphi}^{-1}\frac{\partial \bm{\varphi}}{\partial \rho_i}\mathbf{\bar{d}} - \frac{1}{2}\mathbf{\bar{d}}^T\frac{\partial^2 \bm{\varphi}}{\partial \rho_i^2}\mathbf{\bar{d}}
\end{eqnarray}

\begin{eqnarray}
\label{Eq:Laplace2}
\frac{\partial^2 \Psi}{\partial \rho_i\partial\rho_j} &=& \frac{1}{2}\mathrm{Tr}\left(-\mathbf{\Sigma}^{-1}\bm{\varphi}^{-1}\frac{\partial \bm{\varphi}}{\partial \rho_i}\bm{\varphi}^{-1}\mathbf{\Sigma}^{-1}\bm{\varphi}^{-1}\frac{\partial \bm{\varphi}}{\partial \rho_j}\bm{\varphi}^{-1}\right) \nonumber \\
&-&\mathbf{\bar{d}}^T\frac{\partial \bm{\varphi}}{\partial \rho_i}\bm{\varphi}^{-1}\mathbf{\Sigma}^{-1}\bm{\varphi}^{-1}\frac{\partial \bm{\varphi}}{\partial \rho_j}\mathbf{\bar{d}} 
\end{eqnarray}
We can thus use Eqns \ref{Eq:Laplace1} and \ref{Eq:Laplace2} to evaluate expression \ref{Eq:Laplace} and approximate the evidence.  Whilst this calculation requires that we calculate the maximum likelihood values for incremental numbers of parameters $m$, we believe that in any practical dataset, this will still prove less costly than performing the analysis using the full set of Fourier coefficients present in the data.

\subsection{Approximating the Likelihood}
\label{Section:PaperApproxLike}

We now take a second alternate approach to the subject of model selection, by considering a simpler problem for which we can calculate the evidence directly.  We begin with a simple example where for some time series data $\mathbf{d}$ of length $N$ with uniform white noise we would like to determine the number of basis functions that the data can support as derived in \citet{Bretthorst1988}.  We include the complete derivation of the results given in this section in Appendix \ref{Section:Appendix2}, however below we include only a brief outline.

\subsubsection{Uniform White Noise}

Suppose we have a single realisation of some time series data $\mathbf{d}$ of length $N$.  We then define a set of hypotheses $\{H\}$ such that each $H_i$ purports that our data $\mathbf{d}$ is described by some function $f_i$ where:

\begin{equation}
f_i(t) = \sum_{k=1}^m b_kM_k(t,\mathbf{w})
\end{equation} 
with $M_k$ a set of general basis functions.   The number of functions $m$, the parameters that describe them (e.g. their frequencies) $\mathbf{w}$, and the model coefficients $b_k$ are allowed to vary for each $f_i$.  
We then transform this set of basis functions into an orthonormal set $F_k$ through the transformation:
\begin{equation}
F_k(t) = \frac{1}{\sqrt{\lambda_k}}\sum_{j=1}^me_{kj}M_j(t)
\end{equation}
where $e_{kj}$ is the $k$th element of the $j$th eigenvector and $\lambda_k$ is the $k$th eigenvalue of the covariance matrix $\mathbf{M}^\mathrm{T}\mathbf{M}$. 
Our function $f_i$ can now be written in terms of these new basis vectors:
\begin{equation}
f_i(t) = \sum_{k=1}^m a_kF_k(t,\mathbf{w})
\end{equation} 
where the coefficients $a$ in the orthonormal basis are related to the coefficients $b$ in the original basis through:
\begin{equation}
b_k=\sum_{j=1}^m\frac{a_ke_{jk}}{\sqrt{\lambda_j}}
\end{equation}
The probability of the data given a model $f_i$, assuming that the noise is described by a zero mean random Gaussian process with variance $\sigma$, is given by:

\begin{equation}
\label{Eq:PaperDataprob}
\mathrm{Pr}(\mathbf{d} | \mathbf{a}, \mathbf{w},\sigma,f_i) = (2\pi\sigma^2)^{-N/2}\exp\left[\frac{1}{2\sigma^2}\sum_{k=1}^N\left[d_k - f_i(t_k)\right]^2\right].
\end{equation}
We begin by integrating over both the set of coefficients $\mathbf{a}$ and frequencies $\mathbf{w}$.  We assume that the two parameters are logically independent, in so far as we can write the priors:

\begin{equation}
\mathrm{Pr}(\mathbf{a},\mathbf{w}) = \mathrm{Pr}(\mathbf{a})\mathrm{Pr}(\mathbf{w}) 
\end{equation}
For the amplitude coefficients, we choose an uninformative Gaussian prior given by:
\begin{equation}
\label{Eq:PaperGaussPrior}
\mathrm{Pr}(\mathbf{a} | \delta) = (2\pi\delta^2)^{-m/2}\exp\left[-\sum_{k=1}^m\frac{a_k^2}{2\delta^2}\right]
\end{equation}
with $\delta >> \sigma$.  For our frequencies, we consider that for any given model $f_i$ we are selecting a set of frequencies chosen from an evenly spaced grid.  Therefore we will have a delta function prior for each frequency $w_j$ in the set $\mathbf{w}$ and thus arrive at the expression:
\begin{eqnarray}
\label{Eq:PaperFirstProb}
\mathrm{Pr}(\mathbf{d} | \delta,\sigma,f_i) &=& (2\pi\delta^2)^{-m/2} (2\pi\sigma^2)^{-(N-m)/2}  \nonumber \\
&\times & \exp\left[\frac{\mathbf{d}^2 - \mathbf{h(w_i)}^2}{2\sigma^2}\right] \exp\left[\frac{\mathbf{h(w_i)}^2}{2\delta^2}\right].
\end{eqnarray}
We are now in a position to integrate over our unknown variances $\sigma$ and $\delta$.   As in \citet{Bretthorst1988} we set an upper bound $H$ and lower bound $L$ to this integral, which will therefore be of the form:
\begin{equation}
\frac{1}{\log(H/L)}\int^H_L \mathrm{d}s\frac{s^{-a}\exp\left[-\frac{Q}{s^2}\right]}{s}
\end{equation}
making a substitution $u = Q/s^2$ this becomes:
\begin{equation}
\frac{Q^{-a/2}}{2\log(H/L)}\int^{Q/L^2}_{Q/H^2} \mathrm{d}u\;u^{a/2-1}\exp\left[-u\right]
\end{equation}
If we assume that $H$ is sufficiently large, and $L$ is sufficiently small that we may write $Q/H^2 << 1$ and $a/2 -1 << Q/L^2$ then the integral will evaluate to approximately $\Gamma(a/2)$.  Therefore we can finally write the probability of the data $D$ given a model $f_i$ as:
\begin{eqnarray}
\mathrm{Pr}(\mathbf{d} | f_i) &=&  \frac{\Gamma(m/2)}{2\log(R_{\delta})}\left[\frac{\mathbf{h(w)^2}}{2}\right]^{-m/2}\nonumber \\
&\times &\frac{\Gamma((N-m)/2)}{2\log(R_{\sigma})}\left[\frac{\mathbf{d^2} - \mathbf{h(w)^2}}{2}\right]^{-(N-m)/2}.
\end{eqnarray}

\subsubsection{Non-Uniform White Noise}

In general when dealing with pulsar residuals the white noise level across a dataset for a single pulsar will vary with time, where for example different instruments have been used to collect data for the same pulsar.  In this case the expansion of our likelihood function is not so simple, because the covariance matrix $\mathbf{G}^T\mathbf{N}\mathbf{G}$ will no longer reduce to a diagonal matrix.  If we define $\mathbf{C}=\mathbf{G}^T\mathbf{N}\mathbf{G}$ where we consider $\mathbf{C}$ to be a general dense covariance matrix, Eq. \ref{Eq:PaperDataprob} will take the form:
\begin{eqnarray}
\mathrm{Pr}(\mathbf{d} | \mathbf{a}, \mathbf{w},f_i) &=& (2\pi)^{-N/2}|\mathbf{C}|^{-1/2} \nonumber \\
&\times &\exp\left[\frac{-1}{2}(\mathbf{d} - \mathbf{F}\mathbf{a})^T\mathbf{C}^{-1}(\mathbf{d} - \mathbf{F}\mathbf{a})\right].
\end{eqnarray}
As in Section \ref{Section:White}, we would like to fit for a global scaling factor that modifies the overall noise level in the dataset. I.e. we would like to write $\mathbf{C'}=\mathbf{G}^T(\alpha^2\mathbf{N})\mathbf{G}$ where $\alpha$ is a constant to be determined.  Taking the same priors as the uniform noise case described previously, and following a similar process to integrate over the Fourier coefficients $\bm{a}$, frequencies $\bm{w}$,  and variances $\alpha$ and $\delta$ we arrive at the final probability for a set of $m$ functions $f_i$:
\begin{eqnarray}
\mathrm{Pr}(\mathbf{d}| f_i) &=& \frac{\Gamma(m/2)}{2\log(R_{\delta})}\left[\frac{1}{2}\sum_{k=1}^m\left(\frac{\mathbf{d}^T\mathbf{C'}^{-1}\mathbf{F_i}}{\mathbf{F_i}^T\mathbf{C'}^{-1}\mathbf{F_i}}\right)^2\right]^{-m/2} \nonumber \\
&\times& \frac{\Gamma((N-m)/2)}{2\log(R_{\alpha})}\left[-\frac{1}{2}\left(\mathbf{d}^T\mathbf{\bar{C}}^{-1}\mathbf{d}\right)\right]^{-(N-m)/2}
\end{eqnarray}
where we have defined:
\begin{equation}
\mathbf{\bar{C}}^{-1}=\mathbf{C'}^{-1} - \mathbf{C'}^{-1} \mathbf{F}( \mathbf{F}^T \mathbf{C'}^{-1} \mathbf{F})^{-1} \mathbf{F}^T \mathbf{C'}^{-1}.
\end{equation}

\section{The IPTA Data Challenge}
\label{Section:Results}

We will now apply the three methods discussed thus far to the first IPTA data challenge.  Henceforth we will refer to the numerical marginalisation using the GHS as method (A), the analytical marginalisation using MULTINEST as method (B), and the approach of fitting directly for a model power spectrum, where we use a power law model of the form $\mathrm{P}(f) = Af^{-\gamma}$ as method (C).  Each of these methods will therefore be sampling a different number of parameters, which for clarity we outline explicitly below:

\begin{description}
  \item[Method (A)] \hfill \\
  With the exception of closed dataset 3, we are simultaneously parameterising the white noise for each pulsar ($N_p$ dimensions), a set of $n$ GWB coefficients and ($N_p\times n\times2$) Fourier coefficients.  For closed dataset 3 we also include an additional set of ($N_p\times n$) coefficients to allow for red noise parameterisation such that we allow different pulsars to have different red noise spectra. \\
  \item[Method (B)] \hfill \\
  For method (B) we are parameterising a set of $n$ GWB coefficients only, with the exception of closed 3 where we include an additional $n$ parameters to describe the average red noise across the pulsars, where we assume the dataset has used a single power spectrum model for all pulsar realisations as in the open 3 dataset. In all cases we assume the level of the white noise in the dataset is consistent with that given for the TOAs in the data files.\\
    \item[Method (C)] \hfill \\
	For method (C) we directly parameterise the slope and amplitude of the gravitational wave signal in the data using a power law model of the form $\mathrm{P}(f) = Af^{-\gamma}$, resulting in only 2 dimensions per dataset.  Once more with the exception of closed dataset 3 where we include an additional 2 parameters to describe the average amplitude and slope of the red noise properties in the data. In all cases we assume the level of the white noise in the dataset is consistent with that given for the TOAs in the data files.\\
\end{description}

In total there are 6 datasets in the IPTA data challenge, three of which comprise the `Open' challenge, where the properties of the injected signals are known prior to analysis and three which make up the `Closed' challenge, where at the time of analysis the details were unknown.  We will outline the properties of these datasets below. 

Where present in the data, the injected GWB power spectrum has a characteristic strain spectrum given by:

\begin{equation}
h_c(f) = A_g\left(\frac{f}{1\mathrm{yr}^{-1}}\right)^{\alpha},
\end{equation}
with $A_g$ a dimensionless amplitude at a frequency of $(\mathrm{yr}^{-1})$ and $\alpha$ a power law index.  Parameterising the spectral density as in vHL2013:

\begin{equation}
S(f) = A^2\left(\frac{1}{1\mathrm{yr}^{-1}}\right)\left(\frac{f}{1\mathrm{yr}^{-1}}\right)^{-\gamma},
\end{equation}
the strain spectrum will result in an observed spectral density within the residuals of:
\begin{equation}
S(f) = \frac{A_g^2}{12\pi^2}1\mathrm{yr}^3\left(\frac{f}{1\mathrm{yr}^{-1}}\right)^{-\gamma},
\end{equation}
where in both instances $\gamma=2\alpha-3$.  The parameters of the open and closed datasets are listed below.

\begin{description}
  \item[Open Challenge 1] \hfill \\
  36 Pulsars with 130 observations each evenly sampled in time. Each dataset has white noise with an amplitude of $10^{-7}$s and an injected GWB signal with $A_g = 5\times10^{-14}$ and $\gamma=13/3$. \\
  
  \item[Open Challenge 2] \hfill \\
  As open challenge 1, but the sampling in the time domain is no longer even, and the amplitude of the white noise varies between different pulsars in the range $\sim 10^{-8} \to 10^{-6}\mathrm{s}$ \\
  
  \item[Open Challenge 3] \hfill \\
  As open challenge 2, but now $A_g = 10^{-14}$, and there is additional red noise signal present in each dataset of the form $\mathrm{P}(f) = Af^{-\gamma_{red}}$ where $A=5.77\times10^{-22}\mathrm{seconds}^{1.3}$ and $\gamma_{\mathrm{red}}=1.7$ \\
  
    \item[Closed Challenge 1] \hfill \\
  As open challenge 1, with the injected GWB signal parameters changed to $A_g = 1\times10^{-14}$ and $\gamma=13/3$. \\
  
  \item[Closed Challenge 2] \hfill \\
  As open challenge 2, with the injected GWB signal parameters changed to $A_g = 6\times10^{-14}$ and $\gamma=13/3$. \\
  
  \item[Closed Challenge 3] \hfill \\
  As open challenge 3, but now $A_g = 5\times10^{-15}$, and the red noise signal present in each dataset is given by $A=3.66\times10^{-18}\mathrm{seconds}^{1.8}$ and $\gamma_{\mathrm{red}}=1.2$ \\
\end{description}

In analysing the data we choose a fundamental frequency $f_0$ to be equal to $1/T_{max}$, where $T_{max}$ represents the greatest observing time span for any of the pulsars in the dataset.  Defining $f_n = n/T_{max}$, we then fit for the coefficients corresponding to some set of $\{n\}$ Fourier modes. 

\begin{figure*}[h]
\begin{minipage}{168mm}
\begin{center}$
\begin{array}{ccc}
\includegraphics[width=55mm]{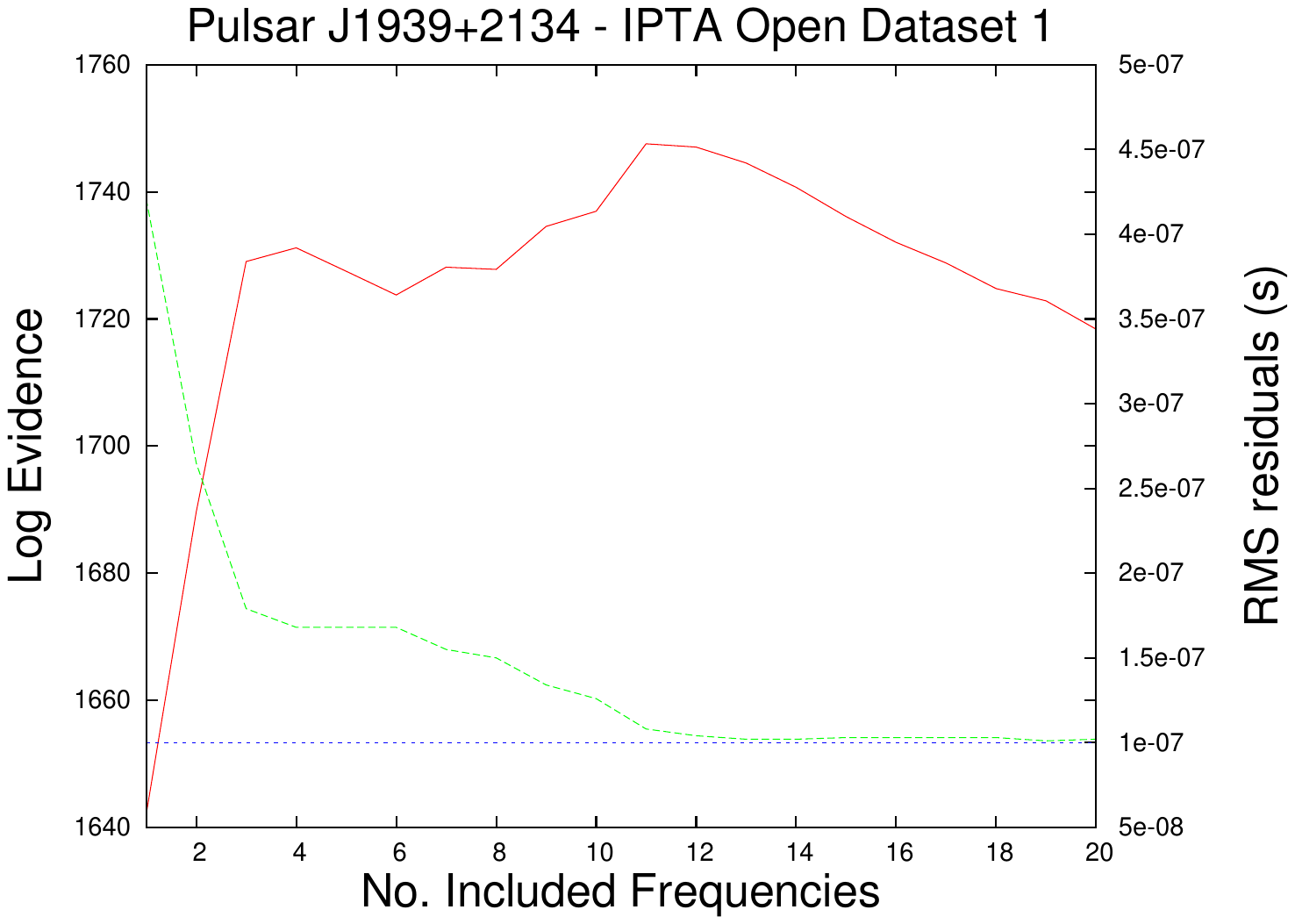} &
\includegraphics[width=55mm]{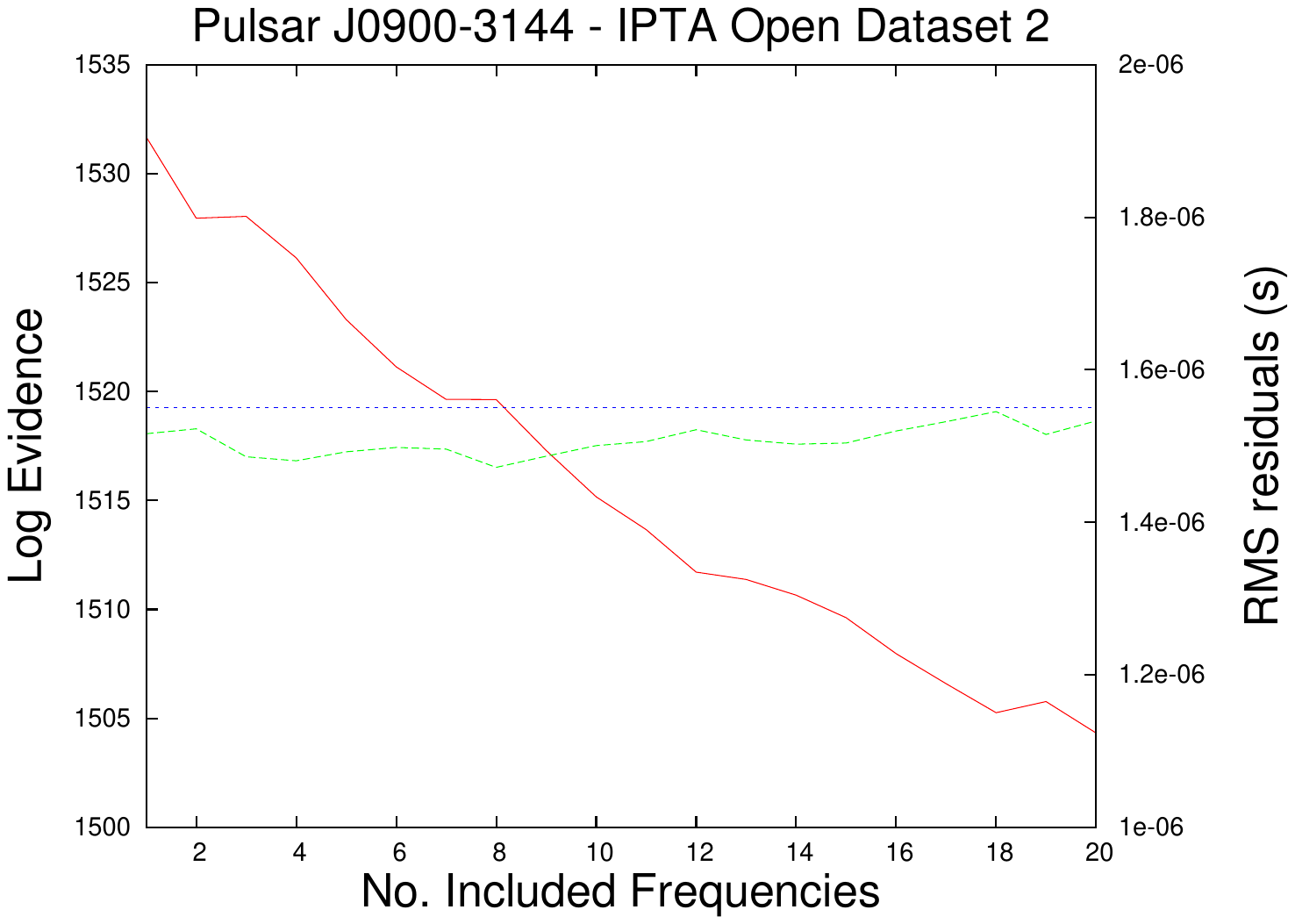} &
\includegraphics[width=55mm]{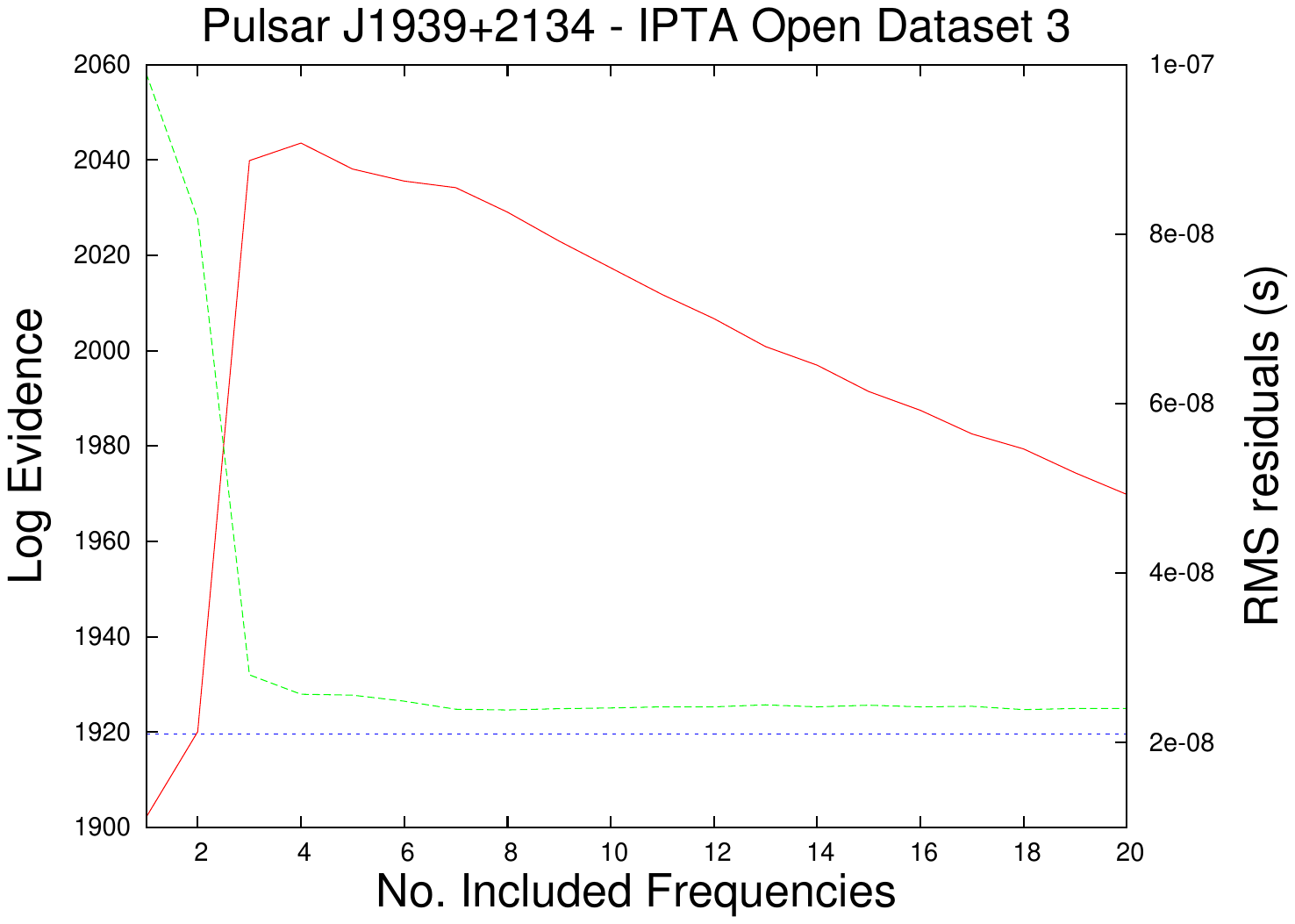}
\end{array}$
\end{center}
\caption{Calculated using the analytical approximation to the likelihood described in Section \ref{Section:PaperApproxLike} we plot the evidence (Red solid line) for models with different numbers of frequency modes, and the RMS residuals (green dashed line) compared with the injected value (blue dotted line) for those models.  Examples are given for open dataset 1 (left), 2 (middle) and 3 (right) where the evidence is maximised for 11, 1 and 4 frequencies for each respectively.}
\label{figure:RMSEvidence}
\end{minipage}
\end{figure*}

In order to determine the optimal set of Fourier modes to include for each dataset for method (A) we use both the Laplace approximation, and analytic approximation methods described in Sections \ref{Section:Laplace} and \ref{Section:PaperApproxLike} respectively.  
Fig.\ref{figure:RMSEvidence} shows an example of the analytic approximation applied to one pulsar from each of the three open datasets.  The red line shows how the evidence changes as the number of frequencies in the model increases, whilst the blue dotted, and green dashed lines show the injected level, and the best estimate of the rms amplitude for the white noise in the data for each model, where the latter is calculated using the expression in Bretthorst (1988) as:
\begin{equation}
\left<\sigma^2\right> = \frac{1}{N - m - 2}(\mathbf{d}^2 - \mathbf{h}^2).
\end{equation}
In all 3 cases the evidence can be seen to reach its maximum when the change in the estimated rms amplitude no longer justifies an increase in the number of model parameters.  Since we wish to include all relevant frequencies, we therefore choose the maximum number of frequencies supported by any single pulsar as the set of frequencies to sample for the GWB. 

The values for these approaches are given in Table \ref{Table:numcoeff} for the three open, and three closed IPTA challenge datasets where those datasets for which the evidence supported the inclusion of additional red noise are marked with an $(r)$.

A comparison of the three methods shows that whilst the analytical estimate performs well in four of the six datasets, for both closed 2 and open 3 there is a marked underestimate in the optimal number of coefficients suggested.  The change in the log evidence calculated using MULTINEST going from 13 to 17 coefficients in closed dataset 2 is $\Delta \log E = 13$ whilst going from 6 to 9 coefficients in open dataset 3 resulted in an increase of $\Delta \log E = 7$ both representing significant losses of information for not including the additional coefficients.  Whilst the analytical approximation to the likelihood would likely hold in the case where the signal is dominated by uncorrelated red noise in the individual pulsars here we see that the additional information gained through the coherence between pulsars is enough to warrant additional Fourier coefficients in the analysis.
In comparison the Laplace approximation agrees well with the results found using MULTINEST in all six datasets.  For the later simulations we will therefore take this approach, however for the IPTA datasets all the results in the following section are derived using the number of Fourier modes found to be optimal via the numerical analysis using MULTINEST.

\begin{table}
\centering
\caption{Number of Frequencies Supported by the Evidence for the IPTA Data Challenges } 
\centering 
\begin{tabular}{c c c c} 
\hline\hline 
Dataset &  \multicolumn{3}{c}{Optimal Number of Frequencies}\\[0.5ex] 
	      &		Laplace & Analytic & MULTINEST 				\\
\hline 
Open 1 &11 & 9&9 \\
Open 2 &15 & 12 &11\\
Open 3 & 9& 6 & 9 \\
Closed 1 &6 & 5 & 6\\
Closed 2 & 17& 13 & 17\\
Closed 3 &9  & 8 & 8 $(r)$\\
\hline
\end{tabular}
\label{Table:numcoeff} 
\end{table}

\subsection{Results}

\begin{table*}
\centering
\caption{IPTA Data Challenge Results} 
\centering 
\hspace*{-1cm}\begin{tabular}{c c c c c c c c c c c} 
\hline\hline 
Dataset & \multicolumn{2}{c}{This Paper (A)} &\multicolumn{2}{c}{This Paper (B)} & \multicolumn{2}{c}{This Paper (C)} & \multicolumn{2}{c}{vHL2013} &\multicolumn{2}{c}{Injected Values}  \\[0.5ex] 
	     &	  $A_g\times 10^{-14}$  &  $\gamma$	&	  $A_g \times 10^{-14}$  &  $\gamma$    	    &$A_g \times 10^{-14}$  &  $\gamma$    	    &	 $A_g \times 10^{-14}$	     &  	$\gamma$	&	 $A_g \times 10^{-14}$	     &  	$\gamma$\\
\hline 
Open 1 &$5.1 \pm 0.2$& $4.34 \pm 0.10$	&$4.62 \pm 0.19 $	& $4.30 \pm 0.08$ &	$4.6\pm0.2$&$4.32\pm0.09$			&$4.82 \pm 0.18 $ & $4.4 \pm 0.08$&	$5 $ &4.333\\
Open 2 & $5.2 \pm 0.3$	& $4.36 \pm 0.12$	& $5.1 \pm 0.3$	& $4.36 \pm 0.11$ & $5.4\pm0.3$		&	$4.29\pm0.12$		&$5.5 \pm 0.3$	&$4.30 \pm 0.09$ &	$5 $ &4.333\\
Open 3 & $1.08 \pm 0.12 $	& $4.2 \pm 0.2$& $1.08 \pm 0.12$	& $4.17 \pm 0.2$   &	$1.09\pm0.13$		&$4.13\pm0.20$			&$1.17 \pm 0.13$	&$4.13 \pm 0.19$ &	$1$ &4.333\\
Closed 1 &$1.07 \pm 0.05$& $4.2 \pm 0.2$&$1.12 \pm 0.13$	& $4.36 \pm 0.08$	      &$1.07\pm0.11$		&$4.25\pm0.19$			&$1.11 \pm 0.09 $ & $4.31 \pm 0.15$&	1 &4.333\\
Closed 2 & $5.6 \pm 0.3 $	& $4.40 \pm 0.12$& $5.6 \pm 0.3$	& $4.36 \pm 0.08$  & 	$5.59\pm0.28$		&	$4.4\pm0.11$		&$6.32 \pm 0.15$	&$4.27 \pm 0.05$ &	6 &4.333\\
Closed 3 & $0.32 \pm 0.09$	& $4.5 \pm 0.4$	& $0.32 \pm 0.09 $	& $4.2 \pm 0.3$&$0.44\pm0.08$	&	$4.0\pm0.3$		&$0.5 \pm 0.16 $	&$4.2 \pm 0.4$ &	0.5 &4.333\\
\hline
\end{tabular}
\label{Table:OpenResults} 
\end{table*}

\begin{figure*}

\begin{minipage}{.5\linewidth}
\centering
\subfloat[]{\label{main:a}\includegraphics[scale=.35]{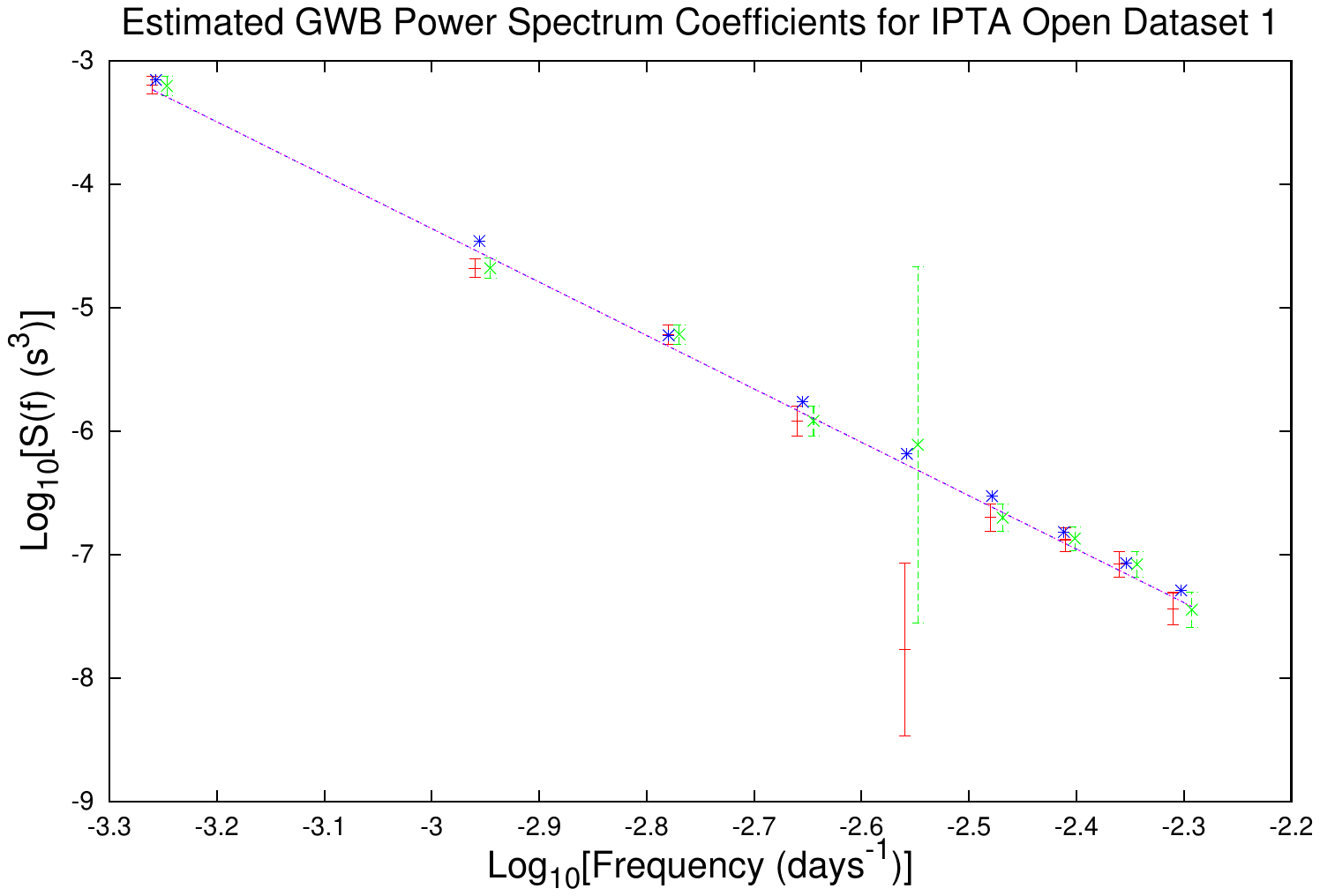}}
\end{minipage}%
\begin{minipage}{.5\linewidth}
\centering
\subfloat[]{\label{main:b}\includegraphics[scale=.35]{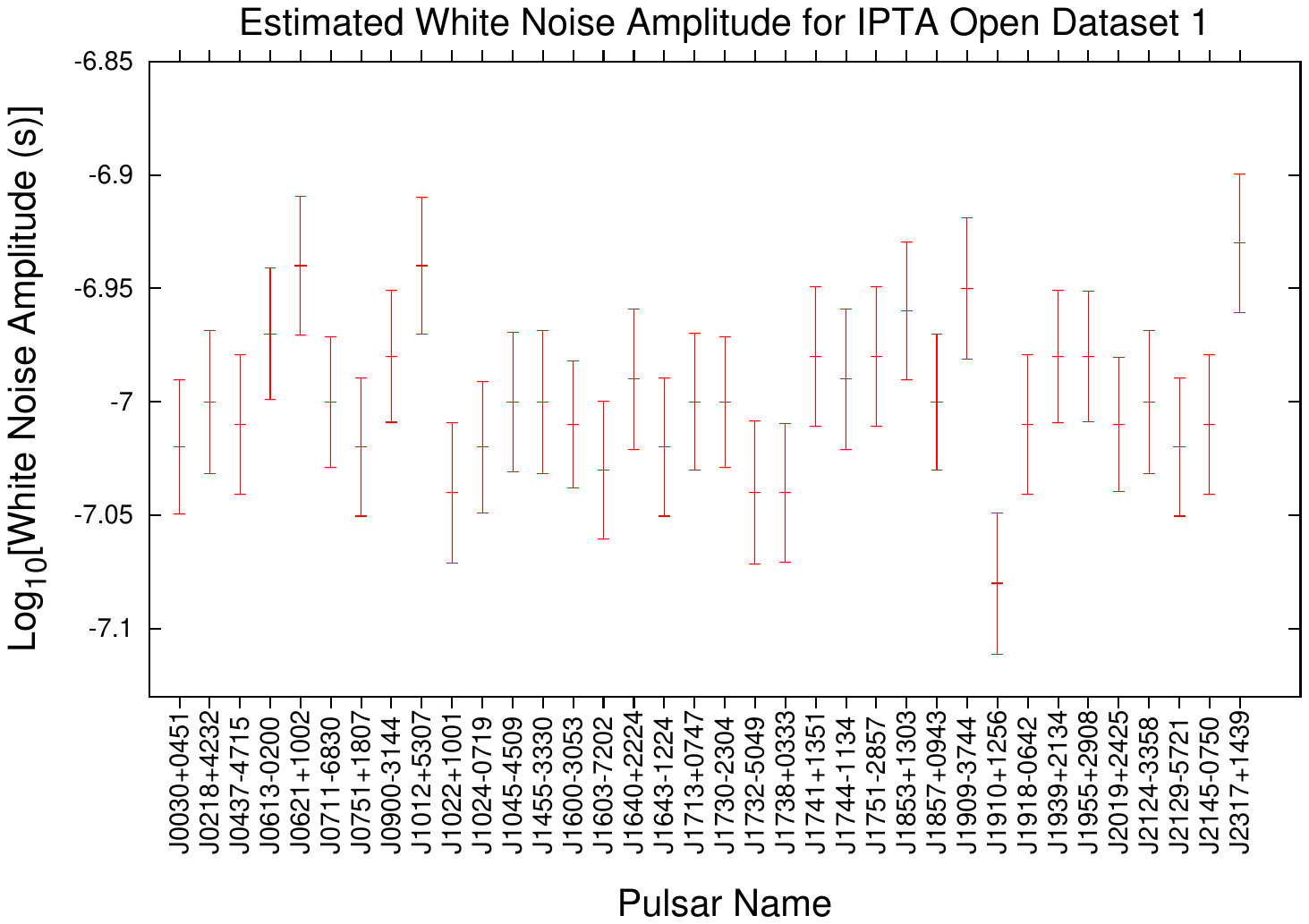}}
\end{minipage}\par\medskip
\centering
\subfloat[]{\label{main:c}\includegraphics[scale=.5]{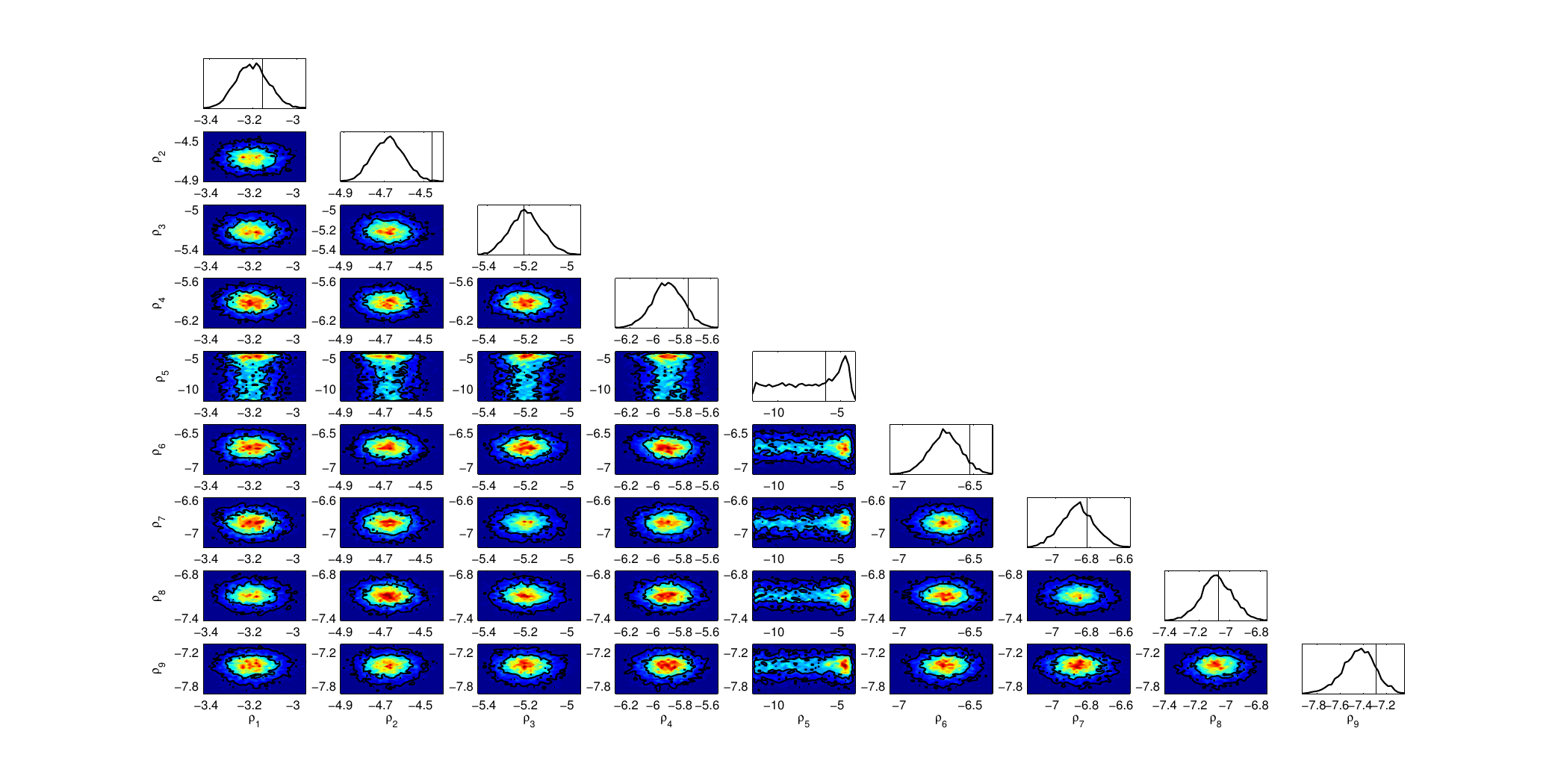}}

\caption{(a) Log-Log Plot of the parameterised GWB power spectrum in open dataset 1.  The red and green bars represent the marginalised values of the fitted GWB power coefficients $\{\rho_i\}$ and their errors for methods A and B respectively.  For clarity we have offset the frequency position for method B however for the analysis both methods were evaluated for the same frequencies.  The blue points represent the power of the injected power spectrum at the sampled frequencies, whilst the dashed blue and purple lines shows the best fit power spectrum to the marginalised coefficients for methods A and B respectively.   (b) Parameterised values for the white noise in each pulsar in open dataset 1 from the IPTA Data Challenge.  Each Pulsar has a white noise component to their residuals with an amplitude of $\sigma_p = 10^{-7}$s.  Averaging across all pulsars we find an rms value for the white noise of $\Sigma_{\mathrm{avg}} = -6.999 \pm$ 0.005 which is thus consistent with the value in the dataset to within $1\sigma$ errors. (c) 1D and 2D marginalised posteriors for the nine GWB Power Spectrum coefficients $\{\rho_i\}$ for method (B).  The vertical line in the 1D distribution represents the power in the injected background at the frequency of that coefficient.  Contours in the 2D plots represent 68 and 95 $\%$ confidence levels.}
\label{fig:Open1}
\end{figure*}

\begin{figure*}

\begin{minipage}{.5\linewidth}
\centering
\subfloat[]{\label{main:a}\includegraphics[scale=.35]{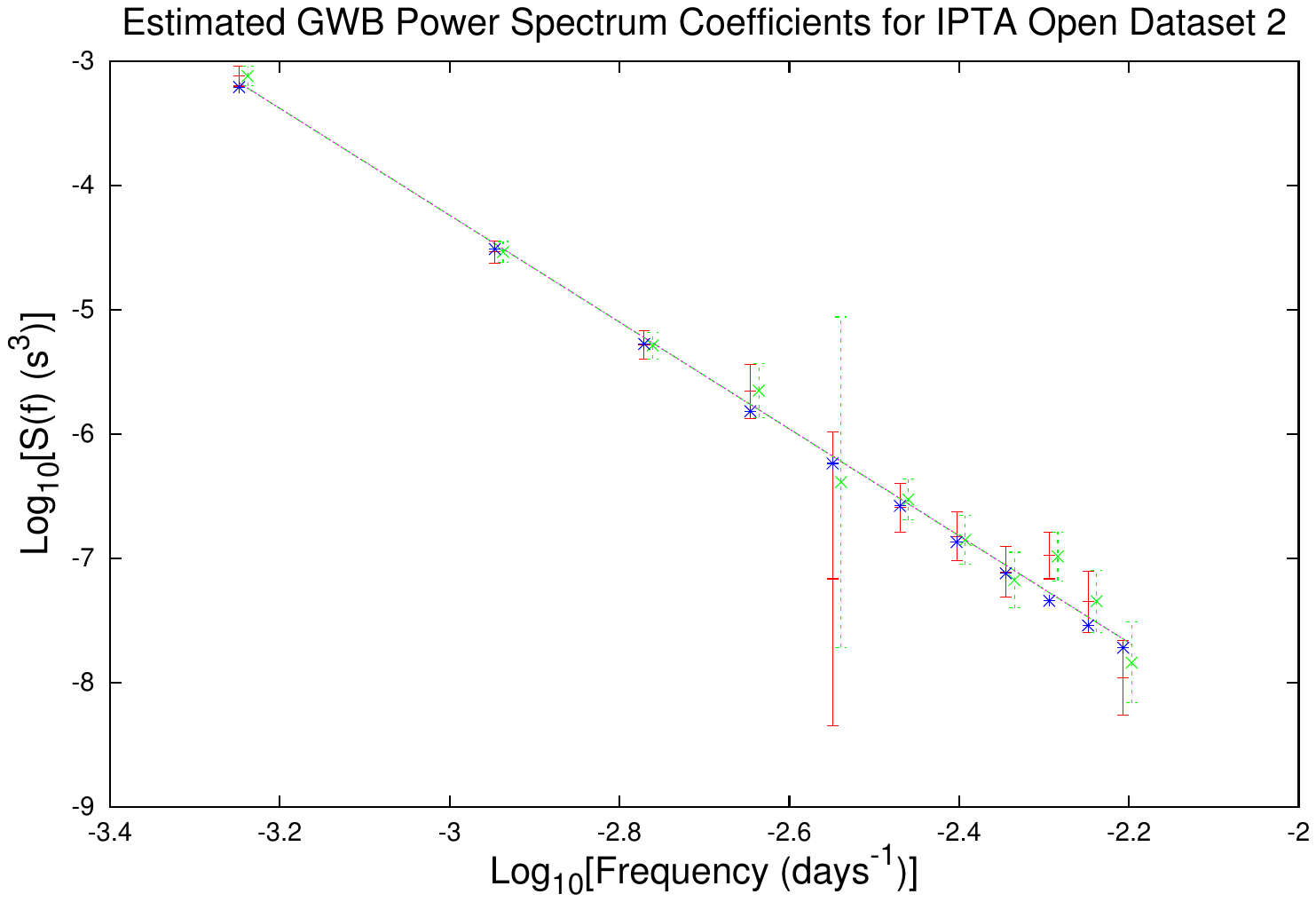}}
\end{minipage}%
\begin{minipage}{.5\linewidth}
\centering
\subfloat[]{\label{main:b}\includegraphics[scale=.35]{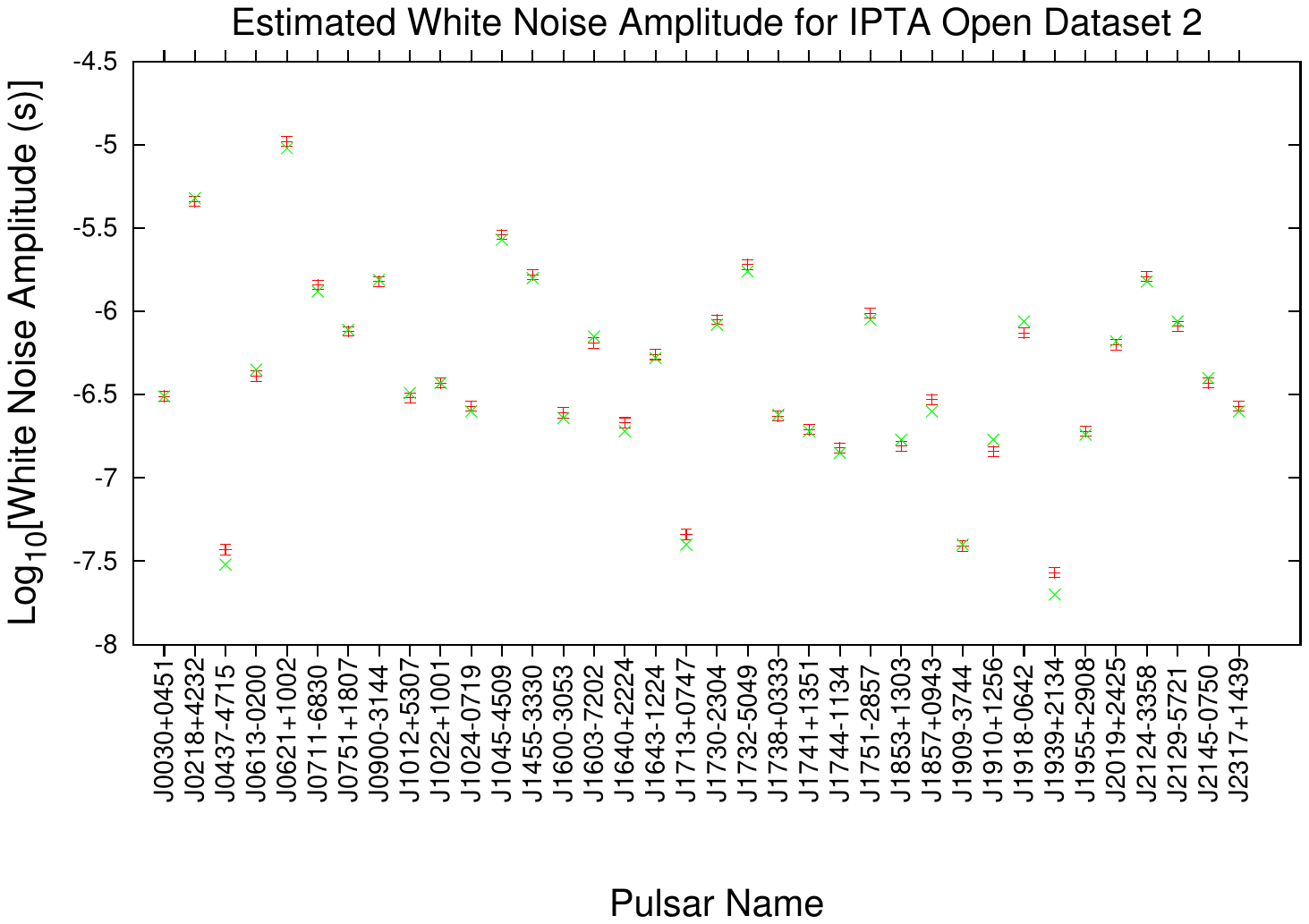}}
\end{minipage}\par\medskip
\centering
\hspace*{-3cm}\subfloat[]{\label{main:c}\includegraphics[scale=.5]{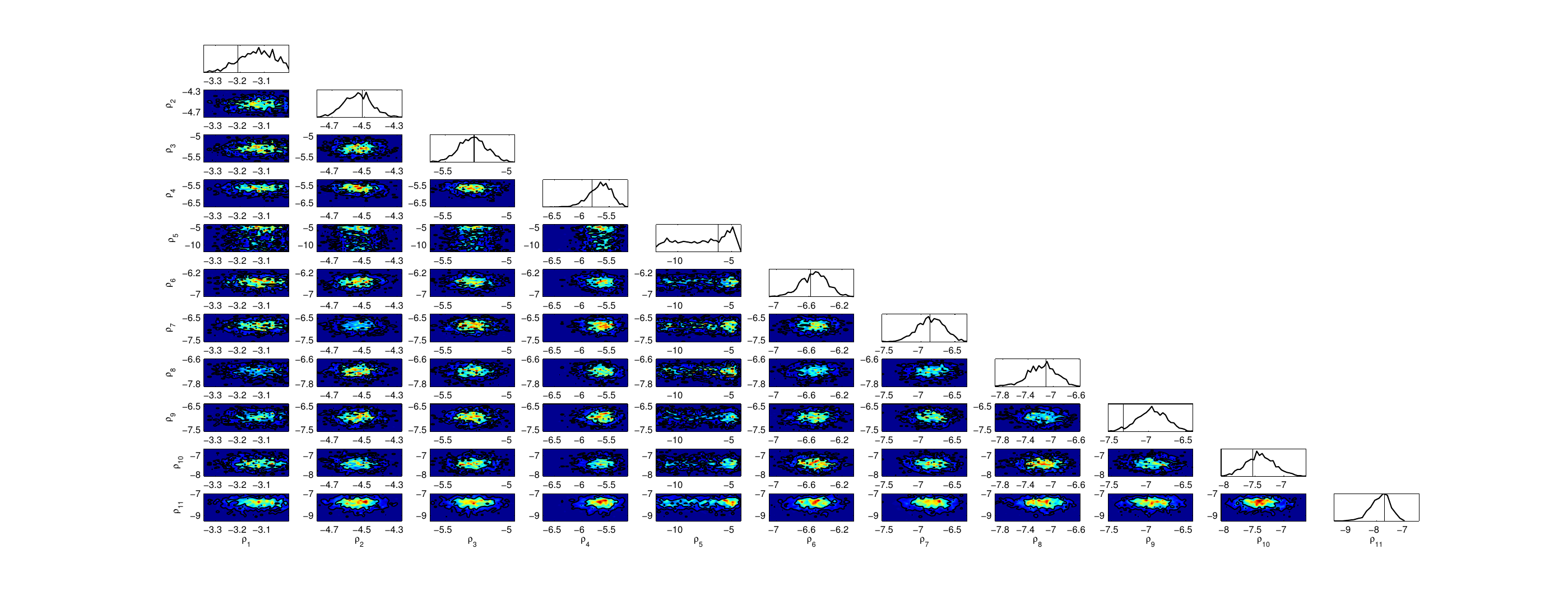}}

\caption{(a)Log-Log Plot of the parameterised GWB power spectrum in open dataset 2.  The red and green bars represent the marginalised values of the fitted GWB power coefficients $\{\rho_i\}$ and their errors for methods A and B respectively.  For clarity we have offset the frequency position for method B however for the analysis both methods were evaluated for the same frequencies.  The blue points represent the power of the injected power spectrum at the sampled frequencies, whilst the dashed blue and purple lines shows the best fit power spectrum to the marginalised coefficients for methods A and B respectively.    (b) Parameterised values for the white noise in each pulsar in open dataset 2 from the IPTA Data Challenge.  Each Pulsar has a different white noise component marked by the green crosses, red data points show the estimated white noise level from the analysis.  (c) 1D and 2D marginalised posteriors for the 11 GWB Power Spectrum coefficients $\{\rho_i\}$ for method (B).  The vertical line in the 1D distribution represents the power in the injected background at the frequency of that coefficient.  Contours in the 2D plots represent 68 and 95 $\%$ confidence levels. }
\label{fig:Open2}
\end{figure*}

\begin{figure*}

\begin{minipage}{.5\linewidth}
\centering
\subfloat[]{\label{main:a}\includegraphics[scale=.35]{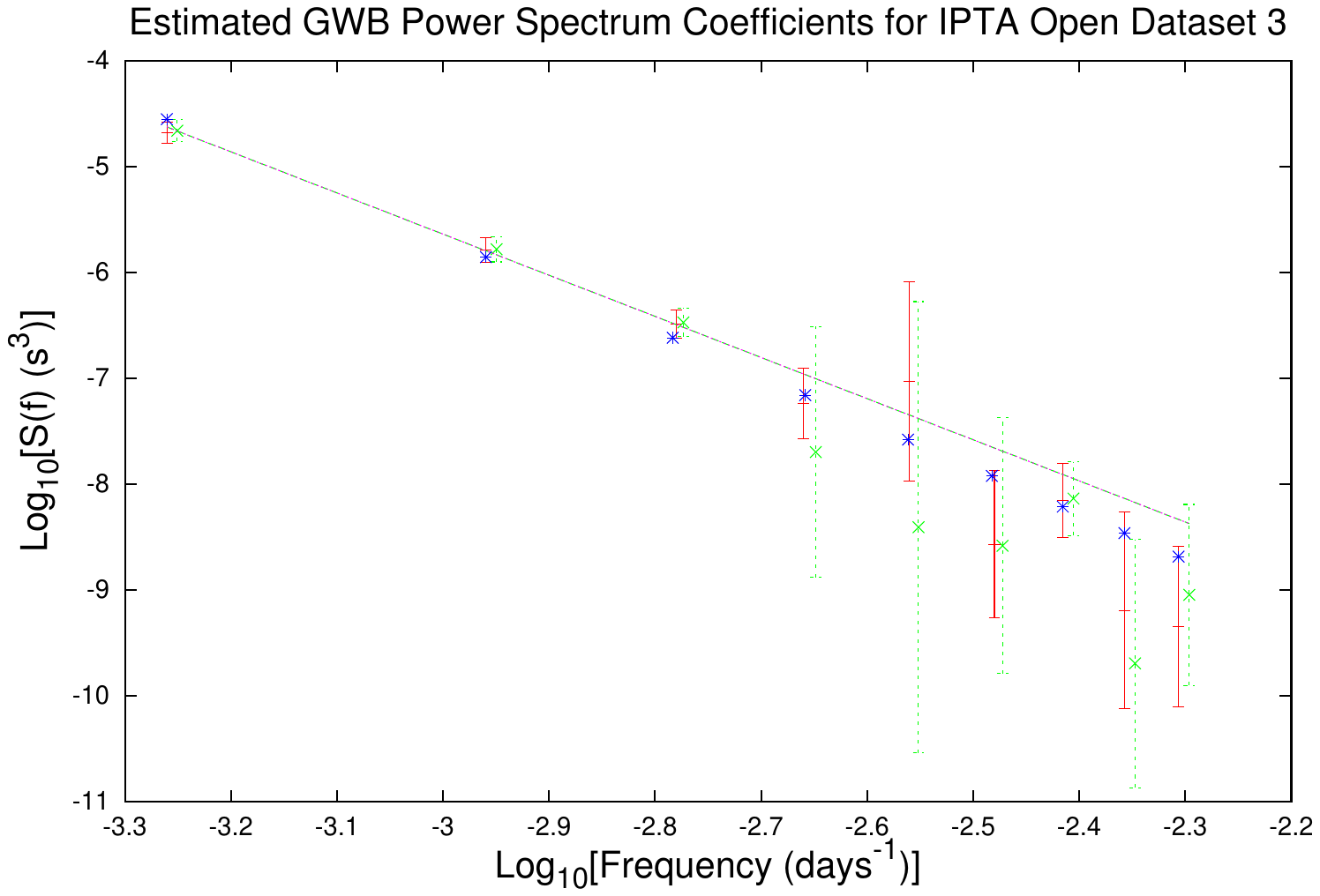}}
\end{minipage}%
\begin{minipage}{.5\linewidth}
\centering
\subfloat[]{\label{main:b}\includegraphics[scale=.35]{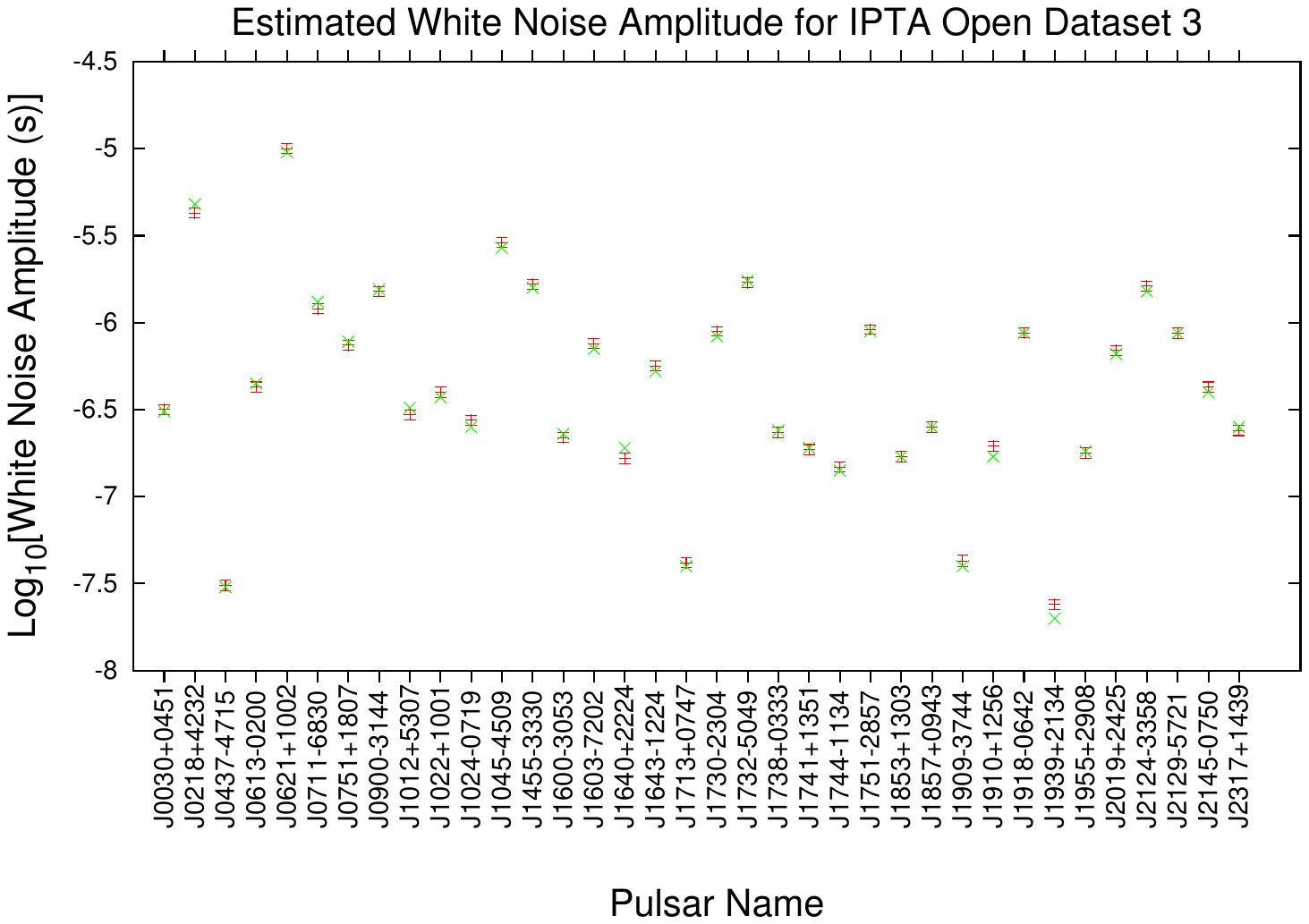}}
\end{minipage}\par\medskip
\centering
\subfloat[]{\label{main:c}\includegraphics[scale=.5]{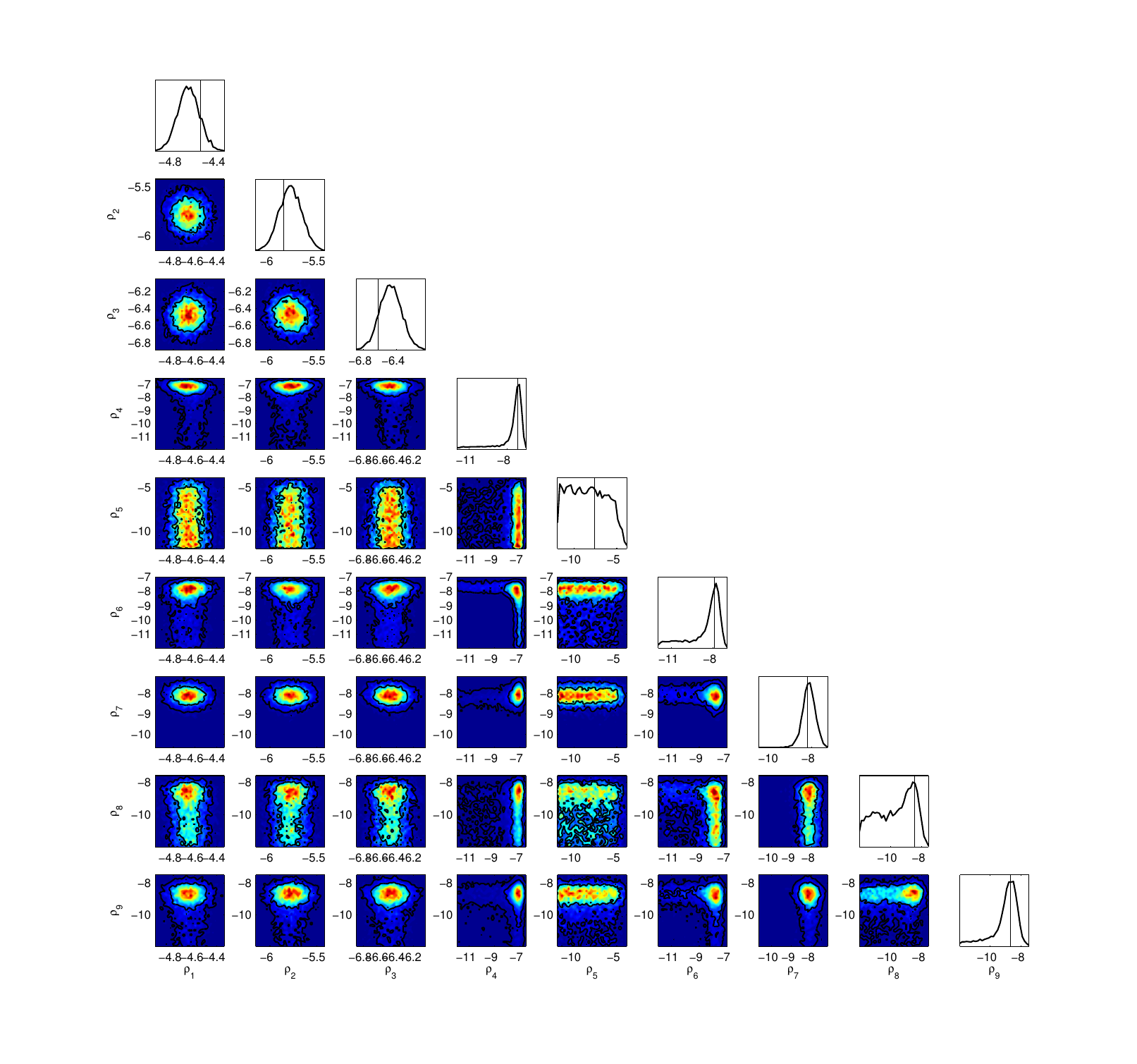}}

\caption{(a) Log-Log Plot of the parameterised GWB power spectrum in open dataset 3.  The red and green bars represent the marginalised values of the fitted GWB power coefficients $\{\rho_i\}$ and their errors for methods A and B respectively.  For clarity we have offset the frequency position for method B however for the analysis both methods were evaluated for the same frequencies.  The blue points represent the power of the injected power spectrum at the sampled frequencies, whilst the dashed blue and purple lines shows the best fit power spectrum to the marginalised coefficients for methods A and B respectively.     (b) Parameterised values for the white noise in each pulsar in open dataset 3 from the IPTA Data Challenge.   Each Pulsar has a different white noise component marked by the green crosses, red data points show the estimated white noise level from the analysis.  (c) 1D and 2D marginalised posteriors for the 9 GWB Power Spectrum coefficients $\{\rho_i\}$.  The vertical line in the 1D distribution represents the power in the injected background at the frequency of that coefficient.  Contours in the 2D plots represent 68 and 95 $\%$ confidence levels.}
\label{fig:Open3}
\end{figure*}

\begin{figure*}

\begin{minipage}{.5\linewidth}

\subfloat[]{\label{main:a}\includegraphics[scale=.35]{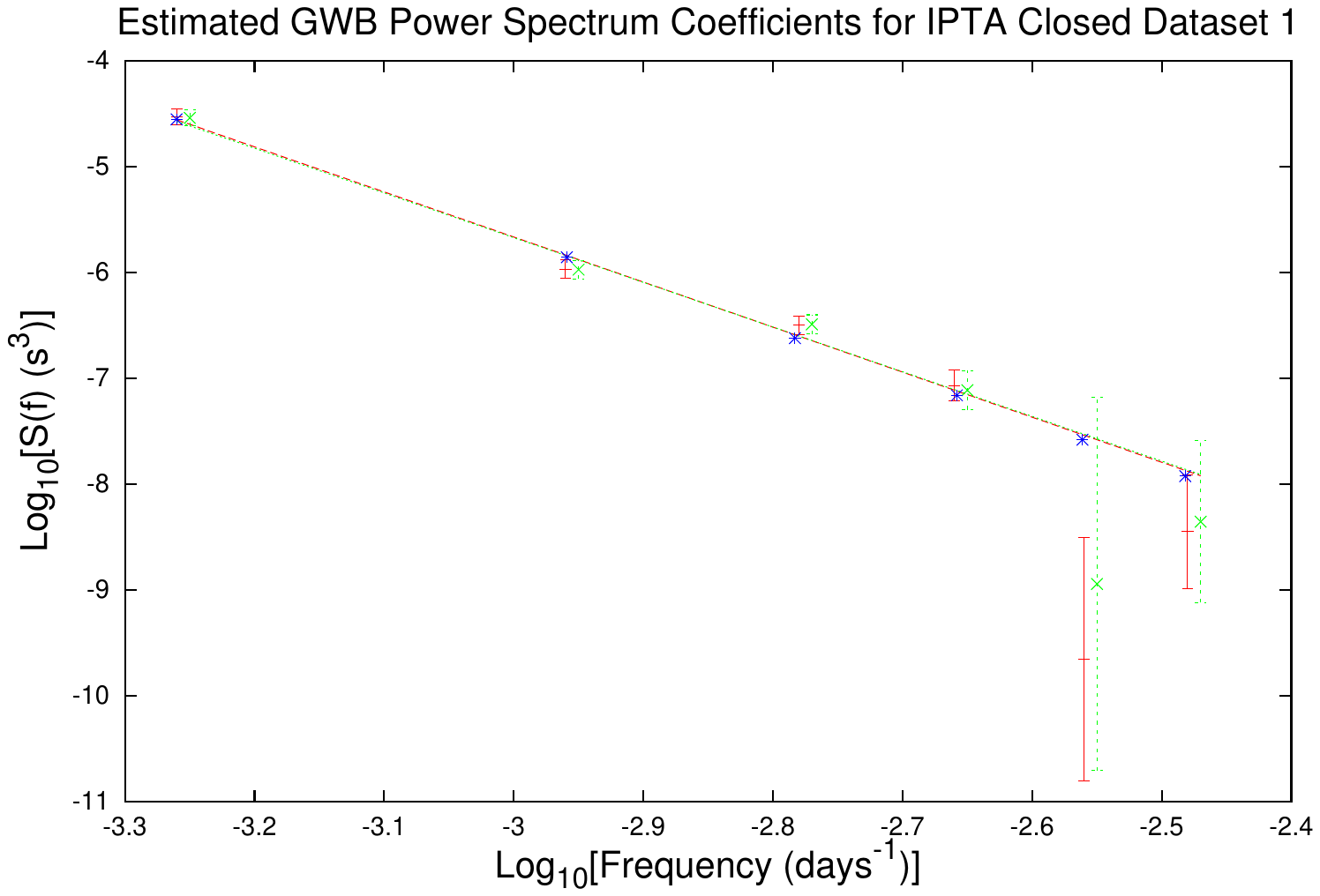}}
\end{minipage}%
\begin{minipage}{.5\linewidth}

\subfloat[]{\label{main:b}\includegraphics[scale=.35]{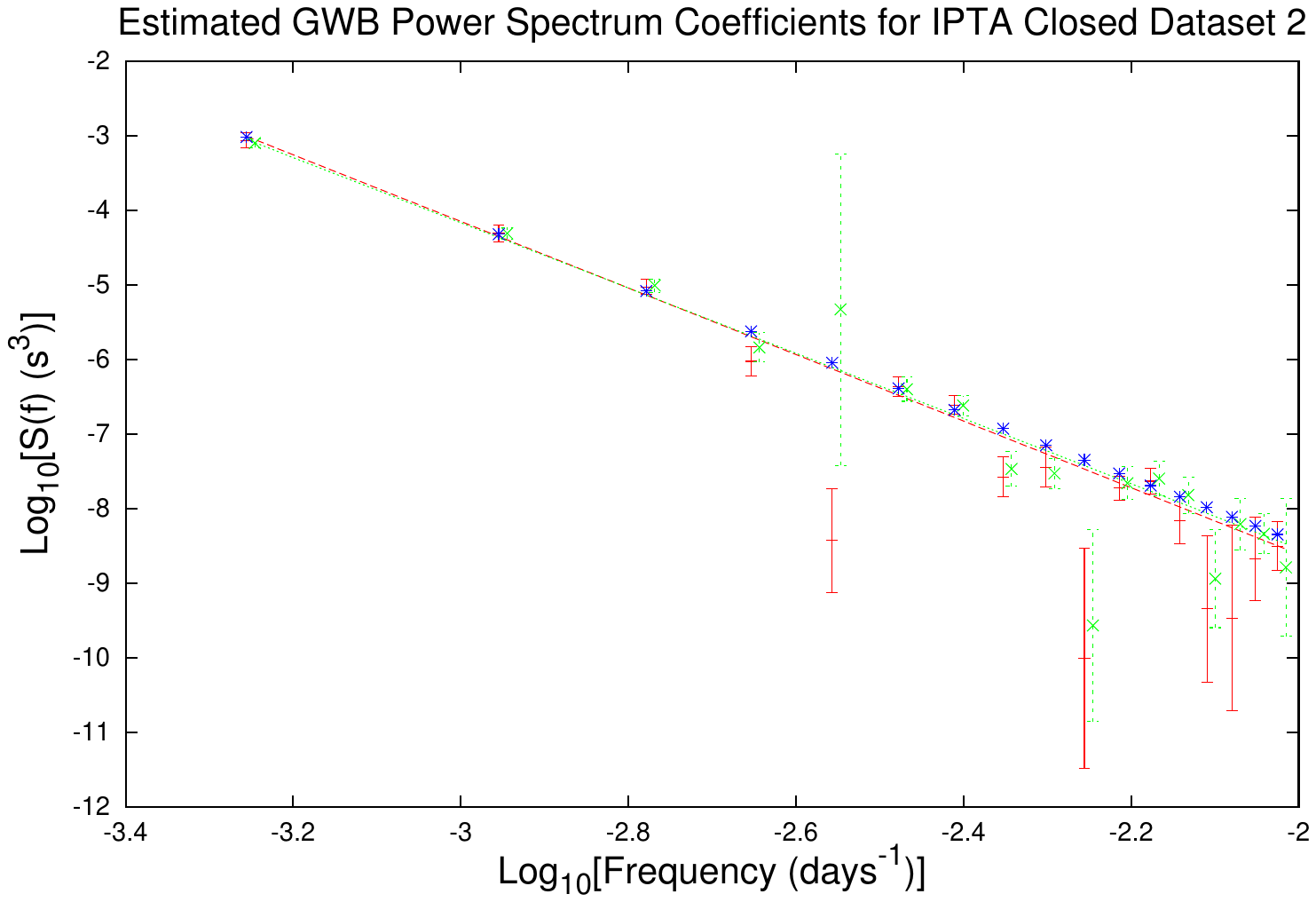}}
\end{minipage}\par\medskip

\subfloat[]{\label{main:c}\includegraphics[scale=.5]{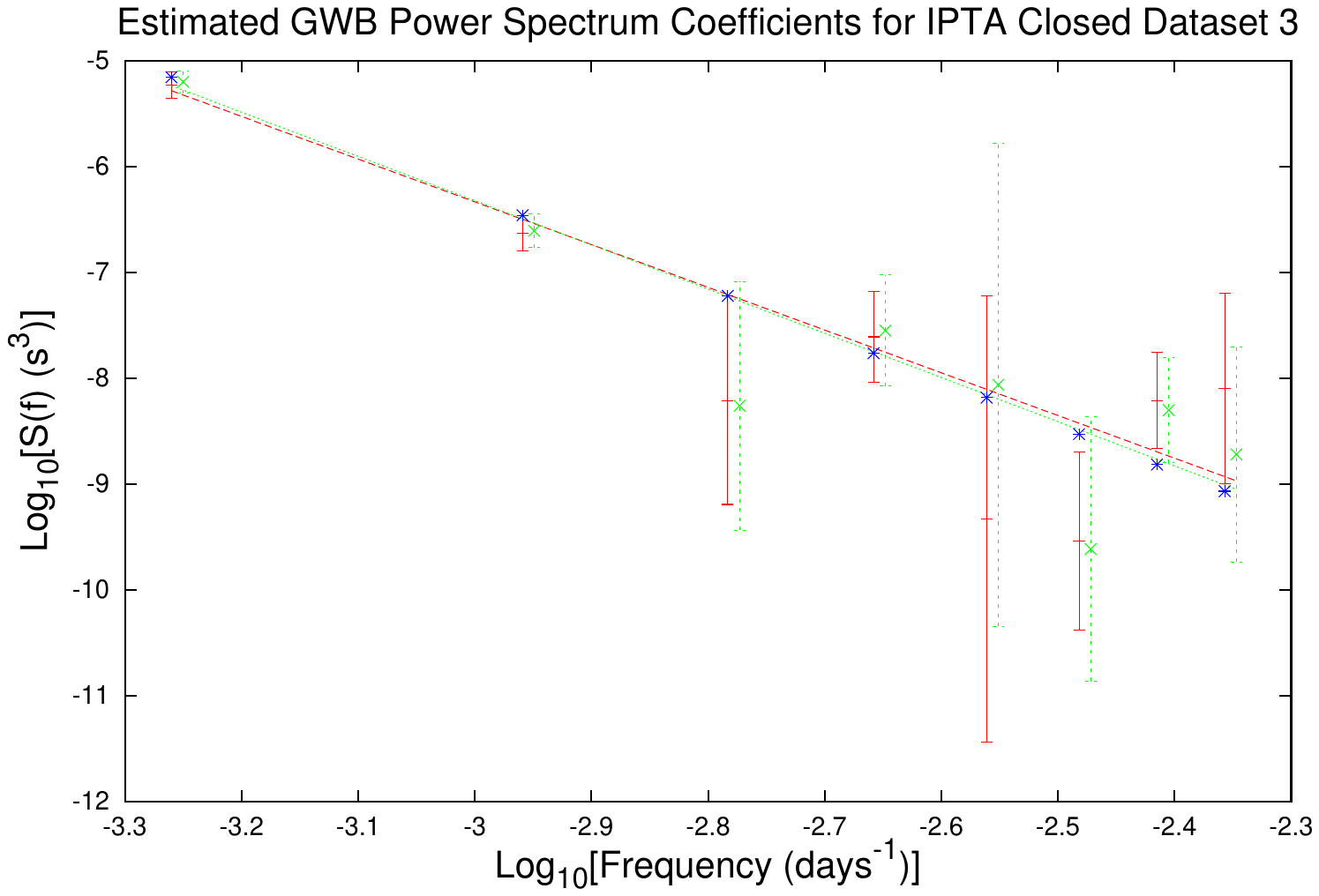}}

\caption{Log-Log Plots of the parameterised GWB power spectrum in closed datasets 1 (a), 2 (b) and 3 (c).  The red and green bars represent the marginalised values of the fitted GWB power coefficients $\{\rho_i\}$ and their errors for methods A and B respectively.  For clarity we have offset the frequency position for method B however for the analysis both methods were evaluated for the same frequencies.  The blue points represent the injected values for those coefficients, whilst the dashed blue and purple lines shows the best fit power spectrum to the marginalised coefficients for methods A and B respectively.     }
\label{fig:ClosedGWB}
\end{figure*}

Table \ref{Table:OpenResults} summarises the results for the six IPTA datasets for methods A, B and C described in this paper, and also the method described in vHL2013.  For methods A and B we give the best fit values and errors for both the dimensionless amplitude $A_g$ and the power law index $\gamma$ that results from a weighted least squares fit to the 1D GWB power coefficients for each of the IPTA datasets, whilst for method C and that from vHL2013 we give the values of $A_g$ and $\gamma$ estimated directly from the data and the errors returned by MULTINEST. For comparison we also include the injected values of the GWB spectrum for each dataset. Figures \ref{fig:Open1} to \ref{fig:Open3} then show a more detailed representation of the results from the open data challenges.  In each figure the top left panel shows a log-log plot of the parameterised GWB power spectrum coefficients for that dataset.  The red and green bars represent the marginalised values of the fitted GWB power coefficients $\{\rho_i\}$ and their errors for methods A and B respectively.  For clarity we have offset the frequency position for method B but for the analysis both methods were evaluated for the same frequencies.  The blue points represent the injected values for those coefficients, whilst the dashed blue and purple lines shows the best fit power spectrum to the marginalised coefficients for methods A and B respectively. The top right panel then shows the parameterised values for the white noise in each pulsar in that dataset.  For open dataset two and three, where the pulsars each have a different white noise level, the injected value is indicated by the green crosses whilst the parameterised values are shown by the red points with their respective errors.  The lower plot in each figure shows the one and two dimensional marginalised posteriors for the GWB Power Spectrum coefficients $\{\rho_i\}$ from method B fitted for that dataset with vertical lines in the 1D distributions representing the power in the injected background at the frequency of that coefficient.  Contours in the 2D plots represent 68 and 95 $\%$ confidence levels.  For the 3 closed data we show only the parameterised GWB power spectrum coefficients from methods A and B in red and green respectively for each dataset in Fig. \ref{fig:ClosedGWB}, and the injected values for each coefficient in blue.  

The predominant message from these results is that for all the datasets methods A-C are all able to extract the correct power spectrum from the data with the same fidelity as the method in vHL2013.  Comparing our results with those in \cite{2013arXiv1301.6673V}, where the data compression method of vH2013 is applied to the IPTA closed datasets, we likewise see consistency between the values and precision of the inferred parameters.  This is true despite the fact that methods A and B at no stage prescribe any form for the shape of the power spectrum, which we believe is the only correct way to perform an analysis of this kind where the true shape of the power spectrum is unknown.

\subsection{Discussion}

\subsubsection{Run Times}

\begin{table*}
\centering
\caption{Comparison of Run Times for Different Sampling Methods} 
\centering 
\begin{tabular}{c c c c c c c c c} 
\hline\hline 
Dataset &  \multicolumn{8}{c}{Method} \\[0.5ex] 
	     &	     \multicolumn{2}{c}{This Paper(A)} & \multicolumn{2}{c}{This Paper(B)} & \multicolumn{2}{c}{This Paper(C)} & \multicolumn{2}{c}{vHL2013}	\\
	     \hline
	     &		Dimensionality & Run Time   &		Dimensionality & Run Time   &		Dimensionality & Run Time   &		Dimensionality & Run Time \\
	     &	&(minutes)&&(minutes)&&(minutes)&&(minutes)\\
\hline 
Open 1 &702 &35 & 9& 10& 2& $< 1$&2 & 145\\
Open 2 & 839& 55&11 & 35&2 & $< 1$&2 & 130\\
Open 3 &702 &40 &9 & 10&2 & $< 1$&2 & 140\\
Closed 1 & 474&30 &9 & 2& 2&$< 1$ &2 & 140\\
Closed 2 &1277 & 110&17 & 180& 2&4 &2 & 160\\
Closed 3 &908 & 130 & 16&145 &4 &3 & 4& 235\\
\hline
\end{tabular}
\label{Table:TimeCompare} 
\end{table*}

Table \ref{Table:TimeCompare} shows a comparison of the run times for the three different sampling methods presented in this paper, and for the method described in vHL2013, when using a single 16 core Sandy Bridge node on the high performance computer (HPC) `DARWIN'.  For our implementation of the method in vHL2013 we use the same number of free parameters as for method (C) described at the start of Section \ref{Section:Results}.
In every case method (C) is $100-1000$ times faster than the method described in vHL2013, precisely what we would expect given the order of magnitude decrease in the size of the covariance matrix that requires inverting when compared to the time domain analysis.  Comparing the run times between methods (A) and (B) we can see at what point the numerical marginalisation becomes favourable over the analytical form.  Below $\sim$ 15 coefficients performing the marginalisation analytically is clearly the preferred choice, being a factor of a few faster than performing the process numerically, however the increase in the number of calculations required for convergence, combined with the O$(n^3)$ scaling of the matrix inversion means that beyond this point it rapidly begins to lose out, ultimately degrading to become the slowest method with which to perform the analysis for the closed 2 dataset.

Whilst the comparisons in Table \ref{Table:TimeCompare} have all been made with the method of vHL2013, it is of interest to see how the speed up compares with the data compression method presented in vH2013.  We therefore used a dummy likelihood function that contained all the computational overhead associated with the data compression algorithm, and set the number of pulsars, the number of observations, and the level of compression used to represent those values that would be chosen for an analysis of the IPTA open 1 dataset.  This function was then compiled and linked to the same libraries used in the analysis of the previous section, at which point ten sets of one thousand iterations each were performed and timed.
We then used the likelihood function of methods (A) and (C), for which the latter provides the most direct comparison to the approach of vH2013, once again set the model parameters to be the same as those used in our analysis of open dataset 1, and performed the same test.  We found that the average computation time for one thousand evaluations of the three likelihood functions were approximately 45, 1.5 and 47 seconds for vH2013, method (A) and method (C) respectively.  The consistency between vH2013 and method (C) is not surprising, the computational burden for each likelihood evaluation is still in the matrix inversions which are of similar order, with the data compression method resulting in 10 data points per pulsar, and method (C) utilising 9 Fourier coefficients to describe the signal.   

One important consideration when discussing the run times of these different methods is how well they scale with the inclusion of more parameters.  Whilst the method described in vHL2013 is shown here to have comparable run times to method (A) we have only been using it to evaluate a two or four dimensional case.  If for example one increases the dimensionality from two to thirty eight in order to include white noise estimation for each pulsar the run time increases from two hours, to over one hundred.  Including white noise estimation in method (A), where the increase in dimensionality (36) is small compared to the total ($\sim$ 1000) results in a similarly small increase in the total run time of $\sim$ 15 minutes.  This is one of the key advantages of the numerical marginalisation coupled with the guided Hamiltonian sampler and one that we exploit in section \ref{Section:CrossCorr} as we introduce an additional 630 dimensions to parameterise the spatial correlations between pulsars.  Even this though is still an extremely small parameter space compared to the greater than $10^6$ dimensional problems that it has been used to solve in other work (B13).  This therefore leaves a practically unlimited space in which to expand, with the inclusion of additional parameters such as simultaneous dispersion measure correction, or even the full non linear timing model that have previously been not thought feasible.

\subsubsection{Frequencies of 1yr$^{-1}$}

Perhaps one of the most striking features of the 1D and 2D confidence contours in Figures \ref{fig:Open1} to \ref{fig:Open3} is that without exception the GWB coefficient $\rho_5$ is totally unconstrained.  All of the datasets in the IPTA data challenge are approximately 1820 days in length, and so in every case $\rho_5$ corresponds to a frequency of $\sim1\mathrm{yr}^{-1}$.  That this should occur at such a distinct frequency is no coincidence; as part of the timing model fit performed by Tempo2 the pulsar's position and proper motion are all included as free parameters.  Inaccuracies in the fitted values of these parameters can result in power being introduced to the residuals at frequencies of 1yr$^{-1}$ (see e.g. \cite{1997A&A...326..924M}).  When we perform the analytic marginalisation over all the model timing parameters we therefore effectively project out contributions to the signal from components with these periods.  The model Fourier coefficients corresponding to frequencies of 1yr$^{-1}$ therefore have no effect on the likelihood when the linear approximation to the timing model holds and therefore the very way in which we account for the timing models for each pulsar results in us being able to make no inferences on the properties of the power spectrum at this frequency.

That this is so clear in the results is a testament to the success of the method;  by not assuming any form for the power spectrum and simply asking in the most general way how the power is distributed in the signal we are able to infer much more information than simply by fitting for a power law.  In this instance that extra information is that we are unable to constrain anything about the spectrum at frequencies of 1yr$^{-1}$, however where the true  power spectrum is unknown this approach is the only way of ensuring an optimal estimate of that power spectrum and of extacting the maximal amount of information possible.

\section{A More Realistic Simulation}
\label{Section:MySims}

Whilst the IPTA data challenges serve as a good introduction to analysing PTA data, they still represent comparatively simplistic datasets when compared to genuine observations.  For example, whilst some of the challenge datasets featured uneven sampling in the time domain, all pulsars within a dataset shared the same TOAs, and thus also shared the same total time span.  Similarly, when included, the properties of the red noise were the same for all the pulsars in the datasets.  There were also no gaps in the data greater than a few weeks, whereas jumps of more than a year can be expected when analysing real data. We have therefore constructed two simulations designed to represent better a potential future IPTA data release and thus provide a more difficult test for the analysis method presented in this paper.

\subsection{Generating the Residuals}
\label{genres}

The simulations are generated using the time domain covariance matrix $\mathbf{C}^{GW}_{(ai)(bj)}$ between observations $i$ and $j$ and pulsars $a$ and $b$ for a GWB given in vH2009:
\begin{eqnarray}
\mathbf{C}^{GW}_{(ai)(bj)}  &=& \frac{\beta_{ab}A_g^2\mathrm{yr}^{3-\gamma}}{12\pi^2f_L^{\gamma-1}}\left\{\Gamma(1-\gamma)\sin\left(\frac{\pi\gamma}{2}\right)\right.\\
&\times&\left.(f_L\tau)^{\gamma-1} - \sum_{n=0}^{\infty}(-1)^n\frac{(f_l\tau)^{2n}}{(2n)!(2n+1-\gamma)}\right\}. \nonumber
\end{eqnarray}
where $\beta_{ab}$ is the Hellings-Downs coefficient between pulsars $a$ and $b$, $f_L$ is a low frequency cut off, chosen only so that $1/f_L$ is much greater than the observing time span and $\tau = 2\pi(t_{ai}-t_{bj})$ with $t_{ai}$ the $i$th TOA for pulsar $a$.
The covariance matrix for the included red noise $\mathbf{C}^{RN}_{(ai)(bj)}$ is identical, however the term $\beta_{ab}$ is replaced with a delta function $\delta_{ab}$ as it will be uncorrelated between pulsars.  Finally denoting the white noise covariance matrix  $\mathbf{C}^{W}_{(ai)(bj)} = \sigma^2_w\delta_{ab}\delta_{ij}$ we can write the total covariance matrix describing our simulated residuals $\mathbf{C}^{T}_{(ai)(bj)}$ as:

\begin{equation}
 \mathbf{C}^{T}_{(ai)(bj)} = \mathbf{C}^{GW}_{(ai)(bj)} + \mathbf{C}^{RN}_{(ai)(bj)} + \mathbf{C}^{W}_{(ai)(bj)}.
 \end{equation}
We then take the Cholesky decomposition of this matrix and use it to generate the residuals.  A quadratic is then fitted to and subtracted from each of the pulsar residuals independently to mimic the effect of subtracting the timing model.  The design matrix used to generate the matrix $\mathbf{G}$ in Eq \ref{Eq:GEq} and beyond will then simply be that of a quadratic polynomial.

\subsection{The Simulations}

Both simulations use a set of 21 pulsars with observations spanning periods of between 4 and 18 years, with spacings between observations ranging from less than a day up to 5 years.  Simulation one then injects a gravitational wave background with parameters $\gamma=4.33$ and dimensionless amplitude $A_g=10^{-14}$ and white noise with an amplitude $\sigma_w = 10^{-7}s$. The second simulation uses the same sampling times as the first however the background now has an amplitude of $A_g=5\times10^{-15}$, and red noise is included for each pulsar, with $\gamma_{\mathrm{red}}$ covering a range from $1.1 \to 5.1$ and amplitudes extending from $A_g = 10^{-16} \to 5\times10^{-14}$.  Table \ref{Table:Simulations} gives a more complete overview of the simulated data listing the total timespan $T_{\mathrm{span}}$ for each pulsar, the number of observations $N_{\mathrm{obs}}$ in that observation window and the red noise parameters $\gamma_{\mathrm{red}}$ and $A_g$ present in simulation two.

\begin{table*}
\centering
\caption{Parameters for Simulation One and Two} 
\centering 
\begin{tabular}{c c c c c } 
\hline\hline 
Pulsar No. & $T_{\mathrm{span}}$ &$N_{\mathrm{obs}}$ & $\gamma_{\mathrm{red}}$ & $\log_{10}[A_g]$ \\[0.5ex] 
	    	 &	   years 			& 	   			 &		     &  	\\
\hline 
1	&	3.18		&	22		& 3.3		& 14.3		\\
2	&	14.86	&	1057		& 2.1		& 15.1		\\
3	&	17.10	&	343		& 1.6		& 13.8		\\
4	&	14.45	&	814		& 1.1		& 13.3		\\
5	&	15.89	&	692		& 2.3		& 14.6		\\
6	&	17.01	&	368		& 1.5		& 14.2		\\
7	&	9.90		&	721		& 4.2		& 13.8		\\
8	&	15.31	&	289		& 1.8		& 13.5		\\
9	&	14.96	&	427		& 2.4		& 16.0		\\
10	&	17.79	&	940		& 1.9		& 14.5		\\
11	&	18.37	&	1291		& 1.6		& 14.0		\\
12	&	17.80	&	422		& 2.2		& 14.2		\\
13	&	8.04		&	153		& 5.1		& 15.0		\\
14	&	16.96	&	728		& 3.4		& 14.6		\\
15	&	5.75		&	164		& 2.6		& 13.9		\\
16	&	4.75		&	35		& 3.5		& 14.0		\\
17	&	9.02		&	728		& 1.5		& 13.4		\\
18	&	10.46	&	284		& 2.3		& 14.4		\\
19	&	15.42	&	293		& 2.8		& 14.1		\\
20	&	17.54	&	914		& 1.2		& 13.7		\\
21	&	14.95	&	402		& 3.4		& 14.0		\\
\hline
\end{tabular}
\label{Table:Simulations} 
\end{table*}

\begin{figure*}
\begin{minipage}{168mm}
\begin{center}$
\begin{array}{cc}
\includegraphics[width=80mm]{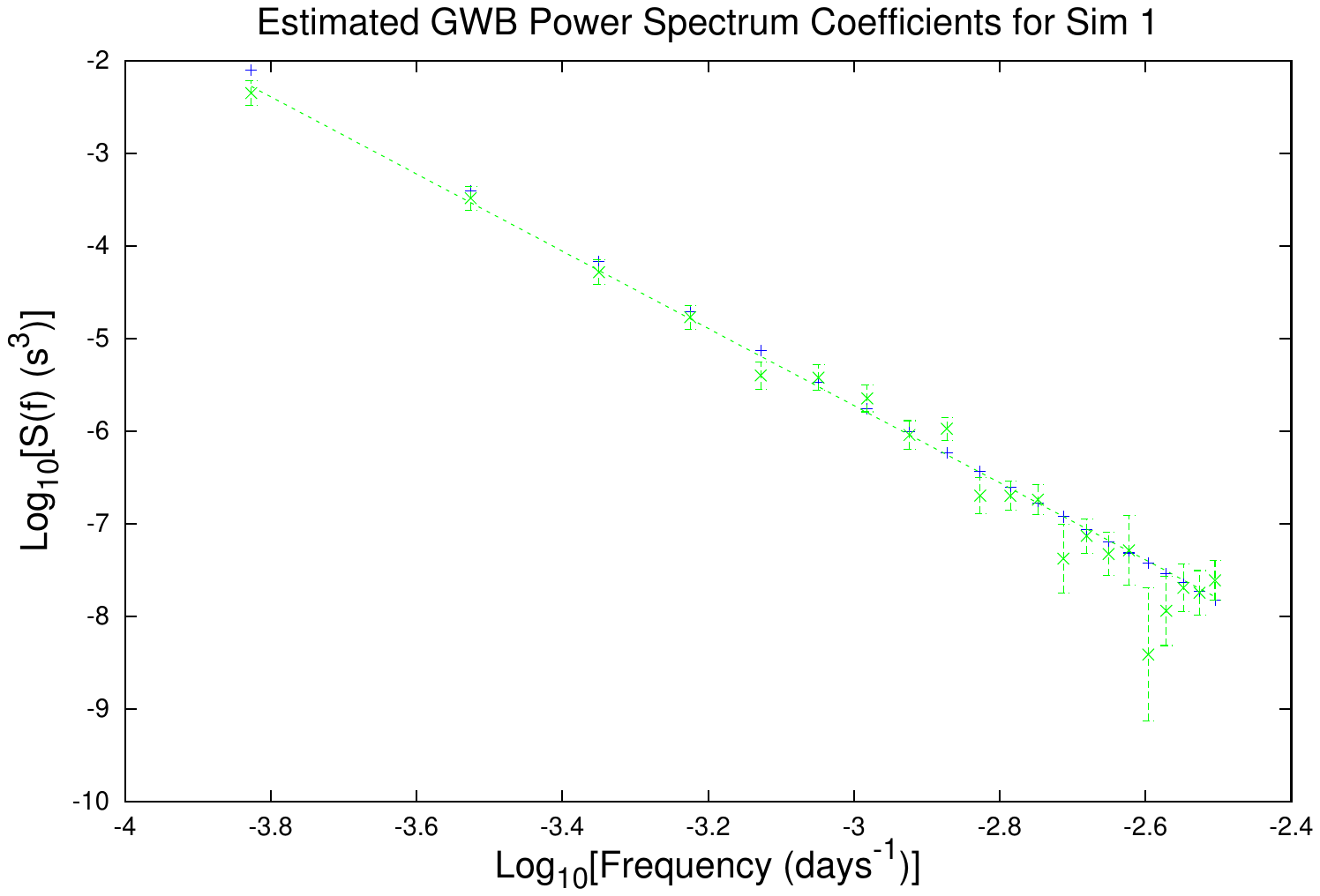} &
\includegraphics[width=80mm]{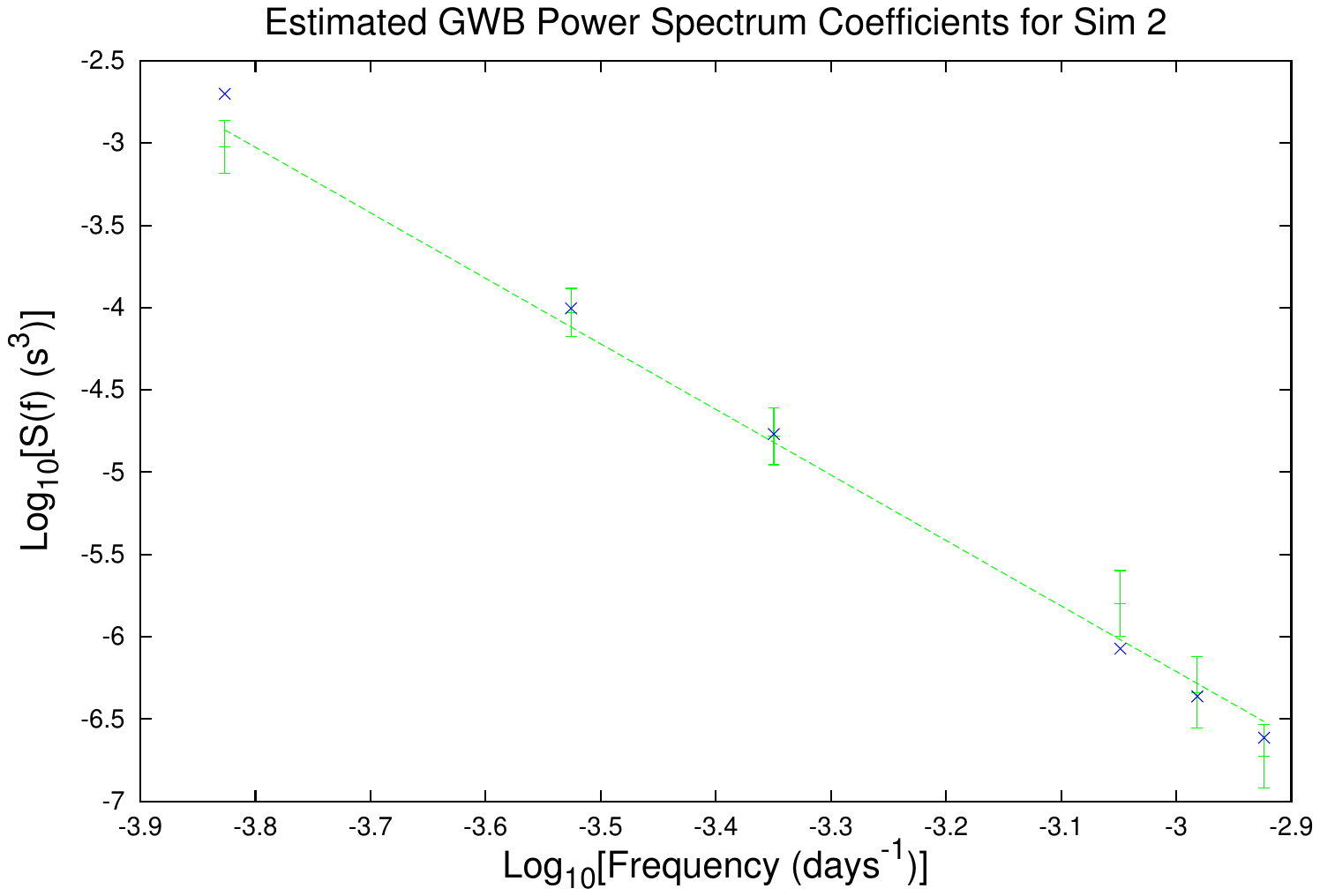}
\end{array}$
\end{center}
\caption{Log-Log Plot of the parameterised GWB power spectrum in simulations one (left) and two (right).  The green bars represent the marginalised values of the fitted GWB power coefficients $\{\rho_i\}$ and their errors derived using method A applied to that dataset.  The blue points represent the injected values for those coefficients, whilst the green line shows the best fit power law spectrum to the marginalised coefficients.}
\label{Fig:IPTASpec}
\end{minipage}
\end{figure*}

In analysing the data we choose a fundamental frequency $f_0$ to be equal to $1/T_{max}$, where $T_{max}$ represents the greatest time span for any of the pulsars in the dataset, which for both simulations is $\sim$ 18.4 years.  We then use the Laplace approximation method described in section \ref{Section:FindCoeffs} to determine the number of frequencies to be used in the analysis.  We find that 21 coefficients should be sufficient to describe the first simulation whilst a maximum of 12 Fourier modes are required for the second.  We then apply method (A) to the two datasets.  In the first case we parameterise only the Fourier coefficients for the 21 pulsars, their white noise and the set of 21 GWB power spectrum coefficients, whilst for the second dataset we also include red noise parameters for each of the pulsars resulting in  264 and 903 dimensional spaces for each respectively.  The results are shown in Table \ref{Table:IPTAResults} while we plot the GWB coefficients in both cases in Fig.\ref{Fig:IPTASpec} with the blue points representing the the theoretical power at the sampled frequency given the injected spectrum.  In the case of simulation 2 we plot only a subset of the frequency coefficients as only those corresponding to frequency modes 1-3 and 6-9 resulted in detections of a correlated signal within the data.  

We see the results are once again consistent with the injected values, demonstrating that even in extremely challenging data where there is a great deal of additional red noise and highly irregular sampling we are able to correctly parameterise the GWB signal.

\begin{table*}
\centering
\caption{Results from the two Simulations in Section \ref{Section:MySims}}
\centering 
\begin{tabular}{c c c c c} 
\hline\hline 
		& \multicolumn{2}{c}{Method (A)} & \multicolumn{2}{c}{Injected Values} \\[0.5ex] 
Dataset & 	$A_g \times 10^{-14}$ & $\gamma$ & 	$A_g \times 10^{-14}$ & $\gamma$ \\
\hline
\hline 
Sim 1 & $1.1 \pm 0.2$ & $4.2 \pm 0.1$ & 1 & 4.333 \\
Sim 2 & $0.61 \pm 0.07$	& $4.0 \pm 0.2$	&0.5	& 4.333 \\
\hline
\end{tabular}
\label{Table:IPTAResults} 
\end{table*}

\section{Fitting for the Cross Correlation}
\label{Section:CrossCorr}

Thus far we have parameterised the angular correlations between different pairs of pulsars using the Hellings-Downs curve; the result derived assuming an isotropic background of gravitational waves when only those polarisation states predicated by general relativity are considered.  Different metric theories of gravity, however, predict different angular correlations, and anisotropies in the background due to bright individual sources can lead to deviations in this description \cite{2012PhRvD..85h2001C}. Furthermore, terrestrial clock errors and inaccuracies in the solar system ephemeris can also generate spatial correlations within pulsar residuals, the latter for example would result in the residuals taking on a dipole signature \cite{2006ChJAS...6b.169H}.  As such, performing the analysis of PTA data assuming the Hellings-Downs curve explicitly could result in a false detection if there is a spatially correlated component, even if the form of that correlation is better described by something other than a GWB.   

Methods for generalising the Hellings-Downs curve at the point of sampling are relatively new, for example \cite{2012arXiv1210.6014T} present two possible approaches.  First they fit for the angular correlation at a set of 5 angular separations, and then use cubic splines to interpolate between those points in order to determine the angular correlations at intervening values and secondly, they use a generalised Hellings-Downs model to parameterise the correlation.  These methods were successfully able to extract the form of the Hellings-Downs curve in the case of the first IPTA open challenge, however we would like to generalise this approach further and fit for the correlations between all pairs of pulsars directly.  This therefore relieves us of the assumption that the background is isotropic, with pairs of pulsars at the same angular separation able to have different correlation coefficients, and still at no point assumes any prescribed form of the correlation that might bias the end result in order to test whether or not the Hellings-Downs curve is distinguishable in simulated data from, for example, a dipole.   

When fitting for the cross correlations between the pulsars, we must ensure that the covariance matrix describing those correlations remains positive definite. Many methods exist where the elements of the upper-triangular elements in the covariance matrix are re-parameterised such that the resultant covariance matrix is ensured to be positive definite \cite{Pinheiro1996}.

For any positive definite covariance matrix $\mathbf{\Sigma}$ we are able to take a Cholesky decomposition such that the matrix can be represented as the product $\mathbf{\Sigma}=\mathbf{L}\mathbf{L}^{\mathrm{T}}$.  In general however such a decomposition is not unique.  If $\mathbf{L}$ is the Cholesky decomposition of $\mathbf{\Sigma}$ then so is any matrix obtained by multiplying a subset  of the  rows  of $\mathbf{L}$  by  -1.  This can therefore give rise to multi-modal distributions that will increase the complexity of the sampling process unnescessarily.
This problem can be circumvented by ensuring that the diagonal elements of $\mathbf{L}$ are positive, in which case $\mathbf{L}$ is unique for a given $\mathbf{\Sigma}$, which can be achieved by fitting for the $\log$ of the diagonal elements.  
In this form however there is no straightforward way of fixing the elements of the matrix $\mathbf{\Sigma}$, such that the diagonal elements are equal to unity.  We therefore use a spherical parameterisation of the elements in $\mathbf{L}$ as in \cite{Pinheiro1996}, which we describe below.

\subsection{Spherical Parameterisation}
\label{Section:CCParam}
If we denote the $j$th element of the $i$th column of the upper triangular matrix $\mathbf{L}$ as $L_{ij}$, and define a second upper triangular matrix $\mathbf{l}$ that contains the spherical  parameterisation of $\mathbf{L}$, we can write any element of $\mathbf{L}$ in the form:

\begin{flalign*}
&L_{i,1} = l_{i,1}\cos({l_{i,2}}) \\
&L_{i,2} = l_{i,1}\sin({l_{i,2}})\cos({l_{i,3}})\\
&L_{i,3} = l_{i,1}\sin({l_{i,2}})\sin({l_{i,3}})\cos({l_{i,4}})\\
&\vdots\\
&L_{i,i-1} = l_{i,1}\sin({l_{i,2}})\ldots\sin({l_{i,i-1}}) \cos({l_{i,i}})\\
&L_{i,i} = l_{i,1}\sin({l_{i,2}})\ldots\sin({l_{i,i-1}})\sin({l_{i,i}})
\end{flalign*}
The diagonal elements of the covariance matrix $\Sigma_{ii}$ are then given by $\Sigma_{ii}=l_{i,1}^2$, and so we can trivially ensure a unit diagonal by setting all $l_{i,1}=1$. Therefore for an $n\times n$ covariance matrix we need only fit for $n(n-1)/2$ elements, which for 36 pulsars, results in an increase of dimensionality of $N_{\mathrm{corr}}=36\times35/2 = 630$.

\begin{figure*}
\hspace*{-3cm}\begin{minipage}{168mm}
\begin{center}$
\begin{array}{ccc}
\includegraphics[width=60mm]{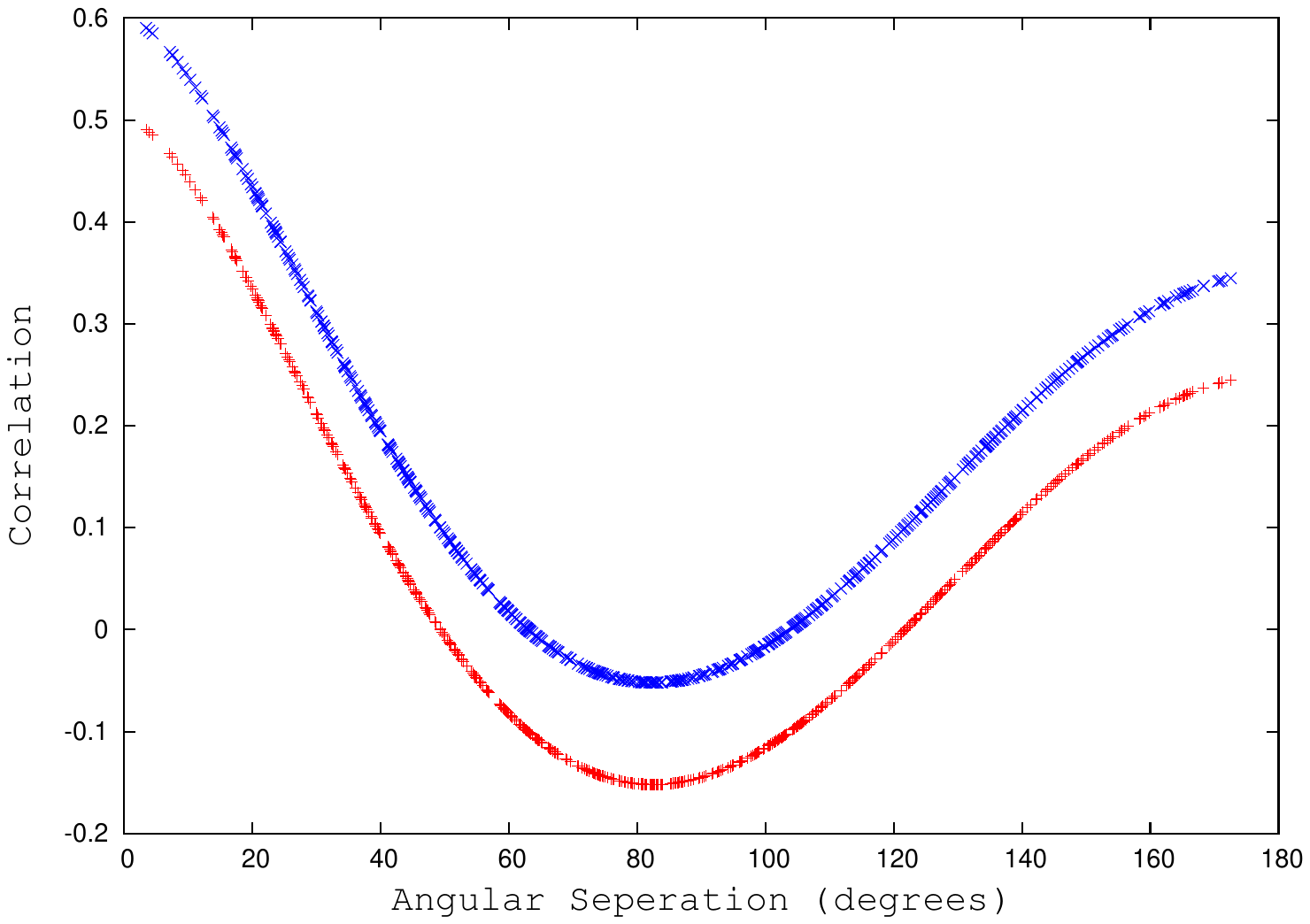} &
\includegraphics[width=60mm]{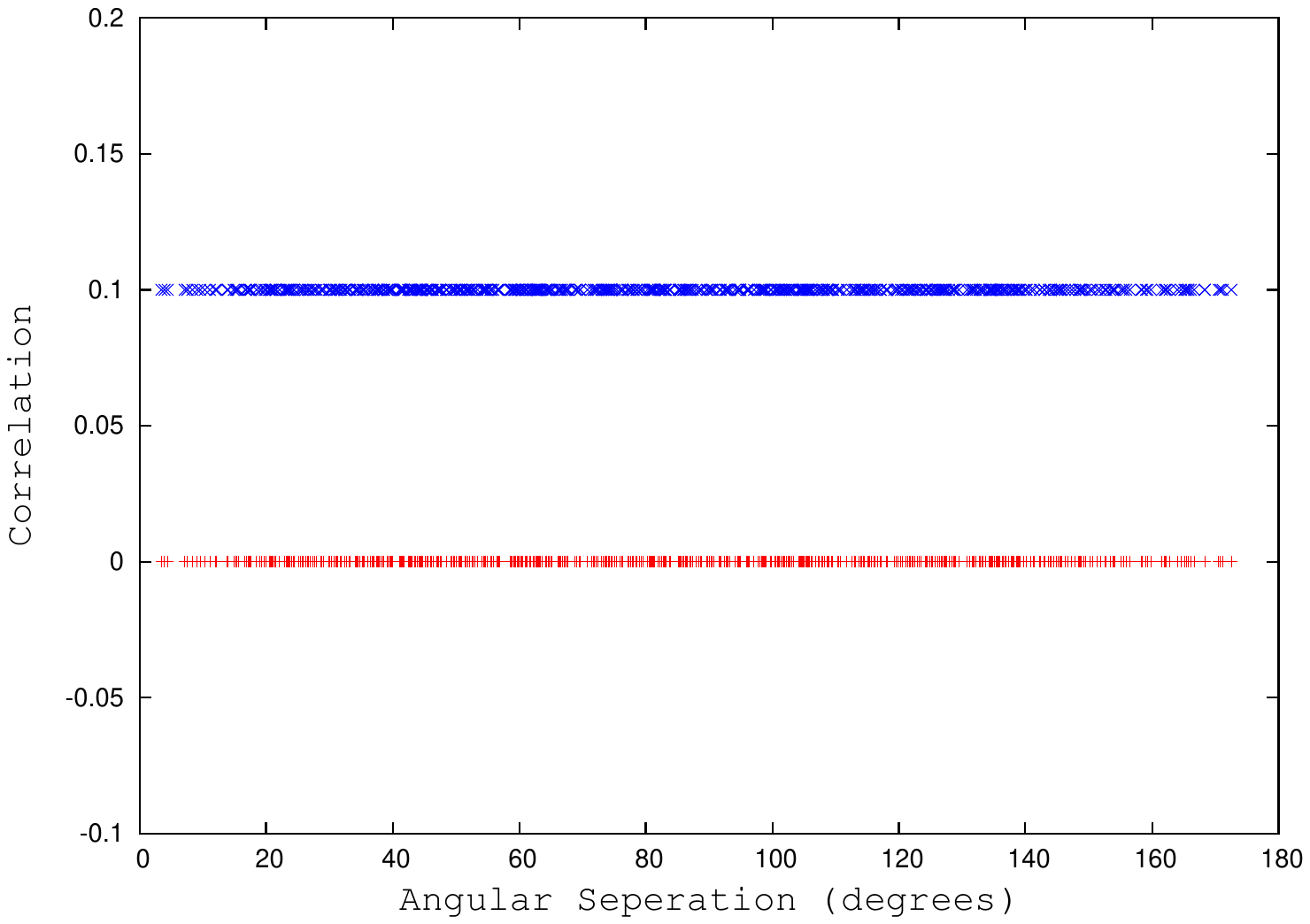} &
\includegraphics[width=60mm]{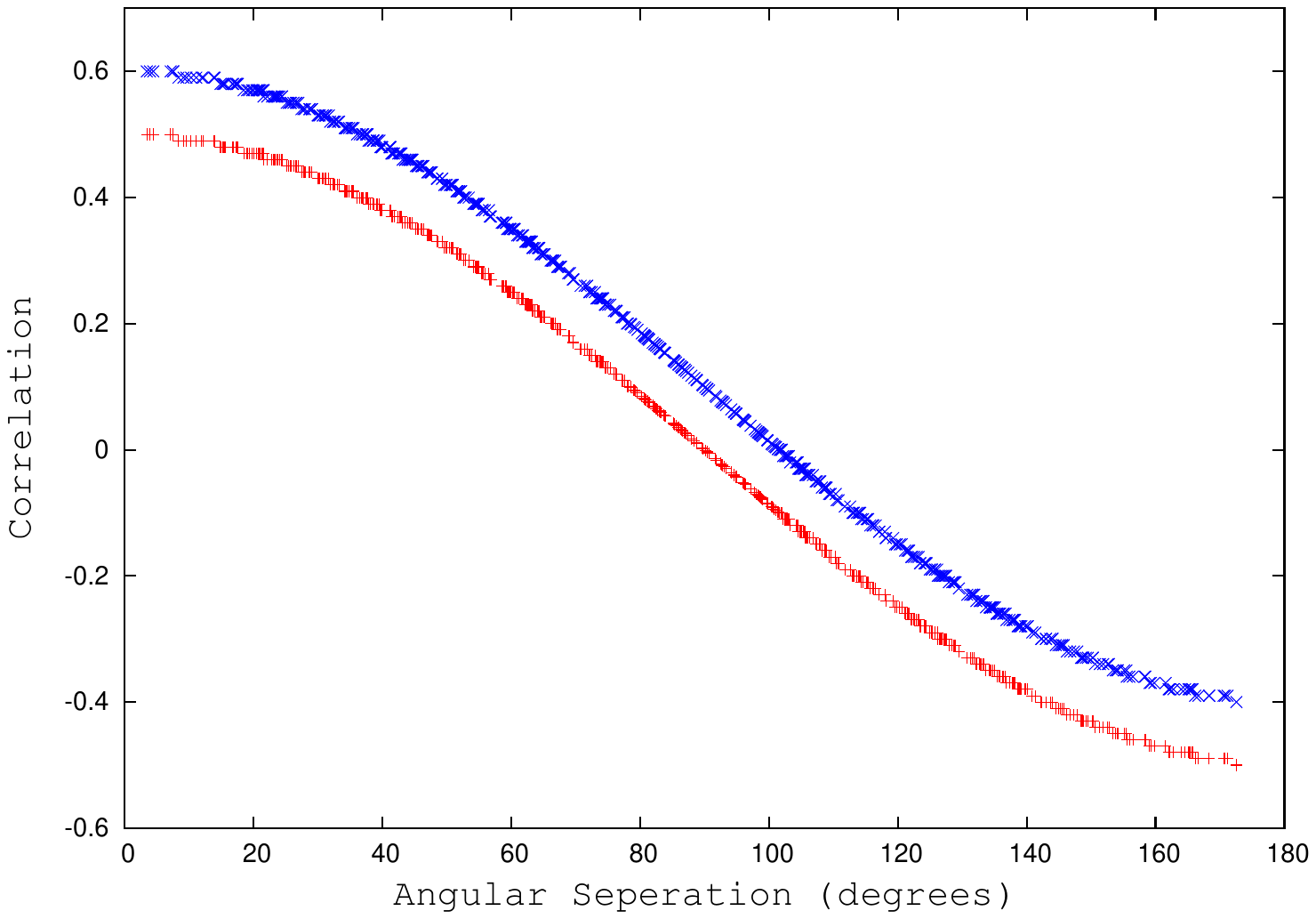} \\
\end{array}$
\end{center}
\hspace*{+3cm}\caption{Demonstration of the parameterisation described in section \ref{Section:CCParam} given the constraints on the parameter space imposed to ensure positive definiteness to reproduce the Hellings-Downs Curve (left), No correlation (middle) and $\cos\theta/2$ (right).  In each case the red line is the analytical evaluation, whilst the blue line is the best fit result.  For clarity we have offset the blue line by 0.1 on the y axis. }
\label{Fig:CCDemo}
\end{minipage}
\end{figure*}

The uniqueness of the spherical parameterisation is then ensured by defining a new set of parameters $\mathbf{\Theta}$ such that:

\begin{equation}
l_{i,j} = \frac{\pi\exp(\Theta_{i,j})}{1+ \exp(\Theta_{i,j})}.
\end{equation}
Whilst in principle this choice of parameterisation should guarantee positive definiteness, in practice machine precision requires that we limit the values that $\Theta_{i,j}$ can take.  Allowing $\mathbf{\Theta}$ to vary beyond $\pm1.5$ results in erroneous behaviour due to this limitation, and so we require that $\mathbf{\Theta}$ lie within the range $\{-1,1\}$, and therefore introduce a final set of parameters $\mathbf{X}$ such that: 

\begin{equation}
\Theta_{i,j} = 2\left(\frac{\exp(X_{i,j})}{1+ \exp(X_{i,j})} - 0.5\right).
\end{equation}
Fig. \ref{Fig:CCDemo} shows the ability for this parameterisation, with these limits in place, to reproduce the Hellings-Downs curve, zero correlation and $\cos\theta/2$ between the pulsars.  We show the analytical expressions in red, whilst the best fit result are in red.  For clarity we have offset the two lines by 0.1 on the y-axis, as the two forms are completely indistinguishable to within machine precision at all points.

\subsection{Performing the sampling using the GHS}

As before in order to perform the sampling with the guided Hamiltonian sampler we will need both the gradients and the hessian for our new likelihood function.  By necessity we are sampling uniformly in the parameter $\mathbf{X}$, however we would like to be sampling uniformly in the parameter space of the correlation coefficients $\mathbf{C}$.  As such we must make a probability transformation so that the prior on our parameters $\mathbf{X}$ will be given by:
\begin{equation}
\mathrm{Pr}\left(\mathbf{X}\right) =  \mathrm{Pr}\left(\mathbf{C})\left|\mathbf{J}(\mathbf{X}\to\mathbf{C}\right)\right|
\end{equation}
where writing the cross-correlation coefficient $C_i$ in terms of its position in the cross-correlation matrix $C_{mn}$ the Jacobian can be written:
\begin{eqnarray}
J_{iq} &=&  \frac{\partial C_i}{\partial X_q}\\
&=&\left[\frac{\partial(\mathbf{L}\mathbf{L}^T)}{\partial l_{q}}\right]_{mn}\frac{\partial l_{q}}{\partial \Theta_{q}}\frac{\partial \Theta_{q}}{\partial X_{q}} \nonumber
\end{eqnarray}
This gives us our new log likelihood expression, which as in Section \ref{Section:Sampling} we write as the negative $\log$, $\Psi$, so that ignoring constant terms:
\begin{eqnarray}
\label{Eq:LogCorr}
\Psi &=& \; \frac{1}{2}\left|\tilde{\mathbf{N}}\right|  + \frac{1}{2}\left|\bm{\varphi}\right| + \frac{1}{2}(\mathbf{\delta t} - F\mathbf{a})^T\tilde{\mathbf{N}}^{-1}(\mathbf{\delta t} - F\mathbf{a})  \nonumber \\
& + &  \frac{1}{2}\mathbf{a}^T\bm{\varphi}^{-1}\mathbf{a} - \left|J\right|
\end{eqnarray}

At first sight calculating the gradient of such an expression with respect to the parameters $\mathbf{X}$ for every likelihood evaluation would seem a formidable computational task.  However, because the $\partial\mathbf{L}/\partial l_{q}$ are all extremely sparse, featuring at most $N_{\mathrm{corr}}$ elements the scaling goes as $\sim$ O$(N_{\mathrm{corr}}^2)$ and thus does not significantly impact the evaluation time.  The gradient and second derivative of $\Psi$ with respect to $\mathbf{X}$  are then of similar form to Eqns \ref{Eq:rhoderiv} and \ref{Eq:rhodderiv} with extra terms corresponding to the derivatives of the Jacobian.

\subsection{Results}
\begin{figure*}
\begin{minipage}{168mm}
\begin{center}$
\begin{array}{c}
\includegraphics[width=100mm]{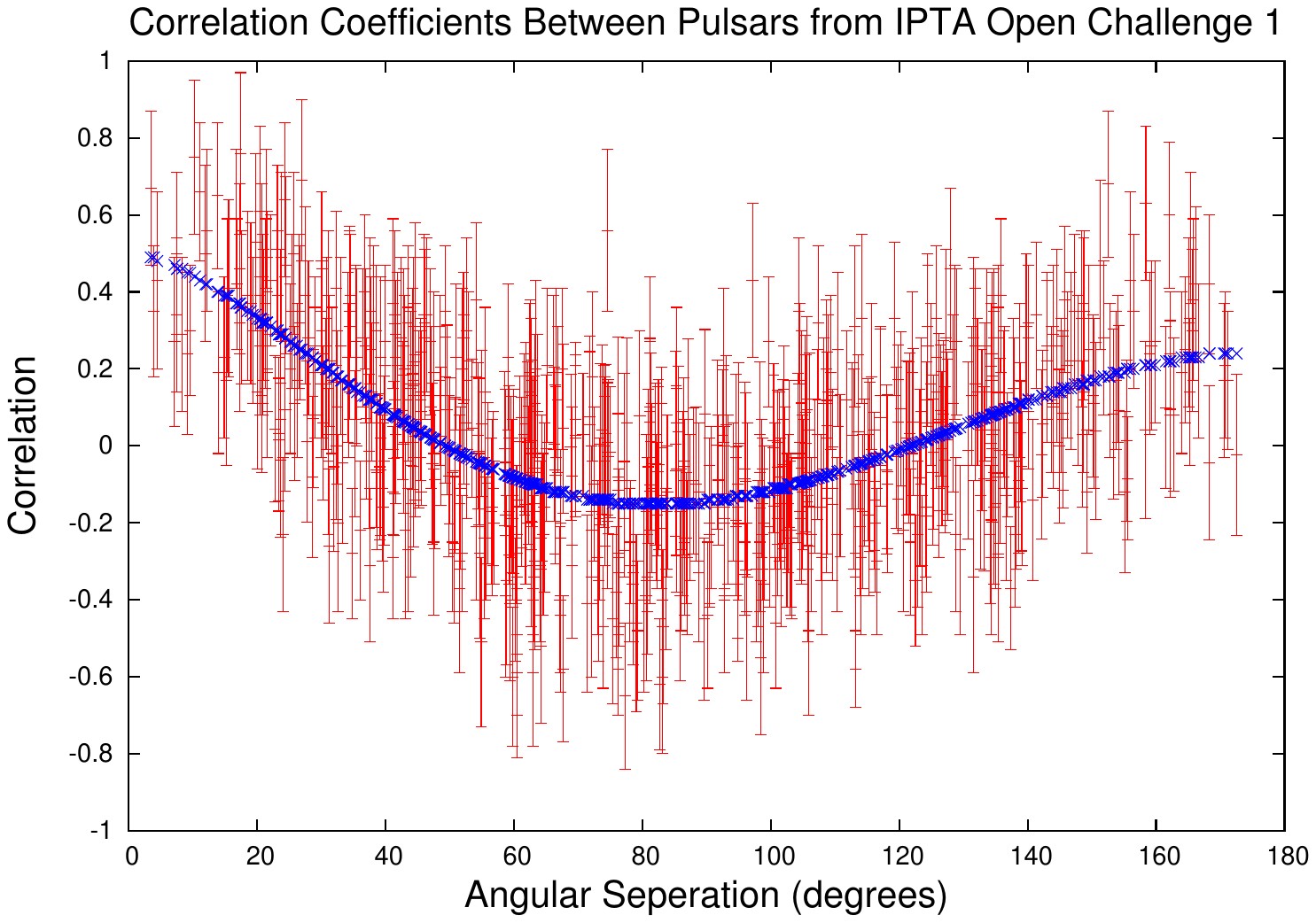} \\
\end{array}$
\end{center}
\caption{Cross correlation coefficients between pairs of pulsars as a function of their angular separation parameterised using the approach in section \ref{Section:CrossCorr}. The blue points represent the analytical values that the Hellings-Downs curve takes for those angular separations. Fitting both the Hellings-Downs curve, and no correlation as potential models results in Chi-sq values of 630 and 1061 respectively, heavily favouring the presence of the Hellings-Downs curve. }
\label{Fig:HDCurve}
\end{minipage}
\end{figure*}
We use this approach on the first open data challenge fitting for both the set of 630 cross-correlation coefficients between the 36 pulsars in the dataset, and 9 GWB coefficients.  Fig. \ref{Fig:HDCurve} shows the cross-correlation coefficients and their associated errors as a function of the angular separation between pairs of pulsars in red, along with the analytical value for the Hellings-Downs curve at those values in blue.  Fitting both the Hellings-Downs curve, and no correlation as potential models results in Chi-sq values of 630 and 1061 respectively, heavily favouring the presence of the Hellings-Downs curve, without having assumed its presence at the point of sampling.  

Clearly this represents the simplest possible case, with no red noise present in the data.  Where red noise is present the ability to recover the Hellings-Downs curve in this manner will inevitably degrade, and it might not prove possible to extract the cross correlation signal in such a completely general way.  In such cases one might wish to reduce the number of free parameters by either assuming a model that has only an angular dependance and binning the coefficients up in angular separation as in \cite{2012arXiv1210.6014T}, or by fitting some more general model that allows for spatial variation and in either case, the extrapolation of this method to these cases is straightforward.

\section{Conclusions}
\label{Section:Conclusions}

We have presented a new model--independent method for analysing pulsar timing array data and estimating the spectral properties of a gravitational wave background. 

We have shown that this method results in a speed up of approximately two orders of magnitude when compared to methods found in vHL2013, and where the signal to noise ratio of the GWB is low, can reduce run times from several hours on a high performance computer to minutes on a regular workstation.  We have accomplished this by sampling either directly from the power spectrum coefficients of the GWB where the number of coefficients to be sampled is small compared to the number of data points in the time domain, or, where the number of coefficients to be sampled increases, from the joint probability density of the power spectrum coefficients for the individual pulsars and the GWB signal realisation, rephrasing the likelihood function to eliminate all matrix-matrix multiplications, and costly dense matrix inversions.  This latter approach therefore scales as O($n\times n_p^3$) where $n$ is the number of frequencies sampled, and $n_p$ is the number of pulsars, as opposed to O($n_o^3$) where $n_o$ is the total number of observations in the dataset across all pulsars.

We have shown this method  requires no prior assumptions to be made regarding the shape of the power spectrum of the GWB.   This is therefore currently the only method that provides a general approach to extracting a GWB signal from pulsar timing data, which we suggest is the only correct way of approaching the problem whilst we have no prior knowledge of the form of the power spectrum.  We have also shown the ability for this method to parameterise correctly the correlation between pairs of pulsars.  This correlation is the defining feature of a GWB signal, and extracting it from the data without first assuming that it is present will thus be a necessary step in any detection process.

Finally we have applied this method both to the first IPTA data challenge, as well as a more realistic pair of simulations and have shown that in all cases it correctly parameterises the properties of the injected signals where they are known, and is consistent with other established methods where they are not known.
\section{Acknowledgements}

This work was performed using the Darwin Supercomputer of the University of Cambridge High Performance Computing Service (http://www.hpc.cam.ac.uk/), provided by Dell Inc. using Strategic Research Infrastructure Funding from the Higher Education Funding Council for England and funding from the Science and Technology Facilities Council.

\appendix
\section{Guided Hamiltonian Sampling}
\label{Appendix:GHS}

The following is a description of both Hamiltonian Monte Carlo and the Guided Hamiltonian Sampler as described in B13.

\subsection{Standard Hamiltonian Monte Carlo (HMC) sampling}

In HMC, one begins by defining the potential energy $\psi(\bm{x})$ 
of the target density \(\pr{\bm{x}}\) as its negative logarithm, namely
\begin{equation}
  \label{eq:psi}
  \psi \left(\bm{x}\right) = - \ln \pr{\bm{x}}.
\end{equation}
For each parameter, \(x_i\) we then introduce a `momentum' parameter \(p_i\) and
a constant `mass' \(m_i\) and construct a kinetic energy term that, when added
to the potential, leads to the Hamiltonian
\begin{equation}
  \label{eq:ham}
  {\cal H}(\bm{x},\bm{p}) 
= \sum_i \frac{p_i^2}{2m_i} + \psi \left(\bm{x}\right).
\end{equation}
Our new objective is to draw samples from a distribution that is
proportional to $\exp \left[ -{\cal H}(\bm{x},\bm{p}) \right]$. 
The form of the Hamiltonian
is such that this distribution is separable into a Gaussian in
\(\bm{p}\) and the target distribution, i.e.
\begin{equation}
  \label{eq:hmchamsep}
  \exp \left[ -{\cal H}(\bm{x},\bm{p}) \right] =  \pr{\bm{x}}
  \prod_i \exp \left( - \frac{p_i^2}{2m_i} \right).
\end{equation}
We can then obtain samples from \(\pr{\bm{x}}\) by marginalising
over \(\bm{p}\).

To find a new sample we first draw a set of momenta from the
distribution defined by our kinetic energy term, i.e.  an \(N\)
dimensional uncorrelated Gaussian with a variance in dimension \(i\)
of \(m_i\). We then allow our system to evolve deterministically, from
our starting point \(\left(\bm{x}, \bm{p}\right)\) in the phase
space for some fixed time \(\tau\) according to Hamilton's equations,
\begin{align}
  \label{eq:hamiltons}
  \frac{\mathrm{d} \bm{x}}{\mathrm{d} t}& =  \nabla_{\!\bm{p}}{\cal H}(\bm{x},\bm{p}) \\
  \frac{\mathrm{d} \bm{p}}{\mathrm{d} t}& = - \nabla_{\!\bm{x}} {\cal H}(\bm{x},\bm{p}) =
  -\nabla_{\!\bm{x}} \psi \left(\bm{x}\right).
\end{align}
At the end of this trajectory we have reached the point
\(\left(\bm{x}^\prime, \bm{p}^\prime \right)\) and we accept
this point with probability
\begin{equation}
  \label{eq:acceptance}
  p_A = \min \left[ 1, \exp \left( -\delta {\cal H} \right) \right],
\end{equation}
where
\begin{equation}
  \label{eq:dH}
  \delta {\cal H} = {\cal H}\left(\bm{x}^\prime, \bm{p}^\prime\right) - 
   {\cal H}\left(\bm{x}, \bm{p}\right).
\end{equation}
This implies that if we are able to integrate Hamilton's equations
exactly then, as energy is conserved along such a trajectory, the
probability of acceptance is unity. In practice, however, numerical
inaccuracies mean that this is not the case.  After a new proposed
sample is generated the momentum variable is discarded and the process
restarts by randomly drawing a new set of momenta as described above.

In fact the method is more general than outlined above since, provided
one uses the Metropolis acceptance criterion (Eq. \ref{eq:acceptance}), it
is permitted to follow any trajectory to generate a new candidate
point. However only trajectories that approximately conserve the value
of the Hamiltonian (Eq. \ref{eq:ham}) will result in high acceptance
rates. For some problems it may be advantageous to generate
trajectories using an approximate Hamiltonian that can be computed
rapidly, and bear the cost of lowering the acceptance probability.

To integrate the equations of motions it is common practice to use the
leapfrog method \citep{Neal1996}. This method has the property of
exact reversibility which is required to ensure the chain satisfies
detailed balance. It is also numerically robust and allows for the
simple propagation of errors. We make \(n\) steps with a finite step
size \(\epsilon\), such that \(n\epsilon = \tau\), as follows,
\begin{align}
\label{eqn::ch1_leap_forg_scalar_1}
\bm{p}\left(t+\frac{\epsilon}{2}\right) & = \bm{p}(t)+\frac{\epsilon}{2}\frac{\mathrm{d}\bm{p}}{\mathrm{d}t}\Big|_{t} \\
\label{eqn::ch1_leap_forg_scalar_2}
\bm{x}(t+\epsilon) &= \bm{x}(t)+\epsilon\frac{\mathrm{d}\bm{x}}{\mathrm{d}t}\Big|_{t+\frac{\epsilon}{2}}\\
\label{eqn::ch1_leap_forg_scalar_3}
\bm{p}(t+\epsilon) & = \bm{p}\left(t+\frac{\epsilon}{2}\right)+\frac{\epsilon}{2}\frac{\mathrm{d}\bm{p}}{\mathrm{d}t}\Big|_{t+\epsilon}.
\end{align}
until \(t = \tau\). Substituting for the time derivatives using
Hamilton's equations (\ref{eq:hamiltons}), one thus obtains explicit
relations for the leapfrog steps, which read
\begin{align}
\label{eqn::ch1_leap_forg_scalar_1a}
\bm{p}\left(t+\frac{\epsilon}{2}\right) & = \bm{p}(t)-\frac{\epsilon}{2}\nabla_{\!\bm{x}} {\cal H}\Big|_{t} \\
\label{eqn::ch1_leap_forg_scalar_2a}
\bm{x}(t+\epsilon) &= \bm{x}(t)+\epsilon\nabla_{\!\bm{p}} {\cal H}\Big|_{t+\frac{\epsilon}{2}}\\
\label{eqn::ch1_leap_forg_scalar_3a}
\bm{p}(t+\epsilon) & = \bm{p}\left(t+\frac{\epsilon}{2}\right)-\frac{\epsilon}{2}\nabla_{\!\bm{x}} {\cal H}\Big|_{t+\epsilon}.
\end{align}
The interval \(\tau\) must be varied, usually by
drawing \(n\) and \(\epsilon\) randomly from uniform distributions, to
avoid resonant trajectories; we therefore draw $n$ and $\epsilon$ from
$\mathrm{U}(1,n_{\rm max})$, $\mathrm{U}(0,\epsilon_{\rm max})$, respectively.
The leapfrog method may be replaced by higher-order integration
schemes provided exact reversibility is
maintained; such methods yield greater accuracy, although
generally incur significant additional computational costs.

\subsection{Setting masses in HMC}

HMC can be extremely sensitive to the choice of masses, in particular
when the marginal distributions of different parameters show
considerable variation in width the masses.  \cite{2001SPIE.4322..456H}
suggests that one should set the mass associated with each parameter
to be approximately equal to the variance of that parameter in the
target density. This is an attempt to circularise the trajectories in
the $(\bm{x}, \bm{p})$ space. Interestingly, \cite{Neal1996}
suggests precisely the opposite approach, where the mass for a
parameter is inversely proportional to the width of the
distribution. 

\cite{2008MNRAS.389.1284T} follow the latter suggestion and justify it by
generalising the framework in \cite{Neal1993} to describe the
application of the leapfrog method. In particular, for the case where
the $N$-dimensional target distribution $\pr{\bm{x}}$ is (well
approximated by) a multivariate Gaussian with covariance matrix
$\mathbf{C}$, they show that the leapfrog method is stable if $\mathbf{M} =
\mathbf{C}^{-1}$ and $\epsilon \le 2$, where $\mathbf{M}$ is the $N\times N$
`mass matrix' that appears in the generalised kinetic term
$\tfrac{1}{2}\bm{p}^{\rm t}\mathbf{M}^{-1}\bm{p}$ of the
Hamiltonian. 

If the dimensionality of the problem is such that it is impractical to
perform the required matrix inversion and decomposition of \(\mathbf{M}\)
(to compute the Hamiltonian and to draw new values for the momentum
variables respectively) then simple approximations must be
employed. Typically one might construct a diagonal mass matrix with
the mass associated with each parameter inversely proportional to the
variance of that parameter.

Moreover, if the target distribution is not Gaussian, it seems
reasonable to use some appropriate measure of the width of the
distribution, such as the curvature at the peak \cite{Neal1996}, to
set the masses.

\subsection{Guided Hamiltonian sampling}

Guided Hamiltonian sampling (GHS) builds on the ideas explored in
\cite{2008MNRAS.389.1284T} to produce an HMC algorithm with just a single
adjustable parameter, thereby eliminating the need for tuning
masses. In particular, GHS takes advantage of, although does not rely
on, the fact that one often wishes to sample from a target
distribution that is unimodal, albeit, in general, non-Gaussian and
high-dimensional. 

In GHS, one first sets the mass matrix in the kinetic term of the
Hamiltonian to the identity, $\mat{M}=\mat{I}$. For the target
distribution $\pr{\bm{x}}$, one then locates the peak
$\hat{\bm{x}}$, typically using some iterative gradient-search
optimisation algorithm starting from, in general, some random initial
point.  One then calculates the Hessian (or curvature) matrix
$\hat{\mat{H}}$ of $\ln\pr{\bm{x}} = -\psi(\bm{x})$ (i.e. the negative of 
the potential energy, for convenience of sign conventions) at the maximum,
either analytically or using numerical differentiation; this thereby
defines a Gaussian approximation to $\pr{\bm{x}}$ in the
neighbourhood of the peak $\hat{\bm{x}}$. 

Once the Hessian at the peak has been calculated, one then determines
its $N$ eigenvalues $\lambda_i$ and $N$ normalised eigenvectors
$\hat{\bm{e}}_i$.  Denoting the matrix containing these normalised
eigenvectors as its columns by $\bm{S}$, one first defines a new
set of variables $\bm{x}' = \bm{S}^{\rm t}\bm{x}$ in which
the Hessian becomes diagonal with the eigenvalues $\lambda_i$ as its
diagonal entries. One then rescales each $x_i'$ to obtain a new set of
variables $y_i = \sqrt{\lambda_i} \,x'_i/\eta$, where the scaling
factor $\eta$ is the single adjustable parameter in GHS, which we will
discuss later. It is straightforward to show that the new variables
are related to the original variables by
\begin{equation}
\bm{y} = \frac{1}{\eta}\hat{\mat{H}}^{1/2}\bm{x}.
\label{eqn:ydef}
\end{equation}
Consequently, in the new variables, the Hessian at the peak has the
trivial form $\eta^2\mat{I}$. One then performs Hamiltonian sampling
employing the standard leapfrog method
(\ref{eqn::ch1_leap_forg_scalar_1a}--\ref{eqn::ch1_leap_forg_scalar_3a}),
but in terms of the new variables $\bm{y}$, rather than
$\bm{x}$. Thus, GHS may be considered simply as standard HMC, but
performed in a set of variables (or coordinates) that are tailored to
the target distribution, namely the scaled eigendirections of the
Hessian at its peak. Consequently, although GHS may take advantage if
$\pr{\bm{x}}$ possesses a single well-defined peak (with zero
gradient), it does not rely on this, since it retains the generality
of standard HMC.

Rather than working in terms of the new variables $\bm{y}$, one
can, if desired, return to using the original variables $\bm{x}$,
in which case the relation (\ref{eqn:ydef}) shows that the leapfrog
steps take the modified form
\begin{align}
\label{eqn::ch1_leap_forg_scalar_1b}
\bm{p}\left(t+\frac{\epsilon}{2}\right) & = \bm{p}(t)-\frac{\epsilon}{2}\eta\hat{\mat{H}}^{-1/2}\nabla_{\!\bm{x}} {\cal H}\Big|_{t} \\
\label{eqn::ch1_leap_forg_scalar_2b}
\bm{x}(t+\epsilon) &= \bm{x}(t)+\epsilon\eta\hat{\mat{H}}^{-1/2}\nabla_{\!\bm{p}} {\cal H}\Big|_{t+\frac{\epsilon}{2}}\\
\label{eqn::ch1_leap_forg_scalar_3b}
\bm{p}(t+\epsilon) & = \bm{p}\left(t+\frac{\epsilon}{2}\right)-\frac{\epsilon}{2}\eta\hat{\mat{H}}^{-1/2}\nabla_{\!\bm{x}} {\cal H}\Big|_{t+\epsilon}.
\end{align}

Using the original variables $\bm{x}$ or the new variables
$\bm{y}$, it is necessary to calculate either the (inverse)
square-root of the $N\times N$ Hessian matrix $\hat{\mat{H}}$ at the
peak, or (equivalently) its eigendecomposition (and, subsequently, the
calculation of the square-roots of its eigenvalues).   Performing the above calculations can be computationally
expensive, particular for large $N$, although it should be noted that
one need only perform these calculations once.

In summary, GHS aims to increase the efficiency of standard HMC,
particularly for high-dimensional, unimodal target distributions, by
performing the sampling in the principal coordinates defined by the
Gaussian approximation at its peak. In this way, one may largely
eliminate the tuning aspect of HMC: the single remaining adjustable
parameter is the scaling $\eta$, the optimal value of which depends on
the dimensionality of the parameter space, and should be chosen such that the acceptance rate is approximately 68$\%$.

\section{Analytical Approximation to the Likelihood}
\label{Section:Appendix2}

\subsubsection{Uniform White Noise}

Suppose we have a single realisation of some time series data $\mathbf{d}$ of length $N$.  We then define a set of hypotheses $\{H\}$ such that each $H_i$ purports that our data $\mathbf{d}$ is described by some function $f_i$ where:

\begin{equation}
f_i(t) = \sum_{k=1}^m b_kM_k(t,\mathbf{w})
\end{equation} 
with $M_k$ a set of general basis functions.   The number of functions $m$, the parameters that describe them (e.g. their frequencies) $\mathbf{w}$, and the model coefficients $b_k$ are allowed to vary for each $f_i$.  
We then transform this set of basis functions into an orthonormal set $F_k$ through the transformation:
\begin{equation}
F_k(t) = \frac{1}{\sqrt{\lambda_k}}\sum_{j=1}^me_{kj}M_j(t)
\end{equation}
where $e_{kj}$ is the $k$th element of the $j$th eigenvector and $\lambda_k$ is the $k$th eigenvalue of the covariance matrix $\mathbf{M}^\mathrm{T}\mathbf{M}$. 
Our function $f_i$ can now be written in terms of these new basis vectors:
\begin{equation}
f_i(t) = \sum_{k=1}^m a_kF_k(t,\mathbf{w})
\end{equation} 
where the coefficients $a$ in the orthonormal basis are related to the coefficients $b$ in the original basis through:
\begin{equation}
b_k=\sum_{j=1}^m\frac{a_ke_{jk}}{\sqrt{\lambda_j}}
\end{equation}
The probability of the data given a model $f_i$, assuming that the noise is described by a zero mean random Gaussian process with variance $\sigma$, is given by:

\begin{equation}
\label{Eq:Dataprob}
\mathrm{Pr}(\mathbf{d} | \mathbf{a}, \mathbf{w},\sigma,f_i) = (2\pi\sigma^2)^{-N/2}\exp\left[\frac{1}{2\sigma^2}\sum_{k=1}^N\left[d_k - f_i(t_k)\right]^2\right].
\end{equation}
Writing the projection of the data onto our basis functions as

\begin{equation}
h_i = \sum_{k=1}^N d_kF_i(t_k),
\end{equation}
and writing $\mathbf{d}^2=\mathbf{d}^T\mathbf{d}$ Eq:\ref{Eq:Dataprob} can be written:

\begin{eqnarray}
\mathrm{Pr}(\mathbf{d} | \mathbf{a}, \mathbf{w},\sigma,f_i) &=& (2\pi\sigma^2)^{-N/2}  \\
& \times& \exp\left[-\frac{1}{2\sigma^2}\left[\mathbf{d}^2 - \sum_{l=1}^m2a_lh_l+ a_l^2\right]\right] \nonumber
\end{eqnarray}
We begin by integrating over both the set of coefficients $\mathbf{a}$ and frequencies $\mathbf{w}$.  We assume that the two parameters are logically independent, in so far as we can write the priors:

\begin{equation}
\mathrm{Pr}(\mathbf{a},\mathbf{w}) = \mathrm{Pr}(\mathbf{a})\mathrm{Pr}(\mathbf{w}) 
\end{equation}
For the amplitude coefficients, we choose an uninformative Gaussian prior given by:
\begin{equation}
\label{Eq:GaussPrior}
\mathrm{Pr}(\mathbf{a} | \delta) = (2\pi\delta^2)^{-m/2}\exp\left[-\sum_{k=1}^m\frac{a_k^2}{2\delta^2}\right]
\end{equation}
with $\delta >> \sigma$.  Therefore, our probability, marginalised over $\mathbf{a}$ and $\mathbf{w}$ can be written:

\begin{eqnarray}
\mathrm{Pr}(\mathbf{d} | \delta,\sigma,f_i) &=& \int \mathrm{d}\mathbf{w}\mathrm{Pr}(\mathbf{w})(2\pi\delta^2)^{-m/2} (2\pi\sigma^2)^{-N/2}  \nonumber \\
&\times& \int_{-\infty}^{+\infty}\mathrm{d}a_1\ldots\mathrm{d}a_m \exp\left[-\sum_{k=1}^m\frac{a_k^2}{2\delta^2}\right]\nonumber\\
&\times & \exp\left[-\frac{1}{2\sigma^2}\left[\mathbf{d}^2 - \sum_{l=1}^m2a_lh_l+ a_l^2\right]\right]
\end{eqnarray}

We have chosen $\delta$ such that the prior term $ \exp\left[-\sum_{k=1}^m a_k^2/2\delta^2\right]$ is constant where the likelihood is large, but goes to zero sufficiently quickly outside this region so as to be normalisable.  Therefore, if we define $\hat{a_i}$ to be the maximum likelihood value for the parameter $a_i$, we can write our probability as:
\begin{eqnarray}
\mathrm{Pr}(\mathbf{d} | \delta,\sigma,f_i) &=& \int \mathrm{d}\mathbf{w}\mathrm{Pr}(\mathbf{w})(2\pi\delta^2)^{-m/2} (2\pi\sigma^2)^{-N/2}  \nonumber \\
&\times& \exp\left[-\sum_{k=1}^m\frac{\hat{a_k}^2}{2\delta^2}\right]\int_{-\infty}^{+\infty}\mathrm{d}\mathbf{a}\nonumber\\
&\times &  \exp\left[-\frac{1}{2\sigma^2}\left[\mathbf{d}^2 - \sum_{l=1}^m2a_lh_l+ a_l^2\right]\right]
\end{eqnarray}
If we take the elements of $\mathbf{a}$ to be independant on our orthnormal basis, then we can write the expectation value of a single element $a_i$ as:
\begin{equation}
\left<a_i\right> = \frac{\int_{-\infty}^{+\infty}\mathrm{d}a_ia_i\exp\left[\frac{-1}{2\sigma^2}\left[ -2a_ih_i+ a_i^2\right]\right]}{\int_{-\infty}^{+\infty}\mathrm{d}a_i\exp\left[\frac{-1}{2\sigma^2}\left[ -2a_ih_i+ a_i^2\right]\right]}
\end{equation}
which evaluates to $\left<a_i\right> =  h_i$.  I.e. the expectation value of the basis vector coefficient is just the projection of the data onto that basis.
Substituting this into our equation for the probability in the place of $\hat{a}$ and performing the Gaussian integral over $\mathbf{a}$ we arrive at the expression:  
\begin{eqnarray}
\mathrm{Pr}(\mathbf{d} | \delta,\sigma,f_i) &=& \int \mathrm{d}\mathbf{w}\mathrm{Pr}(\mathbf{w})(2\pi\delta^2)^{-m/2} (2\pi\sigma^2)^{-(N-m)/2}  \nonumber \\
&\times & \exp\left[\frac{\mathbf{d}^2 - \mathbf{h}^2}{2\sigma^2}\right] \exp\left[\frac{\mathbf{h}^2}{2\delta^2}\right].
\end{eqnarray}
For our integral over our frequencies, we are for any given model $f_i$ considering a set of frequencies chosen from an evenly spaced grid.  Therefore we will have a set of delta function priors for each frequency $w_j$ in the set $\mathbf{w}$ and the integral can be simply evaluated:
\begin{eqnarray}
\mathrm{Pr}(\mathbf{d} | \delta,\sigma,f_i) &=& (2\pi\delta^2)^{-m/2} (2\pi\sigma^2)^{-(N-m)/2}  \nonumber \\
&\times & \exp\left[\frac{\mathbf{d}^2 - \mathbf{h(w_i)}^2}{2\sigma^2}\right] \exp\left[\frac{\mathbf{h(w_i)}^2}{2\delta^2}\right].
\end{eqnarray}
We are now in a position to integrate over our unknown variances $\sigma$ and $\delta$.   As in \citet{Bretthorst1988} we set an upper bound $H$ and lower bound $L$ to this integral, which will therefore be of the form:
\begin{equation}
\frac{1}{\log(H/L)}\int^H_L \mathrm{d}s\frac{s^{-a}\exp\left[-\frac{Q}{s^2}\right]}{s}
\end{equation}
making a substitution $u = Q/s^2$ this becomes:
\begin{equation}
\frac{Q^{-a/2}}{2\log(H/L)}\int^{Q/L^2}_{Q/H^2} \mathrm{d}u\;u^{a/2-1}\exp\left[-u\right]
\end{equation}
If we assume that $H$ is sufficiently large, and $L$ is sufficiently small that we may write $Q/H^2 << 1$ and $a/2 -1 << Q/L^2$ then the integral will evaluate to approximately $\Gamma(a/2)$.  Thus our integral over $\delta$ will become:

\begin{equation}
\frac{1}{\log(H/L)}\int^H_L \mathrm{d}\delta\frac{\delta^{-m}\exp\left[-\frac{\mathbf{h^2}}{2\delta^2}\right]}{\delta} \approx \frac{\Gamma(m/2)}{2\log(R_{\delta})}\left[\frac{\mathbf{h(w)^2}}{2}\right]^{-m/2}
\end{equation}
and similarly for $\sigma$ the integral evaluates to approximately:
\begin{equation}
\frac{\Gamma((N-m)/2)}{2\log(R_{\sigma})}\left[\frac{\mathbf{d^2} - \mathbf{h(w)^2}}{2}\right]^{-(N-m)/2}.
\end{equation}
Therefore we can finally write the probability of the data $D$ given a model $f_i$ as:
\begin{eqnarray}
\mathrm{Pr}(\mathbf{d} | f_i) &=&  \frac{\Gamma(m/2)}{2\log(R_{\delta})}\left[\frac{\mathbf{h(w)^2}}{2}\right]^{-m/2}\nonumber \\
&\times &\frac{\Gamma((N-m)/2)}{2\log(R_{\sigma})}\left[\frac{\mathbf{d^2} - \mathbf{h(w)^2}}{2}\right]^{-(N-m)/2}.
\end{eqnarray}

\subsubsection{Non-Uniform White Noise}

In general when dealing with pulsar residuals the white noise level across a dataset for a single pulsar will vary with time, where for example different instruments have been used to collect data for the same pulsar.  In this case the expansion of our likelihood function is not so simple, because the covariance matrix $\mathbf{G}^T\mathbf{N}\mathbf{G}$ will no longer reduce to a diagonal matrix.  If we define $\mathbf{C}=\mathbf{G}^T\mathbf{N}\mathbf{G}$ where we consider $\mathbf{C}$ to be a general dense covariance matrix, Eq:\ref{Eq:Dataprob} will take the form:
\begin{eqnarray}
\mathrm{Pr}(\mathbf{d} | \mathbf{a}, \mathbf{w},f_i) &=& (2\pi)^{-N/2}|\mathbf{C}|^{-1/2} \nonumber \\
&\times &\exp\left[\frac{-1}{2}(\mathbf{d} - \mathbf{F}\mathbf{a})^T\mathbf{C}^{-1}(\mathbf{d} - \mathbf{F}\mathbf{a})\right].
\end{eqnarray}
In this case, writing $\mathbf{F_i}^T\mathbf{C}^{-1}\mathbf{F_i} = \mathbf{C}_i^{-1}$ the maximum likelihood value of a particular coefficient $a_i$ will be given by
\begin{equation}
\left<a_i\right> = \frac{\int_{-\infty}^{+\infty}\mathrm{d}a_ia_i\exp\left[-\frac{1}{2}\left[ a_i\mathbf{C}_i^{-1}a_i - 2\mathbf{d}^T\mathbf{C}^{-1}\mathbf{F_i}a_i\right]\right]}{\int_{-\infty}^{+\infty}\mathrm{d}a_i\exp\left[-\frac{1}{2}\left[ a_i\mathbf{C}_i^{-1}a_i - 2\mathbf{d}^T\mathbf{C}^{-1}\mathbf{F_i}a_i\right]\right]}
\end{equation}
and evaluates to:
\begin{equation}
\left<a_i\right> = \frac{\mathbf{d}^T\mathbf{C}^{-1}\mathbf{F_i}}{\mathbf{C}_i^{-1}}.
\end{equation}
In the case that $\mathbf{C}$ once again describes uniform white noise across the observation this will reduce to $\left<a_i\right> = \mathbf{d}^T\mathbf{F_i} = h_i$ as before.
Using the same uninformative prior on our coefficients as in Eq.\ref{Eq:GaussPrior} we can then write our integral over the basis coefficients as:
\begin{eqnarray}
\mathrm{Pr}(\mathbf{d} | \delta, f_i) &=& (2\pi)^{-N/2}|\mathbf{C}|^{-1/2}(2\pi\delta^2)^{-m/2} \nonumber \\
&\times& \exp\left[-\frac{1}{2\delta^2}\sum_{k=1}^m\left(\frac{\mathbf{d}^T\mathbf{C}^{-1}\mathbf{F_i}}{\mathbf{F_i}^T\mathbf{C}^{-1}\mathbf{F_i}}\right)^2\right]  \\
&\times&\int_{-\infty}^{+\infty}\mathrm{d}\mathbf{a}\exp\left[\frac{-1}{2}(\mathbf{d} - \mathbf{F}\mathbf{a})^T\mathbf{C}^{-1}(\mathbf{d} - \mathbf{F}\mathbf{a})\right] \nonumber
\end{eqnarray}
If we define:
\begin{equation}
\mathbf{\chi} = (\mathbf{F}^T\mathbf{C}^{-1}\mathbf{F})^{-1}\mathbf{F}^T\mathbf{C}^{-1}\mathbf{d},
\end{equation}
then we can re-express this probability as: 
\begin{eqnarray}
\mathrm{Pr}(\mathbf{d} | \delta, f_i) &=& (2\pi)^{-N/2}|\mathbf{C}|^{-1/2}(2\pi\delta^2)^{-m/2} \nonumber \\
&\times& \exp\left[-\frac{1}{2\delta^2}\sum_{k=1}^m\left(\frac{\mathbf{d}^T\mathbf{C}^{-1}\mathbf{F_i}}{\mathbf{F_i}^T\mathbf{C}^{-1}\mathbf{F_i}}\right)^2\right]  \nonumber\\
&\times&\exp\left[-\frac{1}{2}\mathbf{d}^T\mathbf{C}^{-1}\mathbf{d}\right]\exp\left[\frac{1}{2}\mathbf{\chi}^T\mathbf{F}^T\mathbf{C}^{-1}\mathbf{F}\mathbf{\chi}\right] \nonumber \\
&\times& \int_{-\infty}^{+\infty}\mathrm{d}\mathbf{a}\exp\left[-\frac{1}{2}(\mathbf{a} - \mathbf{\chi})^T\mathbf{F}^T\mathbf{C}^{-1}\mathbf{F}(\mathbf{a} - \mathbf{\chi})\right] ,\nonumber
\end{eqnarray}
which evaluates to:
\begin{eqnarray}
\mathrm{Pr}(\mathbf{d} | \delta, f_i) &=& ((2\pi)^{N-m}|\mathbf{C}||\mathbf{F}^T\mathbf{C}^{-1}\mathbf{F}|)^{-1/2}(2\pi\delta^2)^{-m/2} \nonumber \\
&\times& \exp\left[-\frac{1}{2\delta^2}\sum_{k=1}^m\left(\frac{\mathbf{d}^T\mathbf{C}^{-1}\mathbf{F_i}}{\mathbf{F_i}^T\mathbf{C}^{-1}\mathbf{F_i}}\right)^2\right]  \\
&\times&\exp\left[-\frac{1}{2}\mathbf{d}^T\mathbf{C}^{-1}\mathbf{d}\right]\exp\left[\frac{1}{2}\mathbf{\chi}^T\mathbf{F}^T\mathbf{C}^{-1}\mathbf{F}\mathbf{\chi}\right]. \nonumber
\end{eqnarray}
Thus far we have assumed we know the level of the noise in $\mathbf{C}$ exactly, however in general we would like to fit for a global scaling factor that modifies the overall noise level in the dataset. I.e. we would like to write $\mathbf{C'}=\mathbf{G}^T(\alpha^2\mathbf{N})\mathbf{G}$ where $\alpha$ is a constant to be determined. Including this in our probability we can write:
\begin{eqnarray}
\mathrm{Pr}(\mathbf{d} | \alpha,\delta, f_i) &=& ((2\pi\alpha)^{(N-m)}|\mathbf{C}||\mathbf{F}^T\mathbf{C'}^{-1}\mathbf{F}|)^{-1/2}(2\pi\delta^2)^{-m/2} \nonumber \\
&\times& \exp\left[-\frac{1}{2\delta^2}\sum_{k=1}^m\left(\frac{\mathbf{d}^T\mathbf{C'}^{-1}\mathbf{F_i}}{\mathbf{F_i}^T\mathbf{C'}^{-1}\mathbf{F_i}}\right)^2\right]  \\
&\times&\exp\left[-\frac{1}{2\alpha^2}\left(\mathbf{d}^T\mathbf{C'}^{-1}\mathbf{d} -\mathbf{\chi}^T\mathbf{F}^T\mathbf{C'}^{-1}\mathbf{F}\mathbf{\chi}\right)\right]. \nonumber
\end{eqnarray}
We can then finally proceed as before integrating over both $\alpha$ and $\delta$ to arrive at the final probability
\begin{eqnarray}
\mathrm{Pr}(\mathbf{d}| f_i) &=& \frac{\Gamma(m/2)}{2\log(R_{\delta})}\left[\frac{1}{2}\sum_{k=1}^m\left(\frac{\mathbf{d}^T\mathbf{C'}^{-1}\mathbf{F_i}}{\mathbf{F_i}^T\mathbf{C'}^{-1}\mathbf{F_i}}\right)^2\right]^{-m/2} \nonumber \\
&\times& \frac{\Gamma((N-m)/2)}{2\log(R_{\alpha})}\left[-\frac{1}{2}\left(\mathbf{d}^T\mathbf{\bar{C}}^{-1}\mathbf{d}\right)\right]^{-(N-m)/2}
\end{eqnarray}
where we have defined:
\begin{equation}
\mathbf{\bar{C}}^{-1}=\mathbf{C'}^{-1} - \mathbf{C'}^{-1} \mathbf{F}( \mathbf{F}^T \mathbf{C'}^{-1} \mathbf{F})^{-1} \mathbf{F}^T \mathbf{C'}^{-1}.
\end{equation}

\end{document}